\documentclass[
 reprint,
 onecolumn,
 superscriptaddress,
 amsmath,amssymb,
 aps,
 prx
]{revtex4-2}

\usepackage{graphicx} 
\usepackage{bm} 
\usepackage[english]{babel}
\usepackage{seqsplit} 
\usepackage{braket}
\usepackage{xcolor}
\usepackage{tabularray}
\usepackage[normalem]{ulem}
\usepackage[dvipsnames]{xcolor}

\usepackage{hyperref} 
\hypersetup{
    colorlinks=true,
    linkcolor=blue,
    urlcolor=blue,
    citecolor=blue
}
\renewcommand{\sec}[1]{\hyperref[sec:#1]{Section~\ref*{sec:#1}}}
\newcommand{\app}[1]{\hyperref[app:#1]{Appendix~\ref*{app:#1}}}
\newcommand{\tab}[1]{\hyperref[tab:#1]{Table~\ref*{tab:#1}}}
\newcommand{\fig}[1]{\hyperref[fig:#1]{Figure~\ref*{fig:#1}}}

\newtheorem{zkstatement}{Zero Knowledge Proof Statement}

\begin{document}

\title{Securing Elliptic Curve Cryptocurrencies against Quantum Vulnerabilities:\\ Resource Estimates and Mitigations}

\author{Ryan Babbush}
\email{babbush@google.com}
\affiliation{Google Quantum AI, Santa Barbara, CA 93111, United States}
\author{Adam Zalcman}
\email{viathor@google.com}
\affiliation{Google Quantum AI, Santa Barbara, CA 93111, United States}
\author{Craig Gidney}
\email{craiggidney@google.com}
\affiliation{Google Quantum AI, Santa Barbara, CA 93111, United States}
\author{Michael Broughton}
\affiliation{Google Quantum AI, Santa Barbara, CA 93111, United States}
\author{\mbox{Tanuj Khattar}}
\affiliation{Google Quantum AI, Santa Barbara, CA 93111, United States}
\author{Hartmut Neven}
\affiliation{Google Quantum AI, Santa Barbara, CA 93111, United States}
\author{Thiago Bergamaschi}
\affiliation{Google Quantum AI, Santa Barbara, CA 93111, United States}
\affiliation{Department of Computer Science, University of California Berkeley, Berkeley, CA 94720, United States}
\author{Justin Drake}
\affiliation{Ethereum Foundation, Zeughausgasse 7a, 6300 Zug, Switzerland}
\author{Dan Boneh}
\affiliation{Department of Computer Science, Stanford University, Stanford, CA 94305, United States}

\date{\today}

\begin{abstract}
\vspace{2em}
The expected emergence of cryptographically relevant quantum computers (CRQCs) will represent a singular discontinuity in the history of digital security, with wide ranging impacts. This whitepaper seeks to elucidate specific implications that the capabilities of developing quantum architectures have on blockchain vulnerabilities and potential mitigation strategies. First, we provide new resource estimates for breaking the 256-bit Elliptic Curve Discrete Logarithm Problem over the secp256k1 curve, the core of modern blockchain cryptography. We demonstrate that Shor's algorithm for this problem can execute with either $\leq 1200$ logical qubits and $\leq 90$ million Toffoli gates or $\leq 1450$ logical qubits and $\leq 70$ million Toffoli gates. In the interest of responsible disclosure, we use a zero-knowledge proof to validate these results without disclosing attack vectors. On superconducting architectures with $10^{-3}$ physical error rates and planar connectivity, those circuits can execute in minutes using fewer than half a million physical qubits. We introduce a critical distinction between ``fast-clock'' (such as superconducting and photonic) and ``slow-clock'' (such as neutral atom and ion trap) architectures. Our analysis reveals that the first fast-clock CRQCs would enable ``on-spend'' attacks on public mempool transactions of some cryptocurrencies. We survey major cryptocurrency vulnerabilities through this lens, identifying systemic risks associated with advanced features in some blockchains such as smart contracts, Proof-of-Stake consensus, and Data Availability Sampling mechanism, as well as the enduring concern of ``abandoned'' assets. We argue that technical solutions would benefit from accompanying public policy and discuss various frameworks of ``digital salvage'' to regulate the recovery or destruction of dormant assets while preventing adversarial seizure. We also discuss implications for other digital assets and tokenization as well as challenges and successful examples of the ongoing transition to Post-Quantum Cryptography (PQC). Finally, we urge all vulnerable cryptocurrency communities to join the migration to PQC without delay.
\end{abstract}

\maketitle

\newpage
\tableofcontents
\newpage

\section{Introduction}
\label{sec:intro}

Quantum computers render tractable certain otherwise hard computational problems. This promises an immense benefit to many fields of science and technology~\cite{Feynman1982simulating, Schleich2025quantum, Babbush2025, Montanaro2016, chou2025race}, but also jeopardizes some broadly adopted cryptosystems. Most importantly, cryptographically-relevant quantum computers (CRQCs) pose a threat~\cite{Bernstein2017, Joseph2022, Gidney2025Factoring} to widely deployed public-key cryptography~\cite{Merkle1978secure} including the Rivest-Shamir-Adleman (RSA) cryptosystem~\cite{rivest1978method} and Elliptic Curve Cryptography~\cite{Miller1986Ecc,Koblitz1987Ecc}. The former arises from Shor's efficient quantum algorithm for factoring integers~\cite{Shor1997Polynomial} and the latter from Shor's efficient quantum algorithm for the Elliptic Curve Discrete Logarithm Problem (ECDLP)~\cite{Shor1997Polynomial}. Cryptosystems based on the ECDLP are used to establish authenticity and integrity of software, firmware and microcode updates~\cite{OCP2023GPUFW,Schneider2025Fault}, to ensure authenticity and integrity of operating system images in secure boot~\cite{Schneider2025Fault}, to secure web traffic using the Transport Layer Security (TLS) protocol~\cite{rfc8446}, to ensure integrity and confidentiality in end-to-end encrypted messaging~\cite{Marlinspike2016x3dh}, to secure electronic passports and identification cards~\cite{Bos2013elliptic}, to support passwordless and multi-factor authentication~\cite{yubico2024technical}, to secure administrative access to cloud infrastructure using the Secure Shell (SSH) protocol~\cite{rfc5656}, to authenticate domain name records~\cite{rfc6605}, to implement cryptographic features in systems with low compute resources, such as embedded systems and Internet-of-Things devices~\cite{Nils2004}, and in other critical applications~\cite{Nino2018survey}.

Cryptocurrencies stand out among these applications of quantum-vulnerable cryptography for two reasons. First, in pursuit of efficiency and scale many blockchains depend crucially on ECDLP-based cryptography which uses keys that are almost an order of magnitude smaller than those of RSA at a similar security level~\cite{Barker2020NIST} which means a smaller CRQC is required to break them. Second, unlike traditional finance which generally employs multiple safeguards, blockchains tend to offer no recourse against fraudulent transactions enabling unrecoverable theft with a forgery of a single digital signature.

Despite the critical dependence of cryptocurrencies on quantum-vulnerable ECDLP-based cryptography, the intersection of quantum computing and blockchain technology remains under-explored. Prior works have focused almost exclusively on Bitcoin~\cite{Milton2025bitcoin, Li2025quantum, Deegan2025quantum, Pont2024downtime, Aggarwal2018quantum, Ruffing2025postquantum} with only a few publications exploring the cryptocurrencies landscape more broadly~\cite{Holmes2021assessment, Fukuda2025grand, Costa2025postquantum}. Existing discussions rely on historical timelines for constructing real-world CRQCs and generally do not reflect state-of-the-art resource estimates for quantum algorithms. Furthermore, systemic vulnerabilities with implications for stablecoins and tokenization remain unexplored. This whitepaper seeks to paint a more comprehensive picture by sharing updated resource estimates for quantum attacks while widening the discussion beyond a focus on Bitcoin to other cryptocurrencies and modern developments, such as real-world asset tokenization. In our exploration of cryptographic innovations in modern cryptocurrencies, we come across novel quantum vulnerabilities and attack modes, such as the ability to use a CRQC to manufacture reusable classical exploits---a risk absent from Bitcoin.

We expect this document will be scrutinized by policymakers and stakeholders in the financial sector who will have differing levels of familiarity with these risks. However, we also intend for this to be a wake-up call to our colleagues in the quantum computing community. For many in the quantum community, this may be the first detailed exposure to the practical mechanics of blockchain vulnerabilities and the specific ``onchain'' consequences of the algorithms we study. Indeed, the quantum computing community maintains an odd relationship with cryptanalysis applications. It is seen as scientifically prestigious to develop and refine quantum cryptanalysis algorithms, yet it is often regarded as taboo to discuss the downstream implications of these algorithms or how they will influence the commercial quantum computing landscape. In this sense, the field has developed a culture of detachment from the real-world consequences of its work on breaking quantum-vulnerable cryptosystems.

This paper offers several technical and policy contributions that depart from conventional wisdom. We explain our position that progress in quantum computing has reached the stage where it is prudent to stop publishing details of improved quantum cryptanalysis to avoid misuse. Nevertheless, publishing trustworthy resource estimates remains essential to signal the proximity of quantum threats and motivate timely defenses. This tension between public transparency and the need to deny rogue actors specific attack details is a long-standing debate in computer security~\cite{Maurushat2013disclosure}, leading to the ``Responsible Disclosure'' paradigm~\cite{ISOIEC2018isoiec} where vulnerabilities are disclosed only after a remediation window. Adapting this practice to quantum cryptanalysis --- where patching is exceptionally difficult and the threat timeline is shifting --- we share updated resource estimates while withholding the specific quantum circuits. However, this approach gives rise to issues of trust and confidence.

The value of cryptocurrency assets is based on their digital security (assured by code and cryptography) and public confidence in the decentralized system (rather than in third party intermediaries as in traditional banking). The importance of public confidence in the system reflects the fact that cryptocurrencies, like traditional forms of money, are social institutions~\cite{North1991institutions,North1990political} built on trust in the system and shared understanding of its rules and capabilities. While the digital security of cryptocurrencies can be attacked using CRQCs, public confidence can be undermined using Fear, Uncertainty and Doubt (FUD) techniques~\cite{CMCGlossaryFUD, BTSE2023FUD, Harris1998complete}. In this sense, unscientific and unsubstantiated resource estimates may themselves constitute a genuine or apparent attack on the system, which complicates responsible disclosure and technical discussion of quantum vulnerabilities in blockchain technologies. We address this communication challenge in two ways.

First, we reduce the FUD potential of our discussion by explicitly emphasizing and clarifying areas where cryptocurrencies are resilient against quantum attacks, sometimes despite widespread belief to the contrary, such as in the case of Bitcoin's Proof-of-Work consensus mechanism's immunity to quantum attacks using Grover's algorithm, Zcash's newest shielded pool's resilience against quantum attacks on protocol parameters, and the widespread protection of public keys behind cryptographic hash functions on many blockchains, including Bitcoin. We also discuss successful PQC projects to highlight technical feasibility of continued operation of digital economy in a world with CRQCs.

Second, we rigorously substantiate our resource estimates by sharing a cryptographic zero-knowledge (ZK) proof~\cite{Goldwasser1985knowledge, Quisquater1989to} that enables trustless third parties to cryptographically verify the estimates without access to the underlying attack details. Specifically, we publish a ZK proof that we have compiled two quantum circuits for solving the 256-bit ECDLP: one with 1200 logical qubits and 90 million Toffoli gates and one with 1450 logical qubits and 70 million Toffoli gates. ZK proofs have been previously suggested~\cite{CuellarGempeler2025cheesecloth} as a tool to demonstrate the existence of security vulnerabilities (not related to quantum computers) without leaking the details necessary to carry out an attack. However, past work on this topic appears to focus on demonstrating the feasibility of ZK disclosures in principle, by applying techniques to previously reported vulnerabilities rather than using them to disclose a novel or improved attack as we do here.

On a standard superconducting architecture, using surface code error correction, we estimate these computations could be realized with fewer than half a million physical qubits (nearly a 20 fold reduction over prior estimates~\cite{Litinski2023to}). This reduction in quantum resources needed to solve ECDLP on a quantum computer reflects the general pattern of quantum algorithmic improvements, exemplified by the reduction in resources needed to break 2048-bit RSA~\cite{Gidney2021to, Gidney2025to}. While we focus on compilation to a scaled up version of platforms that have been experimentally demonstrated, resource estimates could be reduced substantially by making more aggressive assumptions about hardware capabilities. Furthermore, our analysis gives the first clear indication that superconducting qubits could launch attacks within the average block time of Bitcoin and Bitcoin Cash, thus enabling ``on-spend'' attacks whereby a transaction is intercepted, the key is broken, and a fraudulent transaction is syndicated in the brief period of time before it is recorded on the blockchain. This prospect highlights the importance of migrating to Post-Quantum Cryptography (PQC) and of mitigation measures that thwart on-spend attacks, such as private mempools and commit-reveal schemes~\cite{Habovstiak2025hashed}.

We delineate how superconducting, silicon, and photonic platforms, with their fast gates and short error correction cycle times, pose a distinct threat to active transactions that slower architectures like neutral atoms and ion traps likely cannot match. This architectural distinction informs our discussion on mitigation strategies. We explore two divergent scenarios: one where the first CRQCs are built on a superconducting, silicon, or photonic platform and the other where the first CRQCs are implemented in a neutral atom or ion trap architecture. In the latter case, we expect that the first CRQCs will only be able to implement ``at-rest'' attacks whereby the quantum attacker requires a substantial period of time to solve for the private key and is therefore only able to attack long-exposed public keys securing funds at rest on the blockchain. In this scenario, the first mitigation is not necessarily a full swap of the vulnerable cryptosystems, but rather a rigorous elimination of public key reuse and minimization of public key exposure. We note that while a distinction between attack types has been discussed in prior literature, sometimes under different names such as ``short-range'' and ``long-range'' attacks~\cite{Milton2025bitcoin}, we add to this discussion by associating attack types with specific quantum architectures. The ultimate path towards post-quantum security in blockchain technologies is technically clear, if logistically difficult: a full switch to PQC~\cite{Bernstein2009introduction, Bernstein2017postquantum}. We argue that steps towards this complex migration should begin immediately. In line with the defense-in-depth principle~\cite{NISTGlossary_defense_in_depth}, we recommend that intermediate mitigation measures also be urgently adopted. Such measures are technically simpler than a full upgrade of the underlying cryptosystems which allows them to be deployed earlier.

The need for urgency arises from multiple considerations. Technical and logistical difficulties make migration to post-quantum cryptosystems a slow process. For some blockchains, this may be exacerbated by challenges involved in reaching sufficiently broad consensus. Key obstacles are related to resource costs of PQC which are substantially higher than those of ECDLP-based cryptosystems. In particular, in the Bitcoin community, some historical proposals that increased bandwidth requirements of running network nodes turned out to be controversial and led to disagreements and hard forks. Moreover, some of the current technical and financial trends in cryptocurrencies magnify quantum risks while exposing new funds and assets. Specifically, the quantum attack surface of blockchain-based systems continues to expand due to the introduction of new privacy and scalability features based on quantum-vulnerable cryptography. At the same time, financial developments, such as fiat-backed stablecoins and tokenization of other real-world assets (RWA), are projected to increase the pool of assets governed by smart contracts by nearly an order of magnitude by 2030~\cite{Gretz2025tokenization}. Most of this activity takes place on general-purpose blockchains for smart contracts, primarily Ethereum and to some extent Solana~\cite{Yakovenko2017solana}, with growing issuance on specialized blockchains, such as Algorand~\cite{Chen2016Algorand}, Stellar~\cite{Lokhava2019stellar, Mazieres2015stellar}, and the XRP Ledger~\cite{Chase2018analysis, Developers2026rippled} notable for protocol-level support for RWA tokenization~\cite{Foundation2026asset, Foundation2026rwa, RippleX2025future}. The account model and smart contracts employed by these blockchains introduce new quantum vulnerabilities not present in Bitcoin and its derivatives.

Fortunately, there is a path to achieving post-quantum security for cryptocurrencies. PQC has become a mature cryptographic discipline: post-quantum cryptosystems have been proposed, scrutinized, implemented and deployed. In fact, they are in active use protecting Internet traffic~\cite{OBrien2023Chromium} and indeed securing blockchain transactions~\cite{Young2025technical}. As we discuss in detail in later sections, some blockchains, such as the Quantum Resistant Ledger (QRL)~\cite{Foundation2026quantum, Waterland2016quantum}, Mochimo~\cite{Zweil2018mochimo} and Abelian~\cite{Alice2022abelian}, rely exclusively on PQC. Others, such as Algorand, the XRP Ledger, and Solana, have made early experimental deployments of post-quantum protocols~\cite{Young2025technical, ForkLog2025xrp}.

However, forward-looking migration to PQC is not a panacea. Dormant digital assets, including those abandoned or inaccessible due to lost private keys, pose a distinct and critical challenge. We highlight the example of Bitcoin's Pay-to-Public-Key (P2PK) locking scripts, which secure over 1.7 million BTC. The total amount of dormant quantum-vulnerable bitcoin may reach 2.3 million BTC when all script types are considered. Unlike active wallets that can migrate to new standards, dormant assets cannot be ``fixed'' via forks that enable PQC protocols for future transactions. They represent a fixed target --- tens or hundreds of billions of dollars in value that will eventually become accessible to a quantum attacker. The community will soon face difficult, unprecedented decisions regarding the fate of these assets, forcing tradeoffs between the immutability of cryptographic property rights and the economic stability of the network.

The currently unclear legal status of the use of CRQCs to recover private keys of dormant P2PK assets creates significant risk that they would inevitably be seized by rogue actors. We consider policy responses to address this risk, including options for bringing the resulting financial gains into the formal, taxable economy and preventing them from falling exclusively to criminals or adversarial state actors. For example, governments may consider national security responses or may create a legal framework for ``digital salvage''. This approach classifies the recovery of these assets as a regulated activity, analogous to recovery of sunken treasure. Ultimately, the fate of dormant quantum-vulnerable assets will depend on protocol changes (which may cause blockchains to fork), economic incentives (which will determine the relative value of forked assets) and government policy (which will constrain activities of regulated entities). The Bitcoin community retains control over protocol changes, but faces challenges related to the need for broad consensus. Therefore, it would be consistent for public policy to create a legal backstop by legalizing salvage while simultaneously recommending that the community exercise its autonomy to ``burn'' all salvageable coins.

The remainder of this paper is organized as follows. In \sec{attacks}, we present new quantum resource estimates for breaking the ECDLP together with a zero-knowledge proof that substantiates our claims while withholding technical details necessary to launch an attack. We also define the operational difference between ``on-spend'', ``at-rest'' and ``on-setup'' attacks contrasting the capabilities of ``fast-clock'' quantum computers, such as those based on superconducting, silicon, and photonic qubits, on one hand, and ``slow-clock'' quantum computers, such as those based on neutral atoms and ion traps, on the other. In \sec{bitcoin}, we analyze the specific vulnerabilities of the Bitcoin blockchain, discussing the threat to digital signatures, explaining the detailed mechanics of ``on-spend'' and ``at-rest'' attacks, and the infeasibility of quantum attacks against its Proof-of-Work consensus mechanism. In \sec{protocols}, we step beyond digital signatures and provide a high-level description of other ECDLP-based cryptographic protocols that enable new features that are a source of novel quantum vulnerabilities in modern cryptocurrencies. In \sec{ethereum}, we examine the Ethereum blockchain, identifying distinct risks to its account model, smart contract governance, Proof-of-Stake validators, and Data Availability Sampling mechanism, highlighting how the emergence of a complex digital ecosystem, including Layer 2 solutions, stablecoins, and tokenization, substantially increases the attack surface for quantum threats. In \sec{blockchains}, we discuss smaller blockchains that are a source of intense financial and cryptographic innovation. We highlight novel quantum vulnerabilities they sometimes introduce as well as examples of successful deployment of PQC to protect cryptocurrencies and other digital assets. Then, in \sec{pqc}, we describe risks and challenges associated with the migration to PQC. In \sec{dormant}, we shift from technical analysis to a discussion of public policy around quantum vulnerabilities, focusing on possible approaches to address the challenge of dormant assets. Finally, we conclude in \sec{outlook} with an outlook on the challenge and urgency of migrating to PQC.

\section{Quantum Attacks on the Elliptic Curve Discrete Logarithm Problem}
\label{sec:attacks}

\subsection{Attack Types and Disclosure Models}

At the heart of most cryptocurrency security lies the assumption that the ECDLP is hard to solve. In later sections, we will discuss specific blockchain functions that rely on this hardness assumption, including transaction validation in Bitcoin and Ethereum and the Proof-of-Stake consensus mechanism in Ethereum. Here, we outline briefly three types of quantum attacks on blockchain security and provide new bounds on the quantum resources required to execute such attacks. Based on these estimates, we discuss two scenarios of the emergence of CRQCs capable of exploiting cryptocurrency vulnerabilities. These scenarios will later anchor our discussion of possible mitigation strategies. In particular, the speed of attacks that early CRQCs can execute determines which blockchain vulnerabilities are exploitable and hence, informs the choice of intermediate mitigation measures.

We classify quantum attacks into three categories based on the necessary execution speed and on whether access to a CRQC is required for each attack or only once to prepare a universal reusable exploit:

\begin{enumerate}
    \item \textbf{On-Spend Attacks:} Attacks targeting transactions in transit. When a blockchain user broadcasts a transaction, an attacker must derive the private key within the window of time allowed before the transaction is recorded on the blockchain. This requires a quantum computer fast enough to solve ECDLP within the transaction settlement time of the target blockchain which ranges from hundreds of milliseconds to a few minutes (e.g., about 400 milliseconds for Solana, about 12 seconds for Ethereum, about 10 minutes on average for Bitcoin). On-spend attacks are also known as ``short-range'' or ``just-in-time'' attacks~\cite{Milton2025bitcoin, Deegan2025look}.

    \item \textbf{At-Rest Attacks:} Attacks targeting public keys that remain exposed onchain or offchain for long periods of time, such as dormant wallets with reused keys. The attacker has days (or more) to derive the private key. At-rest attacks are also known as ``long-range'' or ``long-exposure'' attacks~\cite{Milton2025bitcoin}.

    \item \textbf{On-Setup Attacks:} Attacks targeting fixed public protocol parameters that produce a universal reusable backdoor into a cryptographic protocol. The backdoor is created by means of a one-time off-line quantum computation on a CRQC and subsequent attacks utilizing it are executed on a classical computer. For example, an on-setup attack may involve the use of Shor's algorithm to recover the so-called ``toxic waste'' discarded in a powers-of-tau trusted setup ceremony~\cite{Wang2025trusted}. While the Bitcoin blockchain is immune to on-setup attacks, some scaling solutions, such as Ethereum's Data Availability Sampling mechanism, and privacy protocols, such as Tornado Cash~\cite{Pertsev2019Tornado}, are vulnerable to this especially insidious attack mode.
\end{enumerate}

We note that there are other ways to classify quantum attacks on cryptocurrencies. In particular, one could categorize them by the security property they breach yielding a finer classification into attacks on asset ownership (e.g., transaction forgery in Bitcoin), consensus (e.g., deep blockchain reorganizations after compromising more than half of Ethereum validators), confidentiality (e.g., attacks on unlinkability of diversified addresses in Zcash), monetary integrity (e.g., creation of new coin in Mimblewimble or a vulnerable zk-rollup of Ethereum), solvency (e.g., peg collapse due to illegitimate Rootstock or stablecoin redemptions, stealthy drain of an anonymity pool in Tornado Cash), data availability (e.g., forging a data availability sampling proof in Ethereum), or governance (e.g., vote forgery in Cardano or in a Decentralized Autonomous Organization on Ethereum, asset takeover in a multisig bridge). In later sections, we discuss quantum attacks in terms of their onchain consequences, returning to all of the above examples. Here, we focus on the execution speed, because it has direct implications for the type of CRQC necessary to launch an attack.

Even though all universal quantum computer architectures are in principle capable of unlocking the same asymptotic quantum speedups for certain problems, such as the exponential speedup for ECDLP provided by Shor's algorithm, the actual wall clock time they incur varies by  orders of magnitude. These constant factors govern the ability of a given type of CRQC to launch on-spend attacks. The resource estimates we describe below indicate that superconducting~\cite{Acharya2024quantum}, photonic~\cite{Chan2025practical, Scott2023timing}, and silicon spin qubit~\cite{Stano2022review} CRQCs, with their fast gates and short quantum error correction cycles, will be able to solve ECDLP in the span of a few minutes and thus, to launch on-spend attacks. By contrast, the elementary operations on neutral atom and ion trap devices are about two to three orders of magnitude slower~\cite{Bluvstein2025faulttolerant, Schafer2018fast}. As a consequence, we do not expect CRQCs in these slower architectures to be able to launch on-spend attacks. We will refer to the former as fast-clock CRQCs and to the latter as slow-clock CRQCs.

Traditionally, the gold standard for quantum resource estimation has been full transparency: publishing algorithmic innovations, logical circuits, and error-correction optimizations to ensure results are openly verifiable. Our team has historically adhered to this standard --- for instance, in establishing the most efficient quantum algorithms and circuits for breaking 2048-bit RSA~\cite{Gidney2021to, Gidney2025to}. However, the escalating risk that detailed cryptanalytic blueprints could be weaponized by adversarial actors necessitates a shift in disclosure practices. Accordingly, we believe it is now a matter of public responsibility to share refined resource estimates while withholding the precise mechanics of the underlying attacks. Transparency regarding the overall resource costs of quantum attacks is essential; if the community overestimates the resources required, the perceived ``safety margin'' may lead to a dangerous ``wait-and-see'' attitude. Given the technical trade-offs of PQC, such delays could make it difficult for vulnerable cryptocurrencies to execute an orderly transition.

This tension has sparked debate within the community. For example, Scott Aaronson, a prominent researcher and blogger in quantum computing, has recently oscillated between advocating for total non-disclosure~\cite{Aaronson2025more} and full transparency~\cite{Aaronson2026reducing}. We regard this debate as a quantum iteration of the older vulnerability disclosure controversy~\cite{Maurushat2013disclosure} which has roots in Kerckhoff's Principle~\cite{Kerckhoffs1883la} and which, in modern times, sets the No Disclosure position against the Full Disclosure movement. In computer security, the debate has ultimately yielded a consensus known as the Responsible Disclosure or Coordinated Vulnerability Disclosure (CVD): a public disclosure preceded by an embargo to allow the vulnerable systems to be fixed. Variants of CVD with strict deadlines have been adopted by prominent computer security research institutions, such as CERT/CC at Carnegie Mellon University~\cite{Center2026certcc} and by Google's Project Zero~\cite{Zero2026vulnerability}, and were codified as an international standard ISO/IEC 29147:2018~\cite{ISOIEC2018isoiec}. While ordinary software embargoes typically last only a few months, publication is frequently deferred for hard-to-fix exposures, such as hardware flaws, operating systems bugs, or issues that require updates to technical standards~\cite{Center2026certcc}. Quantum vulnerabilities in ECDLP-based cryptography certainly qualify as hard-to-fix. Consequently, we believe it is consistent with the principles of responsible disclosure to withhold the quantum circuits necessary to launch quantum attacks.

Nevertheless, providing resource estimates without supporting data can appear unscientific, especially to the historically skeptical~\cite{Carter2026Bitcoin} and decentralized cryptocurrency community. We bridge this gap using a zero-knowledge (ZK) proof~\cite{Goldwasser1985knowledge, Quisquater1989to}. By providing a ZK proof, we offer the community a rigorous way to verify the imminent threat to ECDLP without leaking sensitive details to potential quantum adversaries. We produce our ZK proof using a state-of-the-art software tool, the SP1 zero-knowledge virtual machine (zkVM)~\cite{Labs2026sp1}. 

The statement that our ZK proof demonstrates is the following: we possess a classical reversible circuit of a specified size which on most inputs correctly computes point addition on the elliptic curve secp256k1~\cite{Jancar2026secp256k1, Brown2010sec}. This is the primary bottleneck in Shor's quantum algorithm and can be related to the cost of the overall algorithm in a straightforward fashion provided that one is using a ``windowed arithmetic'' technique for multiplying an elliptic curve point by a scalar~\cite{Gidney2021to}. Therefore, verifying circuits for this subroutine is sufficient to substantiate our resource estimates. Furthermore, for the purposes of solving the ECDLP on a quantum computer, it is sufficient for the elliptic curve point addition circuit to yield correct results on most, rather than all, inputs~\cite{Proos2003shors}. This framing does leak two pieces of information about our result: (i) that we are using a classical reversible circuit executed in quantum superposition for the elliptic curve point addition and (ii) that we are using windowed arithmetic. However, neither of these facts is surprising or unexpected. In fact, (i) has been true of all prior work on breaking ECDLP on a quantum computer and (ii) is a common element in all recent works~\cite{Chevignard2026reducing, Litinski2023to, Haner2020improved, Gouzien2023performance, DallaireDemers2025brace}.

In order to produce the ZK proof using the SP1 zkVM, we implement a Rust program that takes a secret classical reversible circuit as input and checks that it correctly computes elliptic curve point addition on nine thousand random inputs. We commit to our secret quantum circuit via its SHA-256 cryptographic hash (a unique ``digital fingerprint''). We generate the random test inputs using a cryptographically secure pseudo-random number generator (CSPRNG) --- a SHAKE256 extendable-output function (XOF) initialized with the raw bytes of our circuit. This approach is closely related to the Fiat-Shamir heuristic~\cite{Fiat1987to}, and can be shown to be sound under certain idealized assumptions about the hash function. More details are provided in the Appendix of this paper. The Rust code and the ZK proof are available in our Zenodo dataset~\cite{ZKP_Zenodo}.

\subsection{Updated Quantum Resource Estimates}

The landscape of published state-of-the-art compilations of Shor's algorithm for $n$-bit ECDLP problem is complicated by the space-time tradeoff between the number of logical qubits (which governs how large the CRQC required for a quantum attack must be) and Toffoli gate count. Toffoli gates dominate the resource costs of fault-tolerant execution of Shor's algorithm dictating its overall runtime~\cite{Jones2013lowoverhead, Fowler2012timeoptimal} and determining whether on-spend attacks are possible. The best published number of logical qubits is from a recent preprint by Chevignard et al.~\cite{Chevignard2026reducing}, where they estimate a logical qubit count of 1100 for $n=256$ (and $3.12n + o(n)$ asymptotically). However, this small logical quantum memory footprint comes at the cost of more than 100 billion Toffoli gates. Such a large number of gates translates to both an exorbitant runtime, as well as a very large number of physical qubits (the number of physical qubits required depends on both the number of logical qubits and the number of gates).

\begin{figure*}[t]
    \centering
        \includegraphics[width=\linewidth]{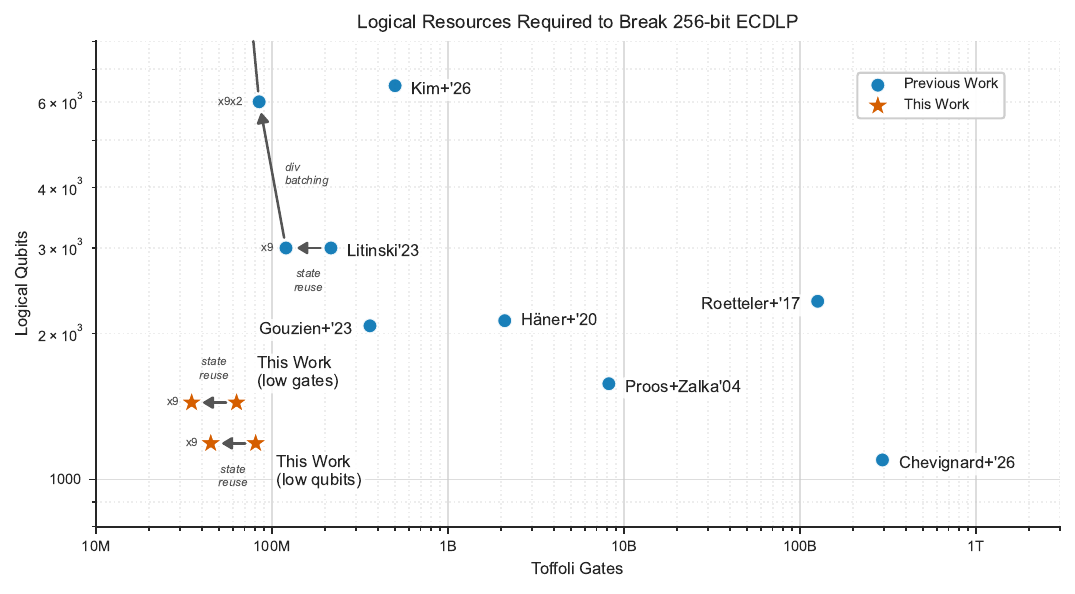}
    \caption{
    Comparison of logical quantum resources (number of logical qubits and Toffoli gates) required to break 256-bit ECDLP for the secp256k1 curve, as reported by various prior works. The arrows illustrate the application of two algorithmic optimizations adopted from Litinski~\cite{Litinski2023to}: state reuse and div batching. State reuse refers to deriving multiple (here: nine) private keys from public keys in a single execution by reusing the initial phase estimation state. Div batching refers to running multiple (here: two for the visible point and nine for the point offscreen which corresponds to $2.7\times 10^4$ logical qubits) instances of the algorithm in parallel and merging the modular division (inversion) operation across those instances; this is known as ``Montgomery's trick'' in the cryptography literature~\cite{Montgomery1987speeding}. Note that the main resource estimates for our approach quoted throughout this work apply to a single instance and do not include such optimizations. The plot represents resource estimates found in~\cite{Chevignard2026reducing, Litinski2023to, Kim2026new, Haner2020improved, Roetteler2017quantum, Proos2003shors, Gouzien2023performance, DallaireDemers2025brace}.
    }
    \label{fig:resource_history}
\end{figure*}

On the other extreme, the best prior published Toffoli count is roughly 200 million by Litinski~\cite{Litinski2023to} (the title of the paper says 50 million not 200 million because it includes optimizations that reduce the amortized cost of solving multiple instances, whereas we are focusing on single instance attacks). Litinski's lower Toffoli count comes at the cost of more than twice the number of logical qubits compared to Chevignard et al. Accounting for error correction overhead, the space required for Litinski's approach is further estimated to be about 9 million physical qubits in a photonic architecture~\cite{Litinski2023to}. We compare these works and others to our own in \fig{resource_history}.

\begin{figure*}[htbp]
    \centering
        \includegraphics[width=\linewidth]{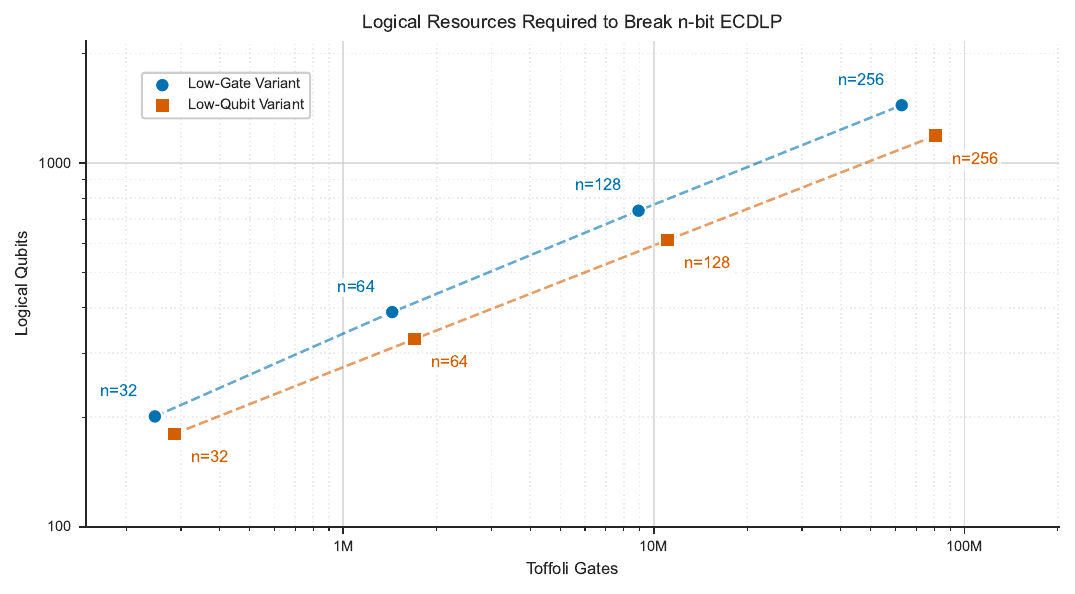}
    \caption{Logical resources required to break $n$-bit ECDLP for curves with bit lengths of $n = 32, 64, 128, 256$.} \label{fig:ecdlp_resources}
\end{figure*}

We are reporting here that our team has developed logical circuits to break ECDLP on elliptic curves over finite fields with $n$-bit prime modulus and $n$-bit group order requiring approximately 4.5n space. In \fig{ecdlp_resources}, we show the resources required for two variants of our approach (one optimized for qubit count, the other optimized for gate count), at different values of n. At $n=256$ bits, the circuits use either 1200 logical qubits and 90 million Toffoli gates or 1450 logical qubits and 70 million Toffoli gates. In terms of the spacetime volume (a key resource which in particular drives the quantum error correction overhead), these estimates represent roughly an order of magnitude improvement over the most efficient prior work when applied to a single ECDLP instance. These are the resource estimates we demonstrate using the ZK proof discussed above. Our findings apply directly to ECDLP on secp256k1 ---  an elliptic curve widely used in digital signatures on popular blockchains, such as Bitcoin and Ethereum.

Using standard assumptions about superconducting qubits hardware, such as planar degree-four connectivity and $10^{-3}$ physical error rates, running variants of the surface code architecture similar to the one assumed in~\cite{Gidney2025to}, we calculate that these circuits can be executed on fewer than half a million physical qubits. This estimate is based on the reduced error correction overhead resulting from the smaller size of the optimized logical circuit and on the use of efficient storage of logical qubits in dense surface code configurations protected by ``yokes''~\cite{Gidney2023yoked, Gidney2025yoked}. In order to estimate the wall clock time of running our circuits, we assume they are executed in a reaction-limited fashion~\cite{Fowler2012timeoptimal}  ---  a style of quantum computation where execution speed is mainly limited by the reaction time of the control system. If we assume a control system reaction time of 10 microseconds (a standard figure for superconducting qubits~\cite{Gidney2019flexible}), and a 50\% overhead per Toffoli gate, then 70 or 90 million Toffoli gates can be resolved in 18 or 23 minutes on a superconducting CRQC  ---  close to the 10 minute average Bitcoin block time.

On-spend attacks, which affect all standard addresses in Bitcoin and similar cryptocurrencies, become even faster when we take into account the fact that a quantum computer can precompute the first half of the algorithm (which only depends on protocol parameters common to all addresses) and then wait in this ``primed'' state until a public key is available. This halves the time between a public key being revealed and its private key being resolved. Hence, we should estimate the time required to launch an on-spend attack starting from this primed state at the moment the public key is learned to be roughly either 9 minutes or 12 minutes. These facts imply that a superconducting CRQC capable of performing at-rest attacks against static holdings recorded on the blockchain would likely also be capable of executing on-spend attacks against active transactions. As we discuss in more detail later on, we do not expect meaningful scaling challenges between a quantum computer with 1200 logical qubits and one with 1450, so, in order to focus and simplify subsequent discussion, we assume that first-generation fast-clock CRQCs may be able to solve ECDLP on secp256k1 and similar elliptic curves in about 9 minutes on average. Note that, if multiple primed machines are available, this duration can be further reduced. For example, according to Table 7.8 of Ref.~\cite{Ekera2024factoring}, an attack with 11 primed machines would involve each machine performing 32 elliptic curve point additions rather than one primed machine performing 208, a 6.5x speedup.

Executing 70 million Toffolis in 9 minutes would require the generation of half a million T states per second. A T state of sufficiently low error can be produced in 50,000 qubit-rounds~\cite{Gidney2024magic}. In a fast-clock architecture with 1 microsecond error correction rounds, this implies allocating 25,000 physical qubits to magic production. This is a small qubit count compared to the half a million needed to run the algorithm at all, and this is why for fast-clock architectures we expect on-spend attacks to become possible at essentially the same time as at-rest attacks. By contrast, in a slow-clock architecture with 100 microseconds rounds or longer~\cite{Bluvstein2025faulttolerant, Schafer2018fast}, two and half million physical qubits would be needed for sufficient T state production. This qubit count is substantially larger than what is needed to run the algorithm slowly, so we currently expect that on slow-clock architectures at-rest attacks become possible before on-spend attacks. Nevertheless, it is plausible that future discoveries of new magic production techniques (e.g., new techniques for producing T states) that can be efficiently executed under the constraints of slow-clock architectures might enable them to launch on-spend attacks, too.

\begin{figure*}[t]
     \centering
           \includegraphics[width=\textwidth]{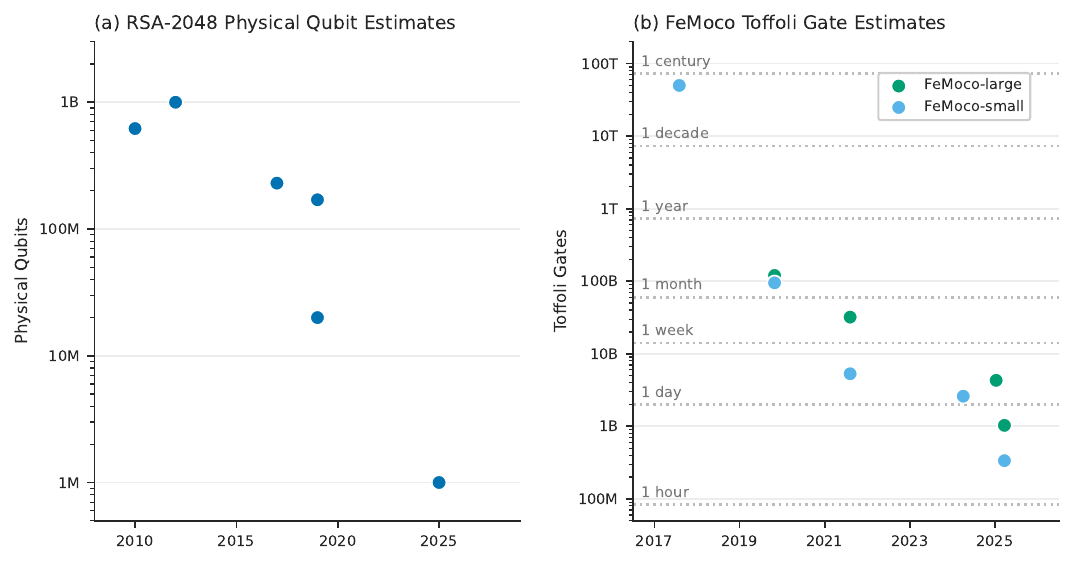}
     \caption{These figures, which first appeared in~\cite{Babbush2025} illustrate that algorithms and error-correction research have dramatically decreased the resources required to solve important problems on quantum computers over the last decade. (Left) The number of physical qubits required of a superconducting qubit architecture running the surface code in the most advanced resource estimates for breaking 2048-bit RSA encryption, as a function of the year the manuscript was published. We note that some papers have claimed even fewer physical qubits by more aggressively changing hardware assumptions (e.g., by analyzing devices with higher connectivity or lower error rates) but here we elect to compare estimates that make comparable hardware assumptions. (b) Number of Toffoli gates required by the most advanced resource estimates for computing the ground state energy of the FeMoco molecule to chemical accuracy, as a function of the year the manuscript was published. Here, resource estimates are reported for the Reiher (``small'')~\cite{Reiher2017Elucidating} and Li (``large'')~\cite{Li2019Electronic} FeMoco active space Hamiltonians.
     The left plot includes  Refs.~\cite{jones2012layered,fowler2012surface,o2017quantum,gheorghiu2019benchmarking,Gidney2025Factoring} and the right plot includes  Refs.~\cite{Reiher2017Elucidating,Lee2020hypercontraction,vonBurg2021Quantum,rocca2024reducing,Low2025Fast}.} \label{fig:reducing_resources}
\end{figure*}

To further put these resource estimates into context we refer the reader to \fig{reducing_resources} which first appeared in a recent perspective article from Google Quantum AI~\cite{Babbush2025grand}. The lesson of \fig{reducing_resources} is that advances in quantum algorithms and quantum error correction architectures continue to dramatically reduce the physical resources required to implement quantum applications (as the saying goes in cryptography: ``attacks always get better''~\cite{Schneier2011new}). While resource estimates cannot drop indefinitely, the applications of focus in these plots (breaking RSA and simulating quantum chemistry) have been the focus of significantly more published research historically than quantum algorithms for breaking ECDLP so it may be the case that algorithms for those applications are closer to optimal than they are for ECDLP.

The high cost of fault-tolerant quantum computation stems largely from the overhead of quantum error correction, which demands a difficult balance of hardware-compatible connectivity, high noise tolerance, fast syndrome decoders, and expressive logical instruction set in order to provide a complete and feasible solution from an engineering perspective. While the well-understood surface code meets these requirements, it suffers from low efficiency  ---  expressed by the logical-to-physical qubit ratio also known as the code's encoding rate or simply rate  ---  compared to other modern codes. For example, recent research into quantum Low-Density Parity-Check (qLDPC) codes~\cite{Bravyi2024highthreshold, Xu2024constantoverhead} offers an intriguing alternative by using long-range connections to increase the encoding rate. Although current resource estimates often default to the surface code for its proven decodability and the feasible demands it makes on hardware design, maturing qLDPC technologies could potentially further reduce physical qubit requirements~\cite{Yoder2025tour, Webster2026pinnacle}. Recently, search for better quantum error correcting codes has been accelerated using AI techniques~\cite{Olle2024Simultaneous,He2025discovering}.

The physical resource estimates we have discussed here (e.g., half a million physical qubits) assume relatively benign hardware capabilities, such as a planar architecture with degree-four connectivity and $10^{-3}$ physical gate error rates (i.e., consistent with a scaled up version of Google's quantum processors that have been demonstrated experimentally~\cite{Acharya2024quantum}). More aggressive hardware assumptions --- such as the ``bicycle'' architecture used for 2-gross qLDPC codes~\cite{Bravyi2024highthreshold, Yoder2025tour} --- could drop qubit counts closer to one hundred thousand physical qubits, but this approach requires non-local degree-seven connectivity that has yet to be demonstrated in actual superconducting qubit devices. Similarly, the recently introduced ``Pinnacle'' architecture~\cite{Webster2026pinnacle} suggests breaking 2048-bit RSA with fewer than one hundred thousand superconducting qubits, but relies on even more challenging degree-ten, non-planar connectivity for codes that currently lack extensive simulations characterizing their performance, or efficient decoders, and would require nearly a month of runtime. The physical complexity of devices meeting these more demanding specifications makes them potentially more difficult to fabricate with sufficiently low physical error rates than larger devices with more benign engineering demands. Consequently, while a Pinnacle-style device would theoretically require even fewer than one-hundred thousand superconducting qubits to break 256-bit ECDLP, a planar degree-four device with half a million qubits remains the better understood and also likely more feasible engineering path. Nevertheless, the search for high rate codes compatible with engineering constraints is an active and promising research area~\cite{Yang2025spatiallycoupled}.

\subsection{The Evolution of Offensive Quantum Capabilities}

Considering the significant speed difference between fast-clock and slow-clock quantum computing platforms, and because potential mitigation strategies depend on whether at-rest attacks become viable significantly earlier than on-spend attacks, we believe it would be prudent for the cryptocurrency community to develop contingency plans for two potential scenarios~\cite{Heilman2025changes}:

\begin{itemize}
    \item \textbf{Scenario 1}: On-spend attacks become viable about the same time as at-rest attacks. This is a possibility if the first CRQCs are fast-clock devices.
    
    \item \textbf{Scenario 2}: At-rest attacks become viable meaningfully sooner than on-spend attacks. This is a possibility if slow-clock quantum platforms advance more rapidly than fast-clock ones.
\end{itemize}

In part, the uncertainty concerning the two scenarios arises from the wide range of quantum computer architectures being actively developed across many research labs, tech companies and startups. For example, Google Quantum AI, IBM Quantum, Amazon, D-Wave, Rigetti and IQM are developing superconducting qubit architectures; PsiQuantum and Xanadu are building photonic quantum computers while Diraq and Intel are working on spin qubit devices. These three architectures feature fast elementary operations and are compatible with Scenario 1. Simultaneously, many companies, including IonQ, Quantinuum (a subsidiary of Honeywell) and Alpine Quantum Technologies are pursuing ion trap quantum processors while others, such as QuEra, Infleqtion, Atom Computing, Pasqal, and Logiqal are developing neutral atom devices. These two platforms have significantly slower elementary operations, so if the fast-clock architectures run into scaling challenges or if neutral atom or ion trap platforms scale rapidly, Scenario 2 could come to pass before Scenario 1.

It has been observed that technological change tends to occur in a two-stage evolutionary process in which an initial ``era of ferment'' characterized by high degree of innovation and technical variation eventually leads to the emergence of a ``dominant design'' and is followed by a period of gradual technological improvement~\cite{Anderson1990technological}. For example, the car industry's ``era of ferment'' (during the late 19th century) saw rapid innovation across a wide variety of cars powered by the steam engine, electric batteries and the internal combustion engine before the latter became the dominant design. In the stage of incremental improvement progress can be measured in terms of gradually increasing performance metrics, such as miles per gallon or the number of integrated circuit components in Moore's law~\cite{Moore1965cramming}. By contrast, quantum computing, with its broad variety of hardware platforms, is still in the ``era of ferment'' where simple models, such as counting physical qubits, fail to adequately capture technological progress. Instead, progress comes in discrete jumps corresponding to development of new internal capabilities and overcoming scaling challenges, e.g., by getting device error rates below the threshold for an error-correcting code~\cite{Acharya2023suppressing} or implementing coherent interconnects for modular architectures~\cite{Awschalom2021development, Escofet2023interconnect, Pattison2024fast}. Therefore, progress in quantum computing is better understood using a threshold model rather than in terms of the number of physical qubits.

Accordingly, other gradual measures of progress, such as the challenge ladder of increasingly difficult ECDLP instances ranging from 6-bit to 256-bit modulus and group order proposed in~\cite{DallaireDemers2025brace}, also fail to adequately measure progress towards a CRQC and may fail to provide a reliable early warning. Indeed, if a leading quantum architecture encounters and overcomes all its scaling challenges before producing a device able to solve (for example) 32-bit ECDLP, then there may be little time between the breaking of 32-bit ECDLP and the breaking of 256-bit ECDLP. Furthermore, the community should not expect to see published demonstrations of the most advanced quantum error-correction architectures and quantum algorithms deployed to cryptanalytic problems. Thus, a successful public demonstration of Shor's algorithm on a 32-bit elliptic curve should not be seen as a wake-up call to adopt PQC as much as a potential signal that PQC adoption has already failed.

Which of the two scenarios outlined above will come to pass depends on the scaling barriers affecting each platform and how quickly human ingenuity may overcome them. Given the uncertainty, we recommend that cryptocurrency communities urgently implement and deploy some of the relatively simple at-rest attack mitigations short of a full transition to PQC, such as removing vulnerable spending paths~\cite{Beast2024bip}, updating wallets to warn against weak addresses, disabling support for Rootstock deposit transactions that expose public keys cross-chain, introducing rotation mechanisms for validator keys in Ethereum and staking and voting keys in Cardano~\cite{Hoskinson2017Why}. Although such measures do not remove quantum vulnerabilities, they make attacks more difficult.

We have seen that while quantum hardware development efforts make progress on scaling quantum computers towards the ``finish line'' of cryptographical relevance, the work on quantum error correction and efficient compilation of quantum algorithms brings the finish line closer. We also argue that the finish line may become progressively more blurry and the current state of quantum capabilities increasingly more opaque as we approach CRQCs. Indeed, as progress in quantum computing lowers the barriers to entry, the final stages of the race to build a large fault-tolerant quantum computer may see ``late-joiners'' attempting a rapid breakout toward a CRQC, possibly accelerated by intellectual property theft or industrial espionage. Simultaneously, transparency of quantum computing research and development programs is likely to decrease as they get closer to large-scale commercially viable quantum computers. These factors may increase the uncertainty regarding the arrival time and the nature of the first CRQCs. Thus, it is conceivable that the existence of early CRQCs may first be detected on the blockchain rather than announced.

As we will argue throughout this piece, \textbf{the safest course of action for the cryptocurrency community is to begin preparing itself against quantum attacks immediately}. We will now turn to the discussion of the details of various quantum attacks against Bitcoin, Ethereum and other cryptocurrencies and how considerations of quantum architecture may come into play in mitigation strategies.

\section{Attacks on Bitcoin's Digital Signature Scheme}
\label{sec:bitcoin}

\subsection{Public Key Exposure in Bitcoin Transactions}
\label{sec:bitcoin_exposure}

The Bitcoin blockchain does not have a native concept of user accounts. Instead, it stores transactions, each of which is connected to other transactions via inputs and outputs. Units of bitcoin (BTC) exist as unspent transaction outputs (UTXOs), i.e., bitcoin that has been received but has not yet been spent, encumbered by a locking script, which ensures that it can only be spent by providing a valid digital signature matching the owner's public key. When bitcoin is transacted from one wallet to another, the sender provides two pieces of cryptographic information: an unlocking script, which proves that they own the bitcoin being spent, and a locking script, which ties the bitcoin to the recipient's wallet. One can think of the locking script as a cryptographic puzzle and of the unlocking script as its solution. Typically the puzzle involves digital signatures and is meant to be too hard for anyone to solve unless they know the private key.

Most common locking scripts are represented in a human-readable form as Bitcoin addresses. The overwhelming majority of bitcoin is locked using one of just seven standard types of scripts with various features, and degrees of vulnerability to quantum attacks~\cite{antonopoulos2014mastering, LearnMeABitcoin, Walker2025script}. A characteristic they all share is the use of ECDLP-based digital signatures to prove ownership of the bitcoin being spent. Transacting bitcoin requires one to present an unlocking script with a valid digital signature and to generate such a digital signature one needs to know the private key. Thus, to own bitcoin means to know the private key associated with the public key to which it is locked. CRQCs enable one to derive the private key from any public key, thereby upending Bitcoin's current cryptographic model of ownership.

The varying degree of quantum vulnerability between various Bitcoin script types primarily arises from the differences in how they manage public keys. For example, P2PK locking scripts simply record the public key of the recipient directly on the blockchain. Later, when the recipient wishes to spend the bitcoin, they supply a digital signature in the unlocking script and the Bitcoin network validates that the signature was generated using the private key matching the public key from the locking script. Since the public key is disclosed directly in the locking script, it remains visible to everyone from the moment the bitcoin is received. Therefore, bitcoin locked by a P2PK script is vulnerable to both at-rest and on-spend attacks. A little over 1.7 million bitcoin (nearly 9\% of all bitcoin) is secured by P2PK locking scripts.

In contrast, Pay-to-Public-Key-Hash (P2PKH) scripts are immune to most vectors of at-rest attack. Here, instead of disclosing the full public key in the locking script, the protocol only records its hash value (a unique ``digital fingerprint''), on the blockchain. Later, when the bitcoin is spent, the unlocking script supplies both the public key and a digital signature. A CRQC can use Shor's algorithm to derive the private key from a public key, but it cannot derive a public key from its hash value (that would require using the quantum computer with Grover's algorithm to invert a hash function - see \sec{bitcoin_pow} for more details on why this is not a credible threat). Therefore, bitcoin locked using a P2PKH script (and for which the public key is not leaked in other ways) is immune to at-rest attacks. However, spending such bitcoin does entail disclosure of the public key in the unlocking script, so it remains vulnerable to on-spend attacks.

Advanced users may require more complex spending conditions, e.g. m out of n signatures. The oldest standard script type supporting such transactions is Pay-to-Multisig (P2MS). Its quantum security properties are similar to P2PK, because it records the n public keys directly onchain in the locking script (and the unlocking script must supply m valid digital signatures corresponding to any m of these n keys). Here, a quantum attacker must derive m private keys rather than just one. However, since the public keys are exposed in the locking script, the attack can be launched offline. Pay-to-Script-Hash (P2SH) scripts potentially allow users to express even more complex spending conditions and hide them, together with any public keys they may contain, behind a hash. Therefore, bitcoin secured by P2SH locking scripts is immune to at-rest attacks, as long as the script is not reused or otherwise exposed before spending. In addition, for scripts that require a large number m of signatures, on-spend attacks by an attacker with a small number of CRQCs are made more difficult by the need to derive a large number of private keys in the ten minutes average block interval.

P2SH scripts and their modern replacements, such as P2WSH, can express spending conditions that do not depend on quantum-vulnerable cryptosystems. For example, in 2013, P2SH scripts were used to lock bounties for identifying weaknesses in hash functions~\cite{Todd2013reward}. However, such scripts fail to cryptographically bind the unlocking script to the spending transaction, so any other Bitcoin user can copy the script to their own transaction before the original is recorded onchain. Thus, the effective spending condition such scripts express is: ``anyone can spend once the puzzle is solved'' rather than ``whoever solved the puzzle can spend''. Therefore, the user must send the transaction to a trusted miner or mine it on their own.

Bitcoin's Segregated Witness (SegWit) upgrade in 2017, introduced two new locking scripts: Pay-to-Witness-Public-Key-Hash (P2WPKH) and Pay-to-Witness-Script-Hash (P2WSH) which hide public keys and scripts behind hash values like the older P2PKH and P2SH scripts~\cite{Developers2026segregated}. The new script types correspond to the modern Bitcoin addresses with the bc1q prefix. More recently, the Taproot soft fork in 2021 upgraded digital signatures from the old Elliptic Curve Digital Signature Algorithm (ECDSA) to the more flexible and efficient Schnorr signatures (both of which are based on ECDLP and thus vulnerable to quantum attacks) and introduced a new script type called Pay-to-Taproot (P2TR). This new versatile script type corresponds to Bitcoin addresses with the \texttt{bc1p} prefix and uses tweaked keys to simultaneously support two spending mechanisms called ``key path spend'' and ``script path spend'' which can provide many alternative complex spending conditions without forcing the user to reveal any of those conditions except the one they actually use at spend time. Thus, P2TR enhances privacy and supports advanced contracts and even promises to enable onchain verification of Turing-complete computations performed offchain~\cite{Linus2023bitvm}. However, P2TR stores the tweaked public key directly in the locking script without hiding it behind a hash value which brings back the quantum vulnerability of P2PK and P2MS addresses and therefore, from the standpoint of quantum security, constitutes a security regression~\cite{Friedenbach2021why, Ruffing2025postquantum}.

\begin{figure*}[t]
    \centering
        \includegraphics[width=\linewidth]{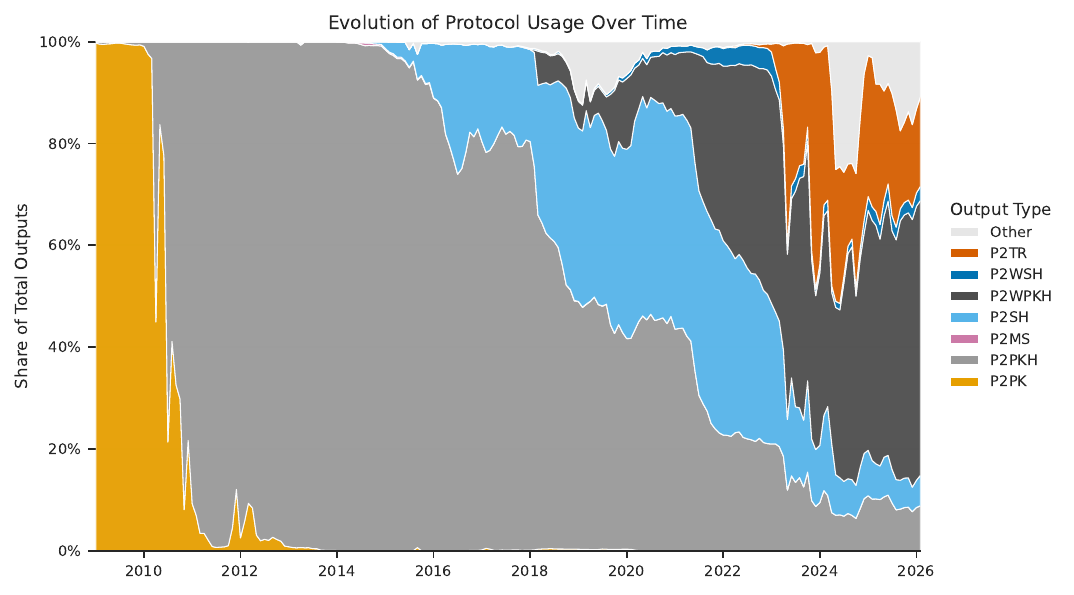}
    \caption{Evolution of Protocol Usage Over Time: The relative market share of transaction output scripts over time, highlighting the network's migration through major protocol upgrades. The early ``Satoshi Era'' (2009--2010) is defined by the Pay-to-Public-Key (P2PK, orange), which was rapidly superseded by the industry-standard Pay-to-Public-Key-Hash (P2PKH, grey). The distinct inflection points in 2017 and 2021 correspond to the activation of the Segregated Witness and Taproot soft forks, respectively. The rapid expansion of P2WPKH (dark grey) and P2TR (red) demonstrates the ecosystem's adoption of modern, weight-efficient cryptographic standards. Plot generated using data from \texttt{bigquery-public-data.crypto\_bitcoin}~\cite{Day2018bitcoin}.} \label{fig:btc_transaction_output_fractions}
\end{figure*}

\begin{figure*}[t]
    \centering
        \includegraphics[width=\linewidth]{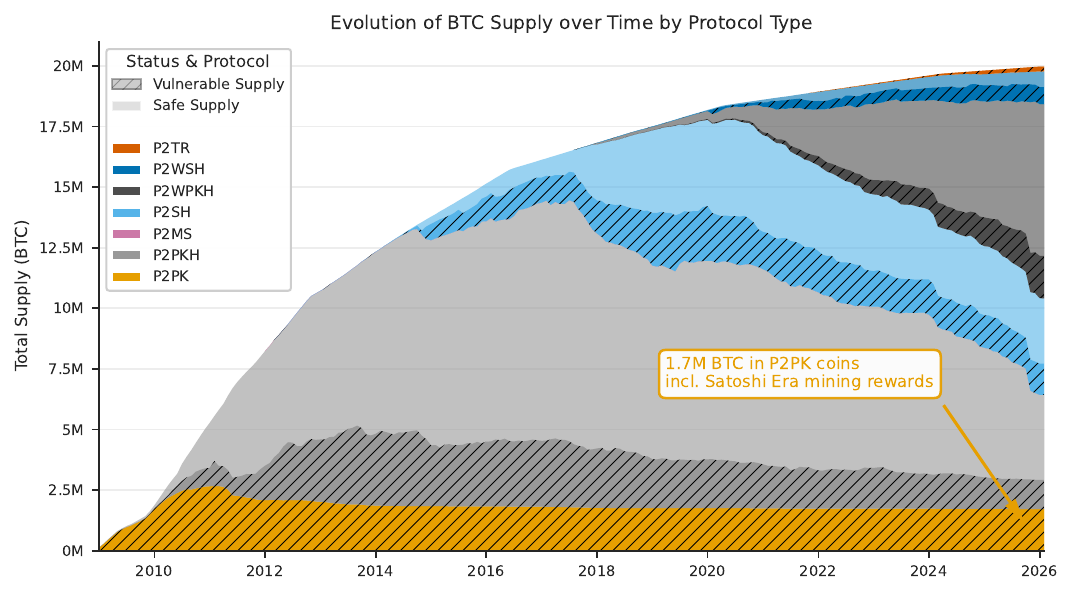}
    \caption{Evolution of BTC supply over time by protocol type. Quantum vulnerable balances are shown in shaded regions for each protocol. P2PK, P2TR and P2MS are considered 100\% vulnerable. The remaining script types are considered vulnerable from having re-used keys if (Addresses that have appeared in an Input) AND (Currently hold a balance in Unspent Outputs). In the case of P2SH and P2WSH, we make the simplifying assumption that if one compromises the script, that it will ultimately equate to being able to steal the bitcoin (in some small number of cases this may not be true). At the time of writing $\sim$6.9M total bitcoin across all protocols are vulnerable. Plot generated using data from \texttt{bigquery-public-data.crypto\_bitcoin}~\cite{Day2018bitcoin}.} \label{fig:btc_supply_over_time}
\end{figure*}

A recent draft Bitcoin Improvement Proposal BIP-360~\cite{Beast2024bip} puts forward a new script type called Pay-to-Merkle-Root (P2MR) which is essentially P2TR with the quantum-vulnerable key path spend removed. This offers many benefits of P2TR without re-introducing vulnerability to at-rest attacks. P2MR, like all other standard script types in Bitcoin, is currently vulnerable to on-spend attacks. Thus, at present, P2MR constitutes a security patch for the Taproot regression. In future, when post-quantum opcodes are introduced in Tapscript~\cite{Wuille2020bip342}, it may become fully quantum-secure. See \fig{btc_transaction_output_fractions} for the relative prevalence of the standard Bitcoin locking scripts over time and \fig{btc_supply_over_time} for the total bitcoin locked by each type of script.

\subsection{The Mechanics of a Quantum Attack on Bitcoin}
\label{sec:bitcoin_mechanics}

An at-rest attack on a Bitcoin address that exposes the public key, such as P2PK or P2TR, proceeds as follows. First, the attacker reads the public key from any past transaction on the blockchain in which the address received bitcoin. Next, the attacker uses their CRQC to derive the private key from the public key. That private key is then used to authenticate a forged transaction locking the bitcoin behind a UTXO that the attacker controls. An at-rest attack on an address that hides the public key behind a hash value, such as P2PKH or P2WPKH, proceeds slightly differently. First, the attacker reads the public key from any past transaction on the blockchain in which the address spent bitcoin. Next, as before, they use their CRQC to derive the private key from the public key. Thus, absence of any spend transactions (and other public records disclosing the public key) denies the attacker the knowledge of the public key, thwarting the at-rest attack.

To understand the on-spend attacks, it is necessary to provide more background on how Bitcoin transactions are executed. First, the spender typically broadcasts their digitally signed transaction to the Bitcoin network where it is placed in a queue called the ``public mempool'' until a miner bundles it together with other transactions into a unit called a ``block''. In order to have the block accepted by the network (and earn the block reward), the miner must solve a cryptographic puzzle - specifically, they must find a number that when included in this block causes its hash to have a value less than a certain difficulty threshold. This takes a considerable amount of computational work. Once the miner has found a solution to the puzzle, they then broadcast the completed block to the network whose nodes check that the block is valid (i.e., correctly formatted, consistent with prior blocks, includes only authentic transaction signatures, contains a correct solution to the cryptographic puzzle, etc.). When other miners construct their own blocks, they choose an ``ancestor'' block to build on top of, and the block reward incentivizes them to choose the tip of the longest valid chain of blocks. As a result, Bitcoin transactions are recorded in a ledger taking the form of a chain of blocks (hence, ``the blockchain''). Due to inevitable network delays and compute power disparities, at any one time, miners might disagree on the last few blocks but their desire to arrest the growth of sunk costs causes them to eventually switch to the longest chain. The more blocks a transaction is buried under, the harder it is to reverse it with the cost of reversal growing exponentially in the number of blocks~\cite{Nakamoto2008bitcoin}. In practice, a transaction recorded a few blocks away from the tip of the blockchain is regarded as ``finalized''. The longest chain rule provides a consensus mechanism while the requirement that blocks include Proofs of Work achieves Sybil resistance, i.e., a means of preventing a single malicious entity from easily gaining disproportionate control or influence by creating a large number of nodes. However, following informal convention we refer to the combination of the longest-chain consensus rule with the Proof-of-Work Sybil resistance as the Proof-of-Work consensus mechanism.

In an on-spend attack, the quantum attacker uses the public mempool as the source of public keys. By the time a transaction is in the mempool, the public key associated with the address it is locked to must be visible, either on the blockchain (as is the case for P2PK and P2TR coins) or in the mempool (as is the case for P2PKH and P2WPKH coins) so that the network can check that the transaction is valid. For example, for a transaction spending P2PKH coins, the public key must be broadcast so that Bitcoin nodes can confirm that the public key indeed hashes to the value that is recorded on the blockchain in order to authenticate the digital signature. Critical for this discussion is the fact that it takes a nontrivial amount of time for a given transaction to be added to a block. On average, a new block is mined about every ten minutes. However, the mining procedure is a stochastic Poisson process, giving rise to an exponential distribution for the time between blocks mined; accordingly, the standard deviation is also about ten minutes. The standard deviation of the mempool wait time is even higher; at times of high network congestion, transactions can take days to process~\cite{CoinPhoton2024bitcoin}. During that time, a quantum attacker can extract the public key from the mempool, use Shor's algorithm to solve for the private key, and then use that private key to broadcast an alternative signed transaction attempting to move the Bitcoin to the attacker's wallet. This ``forged'' transaction could be picked up by miners and added to a block that is ultimately finalized before the original transaction is recorded on the blockchain, thus successfully stealing the bitcoin.

A user spending bitcoin can increase a transaction fee in order to incentivize miners to prioritize their transaction when they construct a new block. Transaction fees are quite important at times of high congestion on the blockchain and, together with a small amount of newly minted bitcoin called ``block subsidy'', form part of the mining reward. The relative importance of transaction fees in mining rewards gradually increases, because the block subsidy shrinks as the system slowly approaches the global monetary supply cap of 20,999,999.9769 BTC~\cite{Walker2025block}. An attacker can flood the mempool with high-fee transactions to gain time for their CRQC to derive the user's private key if needed.

In addition, miners can derive a supplemental revenue stream, known as Miner Extractable Value (MEV)~\cite{Daian2020flash, Chone2025maximal}, from their ability to reorder, insert and remove transactions as they build new blocks. In particular, miners can sometimes extract profit from their control over which transactions are included in the next block, for example by front-running a large trade or benefitting from arbitrage. While MEV is more prevalent on blockchains with more complex financial ecosystems, such as Ethereum, the presence of quantum attackers may provide new MEV opportunities for miners. Bitcoin users might attempt to protect against on-spend attacks by offering miners a high fee to make sure that their transaction has minimal latency. However, a sufficiently fast quantum attacker could choose to make their fee even higher to increase the odds that their transaction is finalized ahead of the original. Such bidding wars might add a new component to the MEV revenue stream.

\begin{figure*}[t]
    \centering
        \includegraphics[width=\linewidth]{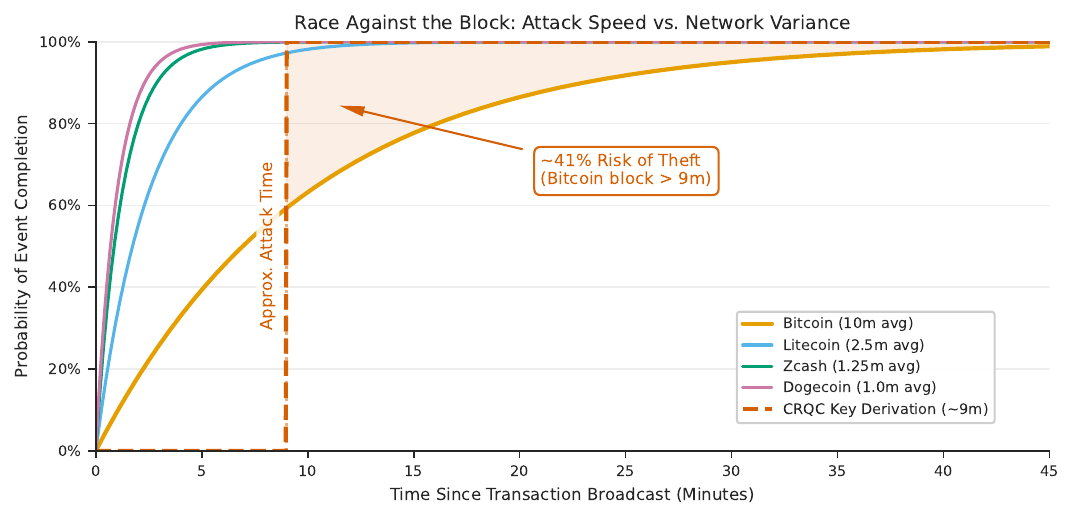}
    \caption{Risk that an on-spend quantum attack using a superconducting qubit CRQC taking approximately 9 minutes to derive a private key succeeds against Bitcoin (with average block time of 10 minutes), Litecoin (2.5 minutes), Zcash (75 seconds), and Dogecoin (1 minute). This plot assumes that the target transaction requires one digital signature (e.g., P2WPKH) and the public key is syndicated to the attacker almost immediately and that once broken (approximately at the ninth minute), the attacker can successfully insert a ``forged'' transaction by offering a higher fee to miners, displacing the original transaction (these assumptions are favorable to the attacker). But we are also assuming zero network congestion, which is favorable to the legitimate cryptocurrency spender. In practice, an attacker can create congestion artificially by flooding the mempool with high-fee transactions.} \label{fig:btc_block_and_network_time}
\end{figure*}

The timing and success probability of on-spend attacks are depicted in \fig{btc_block_and_network_time}. The probability of success is slightly less than 41\% under the idealized assumptions stated in the caption. However, if our current estimate of around 9 minutes per private key derivation on a superconducting qubit CRQC could be brought down by a few minutes, then even high-fee transactions at times of low congestion would be likely to get compromised by a quantum attack. As discussed earlier, it is likely that the quantum attack time will indeed come down further, first due to further theory advances reducing the cost of Shor's algorithm, and then due to increases in the number of qubits available to CRQCs relative to our present assumptions (our current quantum algorithm trades off time in order to fit on as few qubits as possible; doubling the number of qubits available to it would double its speed). The cost of quantum attacks on multi-signature transactions, such as P2WSH, increases linearly with the number of signatures required. However, these attacks parallelize trivially, so an attacker with a sufficient number of CRQCs can achieve the probability shown in \fig{btc_block_and_network_time}. Reducing block interval makes on-spend attacks significantly harder. For example, we estimate the success probability of an on-spend attack using a superconducting qubit CRQC against Litecoin, a Bitcoin derivative which targets 2.5 minutes between blocks, to be less than 3\%. The chance of a successful on-spend attack on Zcash, whose target block time is 75 seconds, is less than one in thirteen hundred and on Dogecoin, whose target block time is 1 minute, is less than one in eight thousand.

Once the private key is derived, the attacker can engage the legitimate owner in a 'Replace-By-Fee' (RBF) bidding war. Because the attacker is stealing funds they do not own, they can rationally bid transaction fees up to incentivize miners to prioritize their theft. The victim might be forced into a 'scorched earth' scenario: to outbid the attacker and prevent the theft, they may have to offer a fee that may consume virtually the entire value of the asset. In this state, the transfer of wealth shifts from the victim to miners, rendering the asset effectively lost even if the theft is technically thwarted. In reality, mining operations are real-world businesses that can set up their own policies for handling such unusual situations. This creates an opportunity to seek outcomes that are more favorable to Bitcoin users than the simple scorched earth scenario suggests. In any case, if a quantum attacker cannot break the cryptography in less than say, 30 minutes or an hour, it seems likely that with a sufficiently high fee (or when the network congestion is sufficiently low), any transaction can be protected from a CRQC.

Legacy Bitcoin software allowed spenders to disable RBF for their transactions and implemented a relay policy known as the ``first-seen'' rule. RBF was also disabled in Bitcoin Cash following its hard fork from Bitcoin. Under the first-seen rule, nodes encountering conflicting transactions would only relay the first one they saw rather than the one with the highest fee. However, the first-seen rule is sensitive to propagation latency and, more importantly, exists outside the consensus rules and is therefore unenforceable by the network. Indeed, miners have a financial incentive to prioritize high-fee transactions over first-seen transactions regardless of whether RBF is enabled or disabled. Consequently, disabling RBF was never a reliable protection against on-spend attacks.

Some Bitcoin mining operations allow users to bypass the public mempool by submitting transactions directly to miners for a fee~\cite{Nichols2024marathon}. Currently, these services are used to enhance privacy and efficiency and to enable users to utilize transactions that adhere to consensus rules, but are non-standard (such transactions can be included in valid blocks but are normally not propagated by the network). If Bitcoin fails to migrate to PQC before on-spend attacks become viable, these services could supplant the public mempool as a trust-based mechanism for achieving a moderate degree of quantum safety. However, like other mitigation measures against on-spend attacks, such as high fees and multi-signature scripts, these services reduce the quantum risk without eliminating it completely. For example, in the event of a blockchain reorganization, the orphaned block produced by the user's trusted miner becomes visible on the network potentially exposing the user's public key to a quantum attacker.

On-spend attacks may also interfere with Bitcoin consensus. The first time that the public key of a high value Bitcoin address appears in a block on chain, a quantum attacker could offer miners a reward to not mine on this block, thereby causing the block to become orphaned. This effectively cancels all transactions in this block. In parallel, the attacker would use its CRQC to break the just published public key included in the orphaned block. Once it has the secret key it issues a transaction to steal the assets from the target address, and this transaction will become part of the next (non-orphaned) block, thereby transferring the assets to the attacker. This reorganization of the chain is caused by the attacker, funded by the stolen funds, and is made possible by the CRQC.

\subsection{The Origins of Quantum Vulnerabilities in Bitcoin}
\label{sec:bitcoin_origins}

All quantum attacks on Bitcoin transactions are ultimately based on the ability of a CRQC to derive the private key from a public key. The most important distinction between different types of quantum vulnerabilities in Bitcoin is how the public keys become known to an attacker. Broadly speaking, there are four ways the attacker can learn the public key: they can retrieve it from a locking script or an unlocking script recorded on the blockchain, from a pending transaction in the mempool, or from offchain records. These possibilities correspond to the following vulnerabilities:

\begin{enumerate}
    \item \textbf{Weak Address} vulnerability arises when the locking script directly reveals the public key of the wallet that is receiving bitcoin by recording it on the blockchain. In this case, the attacker has a long period of time to try to solve for the private key. The locked bitcoin remains vulnerable until it is spent. Weak Address vulnerabilities are caused by the use of locking scripts that leave public keys unprotected, namely P2PK, P2MS (in case of old assets) and P2TR (in recent transactions).

    \item \textbf{Address Reuse} vulnerability arises when the unlocking script recorded in a spend transaction on the blockchain reveals the public key that protects other assets on the ledger. This vulnerability pertains to addresses that hide public keys behind a cryptographic hash, namely P2PKH, P2SH, P2WPKH and P2WSH, and occurs when some but not all bitcoin at an address is spent. The spending transaction reveals the public key, defeating the protection normally afforded by the cryptographic hash. The vulnerability persists until all bitcoin locked to the address is spent. Thus, the vulnerability arises from reuse of Bitcoin addresses.

    \item \textbf{Public Mempool Exposure} vulnerability comes about when the unlocking script exposes the public key of a wallet that is spending bitcoin by revealing the key while broadcasting the transaction to the public Bitcoin mempool. Assuming no reuse or other prior exposure of the public key, the bitcoin remains vulnerable only while it awaits settlement in the public mempool, i.e. until it is permanently recorded on the blockchain. At present, all bitcoin is vulnerable to Public Mempool Exposure or reorganization attacks.

    \item \textbf{Offchain Exposure} vulnerability results from public key disclosure as an oblique consequence of Bitcoin's transaction settlement protocol. For example, the owner may have spent digital currency from the public key on another blockchain, such as Bitcoin Cash or Rootstock, or they may have elected to share their public key with a third party, such as a portfolio tracking app. Offchain Exposure vulnerability arises from user practices and business mechanisms built on top of Bitcoin.
\end{enumerate}

Weak Address and Address Reuse vulnerabilities are two types of Onchain Exposure vulnerability~\cite{Deegan2025quantum}. Public Mempool Exposure vulnerability enables only on-spend attacks. By contrast, Onchain and Offchain Exposure vulnerabilities enable both on-spend and at-rest attacks. The latter can be launched using any CRQC, including slow-clock devices. \tab{bitcoin_vulns} summarizes quantum vulnerabilities of the most prominent existing and proposed script types while Figures 6 and 7 quantify the scale of affected assets.

To understand how Mempool and Onchain Exposure vulnerabilities arise in practice, consider a first time owner who receives their initial bitcoin at an address with the \texttt{bc1p} prefix. These addresses represent P2TR locking scripts that directly record the public key on the blockchain in the receive transaction causing an Onchain Exposure vulnerability that places the user at immediate risk of both on-spend and at-rest attacks from the moment they receive bitcoin. In contrast, a first time owner who receives their initial bitcoin at a previously unused address with the bc1q prefix enjoys potential immunity to at-rest attacks. These addresses represent P2WPKH and P2WSH locking scripts that hide the public key behind a fingerprint, so as long as the user does not share their public key, e.g. by spending some of the bitcoin, there is no public key for an attacker to target with a quantum computer and the user's asset is immune to at-rest attacks. In particular, their bc1q address may continue to receive bitcoin. However, any transaction that spends some, but not all bitcoin at this address introduces Address Reuse vulnerability. From this point on, all bitcoin still locked to this address is vulnerable to both on-spend and at-rest attacks.

Thus, the presence of Address Reuse vulnerability simplifies, but also darkens, this complex situation. If the nominally safer bc1q address has been used previously, its public key is already present on the blockchain in an older transaction that an attacker can easily identify from the public key's has value. Thus, transactions recorded on the blockchain act as a ``cheat-sheet'' that enables an attacker to solve the otherwise completely intractable problem of computing the public key from its hash. Consequently, public key reuse, and indeed any other public key exposure, renders all types of locking scripts equally vulnerable to quantum attacks: the safer scripts, such as P2PKH and P2WPKH, become as vulnerable as the weaker scripts, such as P2PK and P2TR.

Offchain Exposure vulnerabilities arise from user practices and community conventions. Privacy and security recommendations dating back to Satoshi's original Bitcoin whitepaper~\cite{Nakamoto2008bitcoin} dictate that users generate a new address, and hence a new pair of public and private keys, for every transaction. These guidelines are critical in a world with CRQCs if migration to PQC has not yet occurred, especially in Scenario 2. Indeed, these recommendations, when extended with the rule to only use SegWit addresses (corresponding to P2WPKH and P2WSH scripts) and thus to refrain from using Taproot addresses (corresponding to P2TR scripts), can protect users from at-rest attacks which are the only attacks that a slow-clock CRQC can launch. However, the core recommendation against public key reuse pre-dates concerns about quantum cryptanalysis, because even in a world without quantum computers public key reuse weakens the user's security (e.g., by exposing them to implementation flaws and side-channel attacks) and privacy (e.g., by facilitating transaction linking, wealth estimation, and deanonymization).

\definecolor{tableblue}{HTML}{C5D9F1}
\definecolor{tablered}{HTML}{F2C4C4}
\definecolor{tableorange}{HTML}{FDE9D9}

\begin{table*}[t]
\centering
\begin{tblr}{
  colspec = {Q[m,c,font=\bfseries]|Q[m,c]|Q[m,c]|Q[m,l,4cm]},
  row{1} = {bg=tableblue},
  hlines, vlines,
  row{2,4,8} = {bg=tablered}, 
  row{3,5,6,7} = {bg=tableorange},
  row{9} = {bg=tableorange, font=\itshape}, 
}
  Script type & Quantum vulnerability & Address prefix & Remarks \\
  P2PK & {onchain\\mempool\\offchain} & n/a & Includes Satoshi era mining rewards \\
  P2PKH & {mempool\\offchain} & 1... & \\
  P2MS & {onchain\\mempool\\offchain} & n/a & legacy multi-signature script type supplanted by P2SH, P2WSH, and P2TR \\
  P2SH & {mempool\\offchain} & 3... & \\
  P2WPKH & {mempool\\offchain} & \texttt{bc1q...} & currently most popular script type \\
  P2WSH & {mempool\\offchain} & \texttt{bc1q...} & \\
  P2TR & {onchain\\mempool\\offchain} & \texttt{bc1p...} & most recently introduced script type \\
  P2MR & {mempool\\offchain} & \texttt{bc1z...} & at-rest attack mitigation proposal under discussion \\
\end{tblr}
\caption{Quantum vulnerabilities of existing (roman) and proposed (\textit{italics}) Bitcoin script types. Every existing Bitcoin transaction type is vulnerable to on-spend quantum attacks by a potential future fast-clock CRQC, such as one based on superconducting qubits. In addition, bitcoin locked using old P2PK scripts (including $\sim$1.7 million bitcoin in Satoshi era mining rewards) and bitcoin locked using modern P2TR scripts (which moved $\sim$16.8 million BTC in 2025 and which represented 21.68\% of all Bitcoin transactions in 2025) are vulnerable to at-rest attacks by potential future slow-clock CRQCs (per TX outputs in blocks 877259 through 930340). Custom script types, including the unspendable \texttt{OP\_RETURN} (which can be used to permanently destroy coins or to embed arbitrary data on the blockchain), are not listed.}
\label{tab:bitcoin_vulns}
\end{table*}

\begin{figure*}[t]
    \centering
        \includegraphics[width=\linewidth]{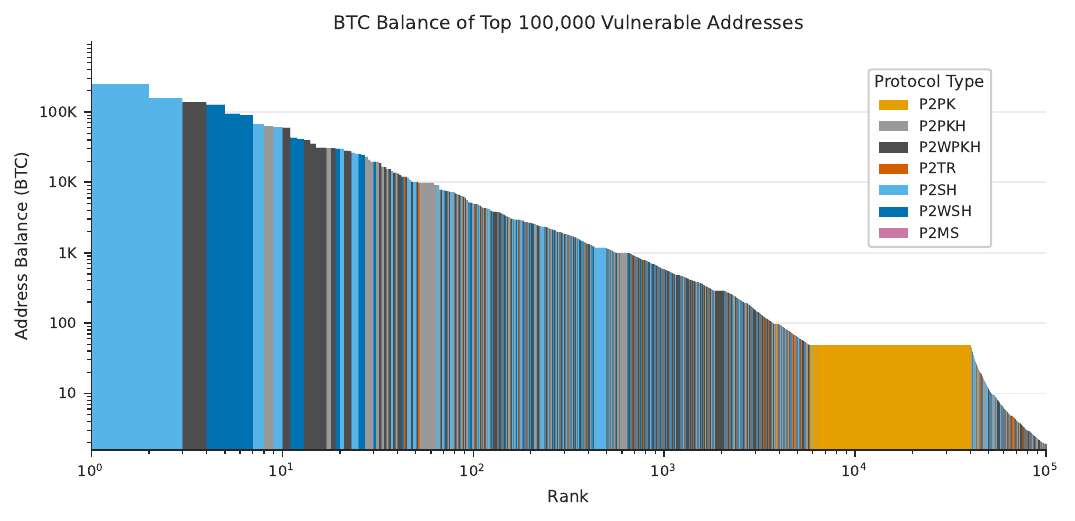}
    \caption{BTC Balance of Top 100,000 Vulnerable addresses. The graph displays the BTC balance of the top 100,000 Bitcoin addresses ranked by value that are vulnerable to at-rest attacks from exposed or re-used keys. In the case of P2PK, P2TR etc. this vulnerability comes from publishing the public key, In the case of other protocols this comes from public key re-use. At around address rank 6000 there is a very large portion of many exposed 50 BTC addresses from the early P2PK mining era because that amount was the mining reward at that point in time. The sum of all BTC in vulnerable addresses is approx. 6.7 million BTC. Plot generated using data from \texttt{bigquery-public-data.crypto\_bitcoin}~\cite{Day2018bitcoin}.}
    \label{fig:btc_vulnerable_addresses}
\end{figure*}

Nevertheless, Bitcoin addresses are often reused for convenience. For example, merchants and exchanges may prefer to publish a single destination address that is stable and recognizable. In fact, public key reuse is extremely pervasive. We quantify the exposure of the largest holdings in \fig{btc_vulnerable_addresses} where all assets are vulnerable due to public key reuse except those locked by P2PK, P2MS, and P2TR scripts (where public keys are exposed even without address reuse). More detailed and comprehensive data is available in the ``Project 11 Risq List''~\cite{112026bitcoin} which is a repository of public keys at risk of at-rest attacks. Some of the large holdings in \fig{btc_vulnerable_addresses} are linked to Binance, Robinhood and Bitfinex, which are also three of the largest cryptocurrency exchanges. In fact, many cryptocurrency services are guilty of address reuse~\cite{Stutz2026reuse}. Reusing public keys helps financial service providers avoid logistical challenges of endless account whitelisting, makes it easy to prove reserves, and helps to reduce fees that they pay in transactions. The key reuse problem also manifests if one uses the same private key on two or more blockchains (e.g., Bitcoin and Rootstock) or if the original blockchain is forked (e.g., Bitcoin and Bitcoin Cash). Spending from an address on one blockchain enables at-rest attacks on the address on all blockchains.

A convenient way of addressing some of the practical challenges of avoiding key reuse is presented by modern Hierarchical Deterministic (HD) wallets~\cite{Wuille2012bip} which make it easy for users to automatically generate a large number of new keys and addresses in a hierarchical structure. At the top of the hierarchy is a secret phrase called a ``seed'' which an HD wallet uses to derive so-called extended keys which can in turn be used to derive further extended keys as well as regular keys used for cryptocurrency transactions. There are two types of key derivations: normal and hardened. In both cases, private keys can only be derived from the parent extended private key. However, normal derivation allows public keys to be derived in two ways: from the parent extended private key and from its corresponding extended public key. By contrast, hardened derivation allows public keys to only be derived from the extended private key. 

In the absence of CRQCs, it is safe to share extended public keys and use normal derivation to obtain keys for cryptocurrency transactions. This can be used for example to enable a third party service to independently derive a large number of a given user's public keys without access to their corresponding private keys  ---  a useful feature that enables automatic monitoring of many Bitcoin addresses. Nevertheless, even without quantum computers, normal derivation allows an attacker to recover the extended private key, and hence any derived private key, from the knowledge of a single derived private key and the extended public key. Therefore, in the presence of CRQCs, reuse or disclosure of an extended public key constitutes a pernicious form of Offchain Exposure vulnerability that enables an attacker with a CRQC to compromise multiple private keys at the cost of one. Third party services, such as wallets, portfolio monitoring apps etc., that require sharing of extended public keys or that build centralized repositories of extended public keys potentially aggravate quantum vulnerabilities in the ecosystem.

Indeed, much of the decentralized financial ecosystem built on top of the Bitcoin blockchain was constructed under the assumption that ECDLP-based public keys are safe to share and reuse. This assumption fails in the presence of CRQCs. At the same time, the system's size and complexity reflect the fact that this assumption often helps to address genuine business needs. However, the possibility of Scenario 2 materializing in near future underscores the importance of greater caution in user and business practices, e.g., by reducing or eliminating reliance on address reuse. These measures will provide a modicum of quantum safety to security-conscious Bitcoin users who are concerned about the possibility of attacks by slow-clock CRQCs and who are willing to eschew the convenience of financial services based on public key reuse. However, such mitigation measures are likely to be seen as insufficient for the broader Bitcoin community. Therefore, the only real long-term remedy to Bitcoin's quantum vulnerabilities lies in migrating to PQC keys for signing transactions.

\subsection{The Infeasibility of Quantum Attacks on Proof-of-Work}
\label{sec:bitcoin_pow}

A significant portion of literature addressing quantum vulnerabilities in Bitcoin seems to touch on the topic of using Grover's algorithm to accelerate mining~\cite{Sattath2020insecurity, Nerem2023conditions, Milton2025bitcoin, Vescovo2025coming} and informal discourse in social and trade media indicates a persistent belief that quantum computers may pose a threat to Bitcoin's Proof-of-Work consensus mechanism~\cite{Musk2025grok, Neagle2025BitcoinQC}. However, we maintain that this is not something to worry about in the next several decades.

Grover-based attacks on Bitcoin mining are not practically relevant for two reasons. First, the quadratic quantum speedup from Grover's algorithm is all but consumed by the overheads of quantum error correction~\cite{Babbush2021focus}. Second, Grover's algorithm does not parallelize well~\cite{Zalka1999grovers}. In Bitcoin mining, hardware acceleration and massive parallelization are a much greater advantage than the modest speedup from Grover's algorithm.

Under the fantastical assumption that a CRQC could compute SHA-256 in a single 1 microsecond error correction cycle, a quantum miner could achieve a hashrate of $0.25\,\,TH/s$ which is more than two orders of magnitude below $110\pm 3\%\,\,TH/s$ of the popular ASIC miner S19 Pro~\cite{Bitmain2021S19Pro}. Under realistic assumptions, rather than fantastical ones, quantum miner's hashrate plummets by over 10 orders of magnitude~\cite{Amy2016estimating}.

Of course, the long arc of technology is difficult to predict. Quantum computers enable efficient simulation of quantum systems at scale, accelerating innovations in materials science. Among the many fields of technology that stand to benefit from this new capability is the development of quantum hardware itself. Thus, in the long term it is possible that currently unanticipated new quantum architectures may emerge which might allow future generations to build quantum computers with very fast error correction cycles on which Grover speedup becomes relevant in Proof-of-Work mining. But for now, quantum mining remains science fiction more than a concrete threat.

Thus, CRQCs pose no direct threat to Bitcoin mining. However, they do pose an indirect one: if the Bitcoin community does not prepare for the arrival of CRQCs in time and quantum attacks on Bitcoin transactions lead to rapid decline in the value of the cryptocurrency, then the fiat value of mining rewards might change faster than the network is able to adjust mining difficulty, potentially rendering mining unprofitable. This could push some mining pools to shut down a portion of their capacity increasing the average block interval. This would reduce transaction rate which would, in turn, affect user experience and make on-spend attacks easier. These issues arise from Bitcoin's fixed mining difficulty adjustment schedule which recomputes difficulty targets once every 2016 blocks (about once every two weeks). By contrast, Bitcoin Cash adjusts mining difficulty after every block~\cite{Ilie2020unstable}.

\section{Quantum Vulnerabilities of other ECDLP-Based Cryptographic Protocols}
\label{sec:protocols}

ECDLP-based digital signatures are a basic cryptographic protocol that introduces quantum vulnerabilities into the digital economy. However, many blockchains, including Ethereum, also take advantage of more advanced ECDLP-based cryptographic primitives, such as signature aggregations, key exchange protocols, commitment schemes, and zero-knowledge arguments, in order to enhance privacy and improve scalability. Here, we give a synopsis of these cryptographic protocols and the novel quantum vulnerabilities they introduce. We also discuss a few special elliptic curves and how they influence quantum vulnerabilities.

In certain applications, such as Ethereum's Proof-of-Stake consensus mechanism, a system may need to aggregate hundreds or thousands of digital signatures into a few dozen. This can be accomplished using ECDLP-based signature aggregation protocol, known as the Boneh-Lynn-Shacham (BLS) scheme~\cite{Boneh2001short}. BLS signatures depend crucially on the ability to efficiently compute a certain non-degenerate bilinear map, called pairing, and therefore require specialized elliptic curves, called pairing-friendly curves, for which pairing can be efficiently computed. This is impossible on the secp256k1 curve, used for most digital signatures in Bitcoin and Ethereum, so protocols that rely on pairings utilize other curves, such as $\text{alt\_bn128}$~\cite{Reitwiessner2017eip196} from the Barreto-Naehrig curve family~\cite{Barreto2005pairingfriendly} and BLS12-381~\cite{Jancar2026bls12381} from the Barreto-Lynn-Scott family~\cite{Barreto2002constructing}.

Key exchange protocols enable two parties in direct communication over a public channel to establish a shared secret that passive eavesdroppers cannot recover from the exchanged messages. Diffie-Hellman (DH) key exchange~\cite{Diffie1976new} and its variant based on elliptic curves called Elliptic Curve Diffie-Hellman (ECDH) key exchange are the most well-known examples. Both DH and ECDH are based on the discrete logarithm problem and allow passive eavesdroppers with a CRQC to recover the shared secret, breaking the privacy normally afforded by the protocol. ECDH is used in certain privacy-preserving blockchains, such as Zcash~\cite{Hopwood2026zcash}, Monero~\cite{koe2020zero}, and Litecoin's Mimblewimble~\cite{Burkett2020lip}.

Commitment schemes are cryptographic protocols that allow one to publish a value that commits them to a choice without revealing that choice. Such schemes have two properties: the choice cannot be altered (binding) and remains secret (hiding). Homomorphic commitments have the additional property that the sum of individual commitments to a set of values equals the commitment to the overall sum. In the context of cryptocurrency transactions, such commitment schemes can be used to hide transaction amounts while enabling the network to verify that inputs and outputs of each transaction balance correctly. Popular examples of homomorphic commitments based on ECDLP are ElGamal~\cite{Elgamal1985public} and Pedersen~\cite{Pedersen1992noninteractive, Pedersen1992noninteractive} commitment schemes. Quantum attacks on the former destroy the hiding property and on the latter destroy the binding property~\cite{Ruffing2017switch, Pointcheval2017commitment}. Consequently, if such a commitment scheme is used to hide transaction values on a privacy-preserving blockchain, a quantum attacker can break the privacy protection in the former case or create new coins in the latter case. An example of a more sophisticated commitment scheme is the Kate-Zaverucha-Goldberg (KZG) protocol~\cite{Kate2010constantsize} which allows a party to commit to a polynomial and subsequently to provide efficient openings that reveal evaluations of the polynomial. KZG scheme is currently used for example in Ethereum's DAS mechanism~\cite{Ryan2024eip7594}.

Related cryptographic protocols are the Zero-Knowledge (ZK)~\cite{Goldwasser1985knowledge} proofs and arguments, including those we use in this paper as the means of verifying quantum resource estimates without disclosing sensitive attack details. These protocols allow one party to convince another that a statement is true without revealing the facts that make it true. In addition to this zero-knowledge property, a secure ZK protocol is supposed to be complete (proofs and arguments of true statements pass verification) and sound (proofs and arguments of false statements are almost guaranteed to fail verification). The distinction between proofs and arguments concerns the computational power of the proving party: in a proof the prover is assumed computationally unbounded while in an argument the prover's computational resources are limited~\cite{Thaler2022proofs}. If the statement to be proven takes the form ``prover knows secret X'', then the protocol is called a proof or argument of knowledge. ZK protocols are typically interactive and involve one party challenging the other to compute values that depend on, but do not reveal, the secret. However, the Fiat-Shamir heuristic~\cite{Fiat1987to} allows one to turn an interactive protocol into a non-interactive one. This transformation is a core component of efficient ZK protocols known as Zero-Knowledge Succinct Non-interactive Arguments of Knowledge (zkSNARKs)~\cite{Bitansky2012from}. Non-interactive protocols that possess the succinctness, soundness and completeness property, but not necessarily the zero-knowledge property, are called SNARKs. In blockchain technologies, zkSNARKs can be used to enhance privacy while SNARKs (zero-knowledge or otherwise) can be used to achieve scalability and improve efficiency.

Many commitment schemes and ZK protocols rely on the assumption that nobody knows certain secrets related to the protocol's fixed public parameters. For some schemes, such as the KZG polynomial commitments, the secrets, sometimes referred to in this context as ``toxic waste'', are known temporarily before being erased during the trusted setup ceremony~\cite{Wang2025trusted} in the course of which the protocol parameters are first established. For other schemes the parameters are produced by methods satisfying the Nothing Up My Sleeve (NUMS) principle~\cite{Koshelev2022generation, Wikipedia2026Nothing, Brown2020rolling} to ensure that nobody learns them in the process of choosing protocol parameters. However, for certain ECDLP-based protocols the secrets can be derived from the fixed public parameters using a CRQC. Moreover, since protocol parameters are fixed, this needs to be done only once and the resulting secrets can be reused to launch multiple attacks later using classical computers. Example protocols vulnerable to such on-setup attacks include Pedersen commitments used in Mimblewimble, KZG commitments used in Ethereum's Data Availability Sampling mechanism, and BulletProofs~\cite{Bunz2017Bulletproofs} used in Monero and Mimblewimble.

Consequences of a quantum attack on an ECDLP-based SNARK generally depend on which cryptographic property of the underlying commitment scheme is quantum vulnerable. For instance, a quantum break of the commitment's binding property collapses the soundness property of the SNARK, allowing forged validity proofs for unauthorized transactions to pass verification. Moreover, in privacy-preserving settings such attacks can remain undetected until systemic effects materialize. For example, a CRQC enables on-setup attacks on the Tornado Cash protocol in which attackers withdraw more funds than they deposited. Due to the privacy properties of the protocol, they can drain an anonymity pool without anybody noticing until the pool's balance falls to zero.

The succinctness of SNARKs can be used to build Layer 2 (L2) protocols~\cite{Gudgeon2019sok} on top of a Layer 1 (L1) blockchain, such as Ethereum, to increase transaction rate and reduce fees. L2 protocols that use SNARKs are known as zk-rollups. They process a large number of transactions offchain and bundle them together into batches that are subsequently committed to the underlying L1 blockchain using SNARKs. Quantum vulnerabilities in these protocols potentially enable attackers to convince the onchain verifier that a batch of L2 transactions is valid even when it includes invalid transactions that, for example, steal assets or inflate monetary supply.  In the long term, these vulnerabilities may be remediated by switching to hash-based or lattice-based protocols which are regarded as post-quantum secure. Indeed, some zk-rollups, such as Starknet, use hash-based protocols which are believed to be resistant to quantum cryptanalysis.

Quantum-vulnerable SNARKs typically employ ECDLP-based commitment schemes which, like BLS signatures, require pairing-friendly elliptic curves. Our quantum resource estimates discussed earlier apply directly to solving ECDLP on curves such as secp256k1 which is not pairing-friendly. The estimates are sensitive to curve parameters and do not apply unchanged to all elliptic curves used in blockchain cryptography. We have not conducted rigorous resource estimates for ECDLP on other elliptic curves, such as those used in SNARKs. However, we expect the cost of solving ECDLP on many curves over finite fields with 256-bit modulus and 256-bit group order to be on the same order of magnitude. This includes the cost of ECDLP on the Pasta curves~\cite{Team2026pasta} employed in zkSNARKs behind the newest type of shielded (private) transactions in Zcash. We expect the cost to be only moderately higher for ECDLP on other 256-bit curves such as $\text{alt\_bn128}$~\cite{Reitwiessner2017eip196} whose efficient implementation is provided as precompiled smart contracts~\cite{Contributors2026precompiled} on Ethereum and used for example in zkSNARKs behind its cross-chain integrations and zk-rollups. However, ECDLP on curves over larger finite fields may require additional quantum memory. In particular, the BLS12-381 curve~\cite{Jancar2026bls12381, Barreto2002constructing} used in older types of shielded transactions in Zcash and in Ethereum's Proof-of-Stake consensus mechanism, requires 381 bits to store point coordinates. This leads to a 50\% increase in the size of some quantum registers and implies that solving ECDLP on BLS12-381 may require a somewhat larger CRQC than the above 256-bit curves. Nevertheless, because of the efficient scaling of Shor's algorithm~\cite{Gidney2025to}, we still expect ECDLP on this curve to be accessible to the first CRQCs.

The amount of time and the number of qubits that an early CRQC will need to solve ECDLP on a specific elliptic curve increases with the modulus and the order of the elliptic group. Asymptotically, this increase is modest (merely polylogarithmic) due to the efficiency of Shor's algorithm. However, the question of whether increasing the modulus and the group order are appropriate stopgap measures to protect ECDLP-based cryptosystems against attacks by early CRQCs is more complex. For example, solving ECDLP on an elliptic curve with 1024-bit modulus and group order may require a CRQC with almost five thousand logical qubits. Moreover, quadrupling the number of bits in the modulus and group order will increase the runtime of Shor's algorithm by roughly a factor of 64, although (as mentioned before) the runtime can be reduced if the number of qubits is increased. Similarly, an attack on the strongest ECDLP-based cryptography in the TLS protocol, which is used to encrypt and authenticate HTTPS traffic and which currently supports a 521-bit elliptic curve~\cite{rfc8446}, may require a slightly larger CRQC than an attack on the BLS12-381 curve and the other curves discussed here.

However, the question whether a switch to a large, e.g. 1024-bit, modulus can provide blockchains with a temporary protection is complex and depends on a detailed understanding of scaling barriers in leading quantum computing platforms within relevant machine sizes. Given broad progress across multiple hardware architectures, the safe assumption is that there may be little time between the breaking of 256-bit ECDLP and the breaking of 1024-bit ECDLP. Therefore, large elliptic curves may provide no real protection against at-rest attacks. At the same time, such curves would, at least initially, increase the time needed to derive private keys, so they may provide a short reprieve from on-spend attacks. In any case, we expect the reprieve to shrink as CRQCs become larger and are able to trade quantum memory for time. Thus, the security benefits of large elliptic curves are, at best, partial and temporary and, at worst, nearly non-existent.

\section{Quantum Vulnerabilities of the Ethereum Blockchain}
\label{sec:ethereum}

Ether (ETH), is the native digital asset stored and traded on the Ethereum blockchain. It is the world's second largest cryptocurrency by market capitalization (approximately 400 billion USD as of February 2026), second only to Bitcoin (approximately 1.9 trillion USD as of February 2026). However, relying solely on cryptocurrency market capitalization obscures the full economic weight of the Ethereum network. When measured by the Total Value Secured (TVS)\cite{Saggese2025towards}, which sums the native ether, over 150 billion USD in fiat-pegged stablecoins, tokenized Real World Assets (RWAs), and the fast-growing Layer 2 ecosystem, Ethereum's aggregate economy amounts to over 600 billion USD. In contrast, Bitcoin's ecosystem (including Layer 2s like Lightning Network~\cite{Poon2016bitcoin} and tokens) adds a relatively modest $\sim$15 billion USD to its base value, resulting in a TVS still largely dominated by the 1.9 trillion USD market cap of the cryptocurrency itself.

While Bitcoin is a world-wide distributed ledger recording transactions in a single native digital asset, Ethereum is effectively a world-wide distributed computer that employs blockchain technology to store value and information~\cite{Buterin2014ethereum, Wood2025ethereum}. It does not merely record balances, but executes code which implements a wide variety of financial, semi-financial and non-financial applications, such as derivatives, hedging contracts, savings accounts, wills, reputation systems, peer-to-peer gambling applications, prediction markets, non-fungible tokens (NFTs) and online voting systems~\cite{Buterin2014ethereum}. By creating trust in a trustless environment through code and cryptographic protocols rather than through centralized institutional intermediaries, the system enables new forms of financial arrangements, such as decentralized exchanges and peer-to-peer lending, known as Decentralized Finance (DeFi). Ethereum smart contracts~\cite{Szabo2018smart, Wood2025ethereum}, which enable realization of these diverse applications, are sufficiently powerful to facilitate implementation of the bylaws of an entire business organization in the form of a stateful long-term smart contract sometimes referred to as a Decentralized Autonomous Organization (DAO)~\cite{Buterin2014ethereum}. The combination of smart contracts and blockchain technology creates a smart legal system for a digital economy in which smart contracts describe and enforce the rules and the blockchain acts as a decentralized notarial service providing a shared source of truth about the economy's current state. Ethereum's grand ambition is to become ``a decentralised secure social operating system'' that enables business contracts to ``algorithmically specify and autonomously enforce rules of interaction''~\cite{Wood2025ethereum}.

Ethereum's quantum security posture is very different from Bitcoin's. For example, Ethereum produces new blocks in deterministic 12-second slots with most transactions processed in less than a minute (see Figure B.3 in~\cite{Zhao2023cost}). Therefore, our estimates indicate that early fast-clock CRQCs are unlikely to be able to launch on-spend attacks against Ethereum. Moreover, the Ethereum ecosystem has sophisticated private mempools, including TEE-based BuilderNet, that allow users to submit transactions directly to builders which provides a mitigation for potential on-spend attacks.

The main quantum threat against Ethereum lies in a variety of at-rest attacks. The Ethereum blockchain has three prominent features differentiating it from Bitcoin that give rise to five types of quantum vulnerabilities. First is the account model: unlike Bitcoin, where currency units exist as UTXOs, Ethereum maintains global state through persistent accounts. While users can rotate addresses, certain ecosystem incentives (maintaining DeFi positions, governance history and stable identifiers) can incentivise the use of longer lived accounts~\cite{Wood2025ethereum}. Once an account initiates a transaction, its public key remains exposed indefinitely. Second is the native support for smart contracts which are autonomous computer programs executed by the Ethereum Virtual Machine (EVM) that can hold assets and enforce complex rules without intermediaries~\cite{Wood2025ethereum}. Third is the Proof-of-Stake (PoS) consensus mechanism in which Sybil resistance based on energy-intensive mining is replaced with a system where ``validators'' secure the network by staking their own capital as collateral~\cite{Kalinin2021eip3675, Buterin2020combining, LewisPye2023permissionless}. 

These three architectural choices introduce five distinct attack vectors: the Account Model creates the Account Vulnerability (due to the use of quantum-vulnerable ECDSA~\cite{Johnson2001elliptic}); Smart Contracts introduce Admin Vulnerability~\cite{Palladino2019erc1967} and Code Vulnerability; and the specific cryptography required for validators gives rise to the Consensus Vulnerability (due to the use of BLS signatures~\cite{Boneh2001short}) and Data Availability Vulnerability (due to the use of KZG commitments~\cite{Kate2010constantsize}). In the following sections, we will define each type of vulnerability, quantify risk (in native ETH and other assets), estimate the cost of quantum attacks and review some mitigation measures. We summarize these vulnerabilities in \tab{eth_vulnerabilities_tab}.

\begin{table*}[t]
    \centering
    \renewcommand{\arraystretch}{1.3}
    \begin{tabular}{ p{2.5cm} p{3.5cm} p{3.2cm} p{3.8cm} p{4.0cm} }
        \hline
        \textbf{Ethereum Component} & \textbf{Vulnerability} & \textbf{Vulnerable Cryptographic Primitives} & \textbf{Assets at Risk} & \textbf{Second Order Effects} \\
        \hline
        
        Account model & Account Vulnerability & ECDSA & User funds: 20.5M ETH & Compromise of Hot wallets, store accounts etc. \\[1.5ex] 
        
        Smart Contracts & Admin Vulnerability & ECDSA & Contract TVS: 2.5M ETH, 200B USD in stablecoins/RWAs & Compromise of Oracles, RWAs, Bridges, Guardians etc. \\
        
        & Code Vulnerability & ECDSA, alt\_bn128, KZG, BLS12-381~\cite{Contributors2026precompiled} & L2/Protocol TVS: 15M ETH & Compromise of L2s with need for admin intervention \\[1.5ex] 
        
        Validators & Consensus Vulnerability & BLS Signatures & Consensus stake: 37M ETH & Compromise trust in the blockchain itself \\
        
        & Data Availability\newline Vulnerability & KZG Commitments & L2/Protocol TVS: 15M ETH & Compromise trust in the blockchain itself \\
        
        \hline
    \end{tabular}
    \caption{Taxonomy of Ethereum Components and Vulnerabilities. This table summarizes the vulnerabilities, cryptographic root causes, assets at risk and second order effects from the subsequent sections. All of these vulnerabilities are at-rest attacks.}
    \label{tab:eth_vulnerabilities_tab}
\end{table*}

\subsection{Account Vulnerability}
\label{sec:ethereum_account}

\begin{figure*}[t]
    \centering
        \includegraphics[width=\linewidth]{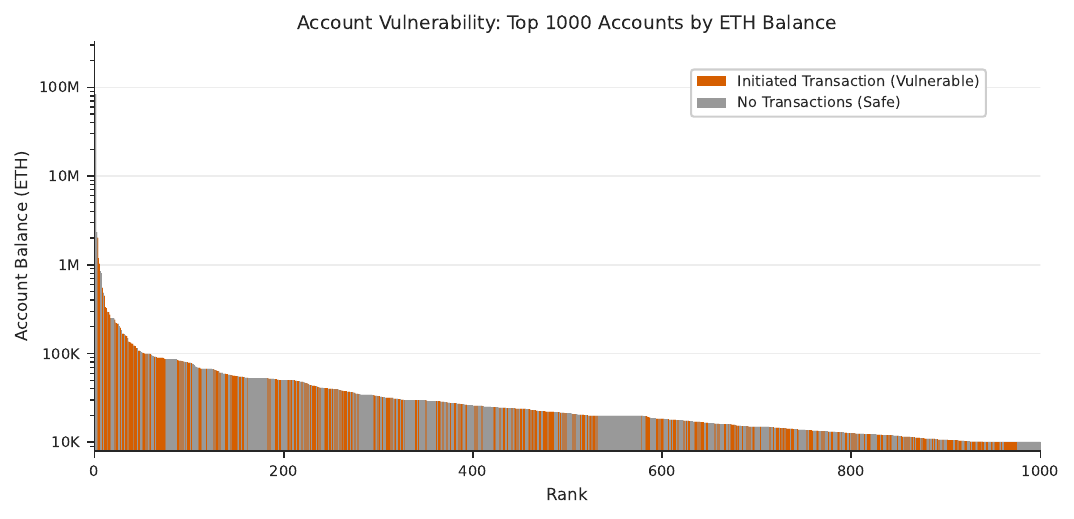}
    \caption{Account Vulnerability of Top 1000 accounts by ETH balance. The graph displays the ETH balance of the top 1000 Ethereum accounts ranked by value. Orange indicates accounts that have initiated a transaction and are therefore vulnerable to at-rest attacks, grey accounts have not. The sum of all ETH in vulnerable accounts in this list is $\sim$20.5 million ETH. Plot generated using data from \texttt{bigquery-public-data.crypto\_ethereum}~\cite{Day2018ethereum}.}
    \label{fig:eth_account_vulnerability_accounts}
\end{figure*}

Bitcoin prioritizes privacy by tracking UTXOs rather than persistent user identities. Unlike Bitcoin where address rotation can avoid at-rest vulnerabilities, Ethereum's account model is structurally prone to at-rest attacks. There are two types of accounts: Externally Owned Accounts (EOA), controlled by a private key and possessing agency to initiate actions on the blockchain, and contract accounts, governed by code and able to react to transactions sent to their address~\cite{Wood2025ethereum}. Unlike Bitcoin, Ethereum users typically maintain the same EOA for an extended period of time to accumulate assets and reputation (identity). Because an EOA's address is derived directly from its public key, the keys cannot be rotated; securing an account with a new key pair requires abandoning the account entirely~\cite{Wood2025ethereum, Buterin2021erc4337}. Crucially, while an address initially masks the public key, the moment a user sends their first transaction, the digital signature reveals the full public key to the network~\cite{Johnson2001elliptic}. This results in Account Vulnerability: a systemic, unavoidable exposure that cannot be mitigated by user behavior, short of a protocol-wide transition to PQC. We estimate that a quantum attacker using a fast-clock CRQC could crack the 1,000 highest-net-worth Ethereum accounts (at the time of writing, holding approximately 20.5 million ETH, see \fig{eth_account_vulnerability_accounts}) in less than nine days.

Account Vulnerability has other potential consequences. For example, it allows attackers to forge votes in decentralized governance of Decentralized Autonomous Organizations (DAOs). It also affects Exchange Hot Wallets: automated, internet-connected accounts used by centralized exchanges to process high throughput withdrawals. Even though ETH balance in these wallets may be low at a given point in time, the overall danger is very high: a breach here does not just represent a loss of funds for the exchange, but could lead to the draining of custodial liquidity that could induce panic across a broader market.

The Ethereum community has deployed several proposals that mitigate Account Vulnerability, most notable of which is Account Abstraction (AA) via ERC-4337~\cite{Buterin2021erc4337}. Introduced in 2023, AA enables users to interact with Ethereum via smart wallets that offer (among other features) customizable authentication logic. This decouples user identity (reputation) from a single static key, reducing the quantum attack surface through more frequent key rotation. Furthermore, AA helps mitigate the future challenge of migrating the ``long tail'' of legacy accounts to PQC without requiring a protocol-level hard fork. More recently, EIP-7702~\cite{Buterin2024eip7702} expanded these capabilities by allowing EOAs to temporarily function as smart contracts. However these enhancements mitigate the symptoms and not the root cause: the only gateways for external agency into Ethereum are the vulnerable EOAs. Thus, while these upgrades improve flexibility, they fall short of providing a fully quantum-secure interface.

\subsection{Admin Vulnerability}
\label{sec:ethereum_admin}

An important distinction between Bitcoin and Ethereum is the latter's extensive support for smart contracts~\cite{Buterin2014ethereum}. While Bitcoin scripts are fundamentally stateless, Ethereum smart contracts are fully fledged distributed applications with persistent state. They can initiate transactions, access blockchain data, and even autonomously deploy other contracts~\cite{Wood2025ethereum}. To facilitate the ability for different smart contracts and decentralized applications to interact with one another, the Ethereum community established standardized application programming interfaces known as Ethereum Request for Comments (ERCs). These include: ERC-20~\cite{Vogelsteller2015erc20} for fungible tokens, ERC-721~\cite{Entriken2018erc721} for Non-Fungible Tokens (NFTs), and ERC-3643~\cite{Lebrun2021erc3643} for compliance-aware Real World Assets (RWAs). These standards have helped fuel a massive ``tokenized'' economy that resides on top of Ethereum but is distinct from the native ether (ETH) asset (see \fig{eth_admin_vulnerability_rwa}). Furthermore, the rapid adoption of digital representations of bonds, stocks, commodities, etc., as tokens~\cite{Fink2025larry} means that the value exposed to Ethereum's security model extends far beyond the native currency. 

\begin{figure*}[t]
    \centering
        \includegraphics[width=\linewidth]{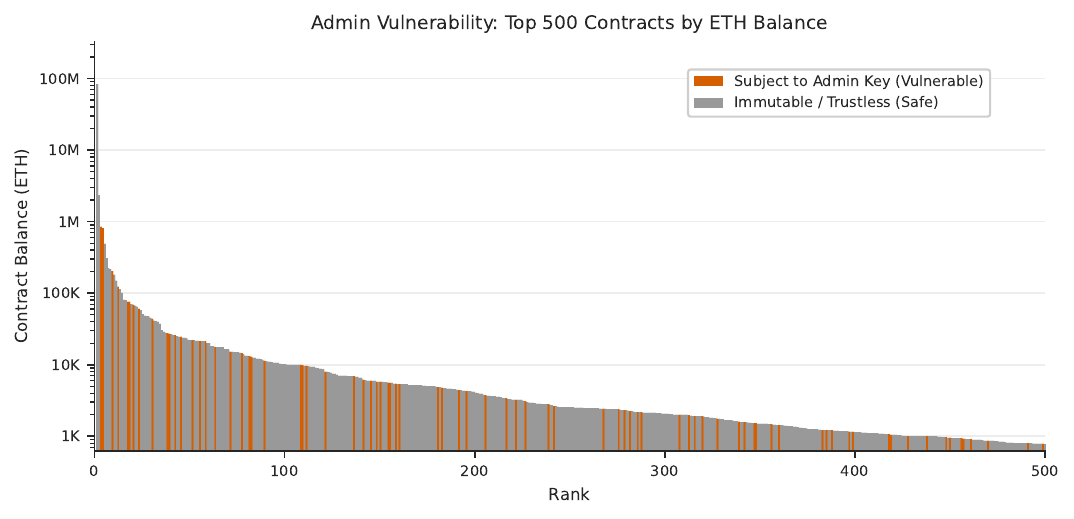}
    \caption{Admin Vulnerability of Top 500 smart contracts by ETH balance. Contracts were classified as ``Subject to Admin Key'' if their event logs contain signatures corresponding to AdminChanged, Upgraded (ERC-1967 proxy standards), or OwnershipTransferred (OpenZeppelin Ownable standard). The sum of all ETH in these vulnerable contracts is $\sim$2.5 million. Plot generated using data from \texttt{bigquery-public-data.crypto\_ethereum}~\cite{Day2018ethereum}.}
    \label{fig:eth_admin_vulnerability_contracts}
\end{figure*}

\begin{figure*}[t]
    \centering
        \includegraphics[width=\linewidth]{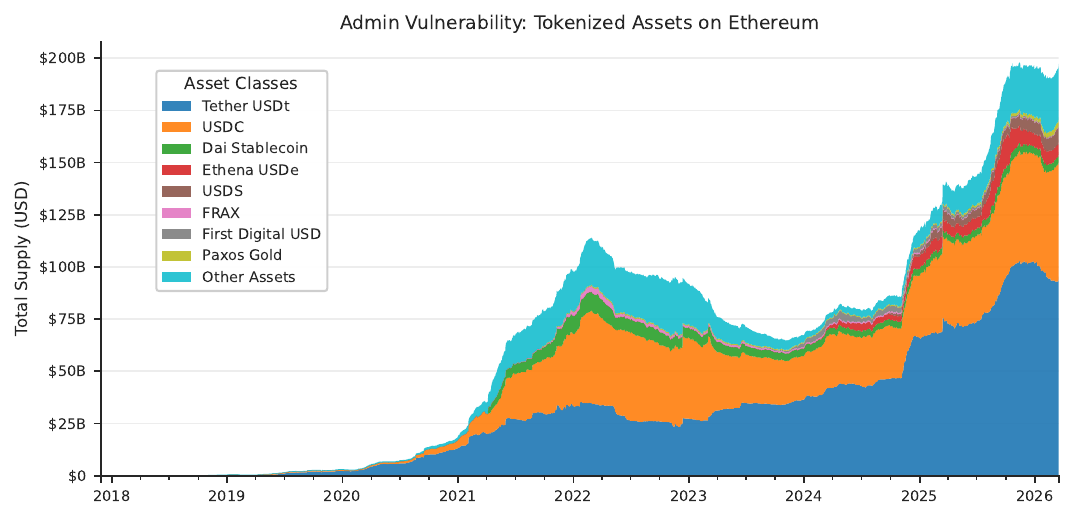}
    \caption{Admin Vulnerability exposure across distributed Real World Assets (RWAs). The chart details the market capitalization of major distributed assets (onchain tokens) on Ethereum, explicitly excluding the underlying offchain represented assets (e.g., fiat reserves, physical properties, repurchase agreements ). These distributed assets rely on quantum-vulnerable ECDSA keys for administrative control of their smart contracts. Within this ecosystem, distributed fiat-backed stablecoins and tokenized treasuries present a systemic risk; a compromise of issuer keys would allow arbitrary minting, effectively depegging the distributed asset and destabilizing the broader DeFi economy, even though the offchain represented reserves remain untouched. Private credit protocols carry institutional risk, where attackers could manipulate onchain loan books to redirect repayment flows or bypass credit underwriting standards. Finally, commodities, real estate, and equities embody legal title risk, where an admin takeover would allow the seizure of digital ownership rights, severing the legal link between the distributed token holder and the underlying represented asset. Ethereum currently captures the vast majority of RWA market share (and is therefore the focus of this chart), but it is not the sole network hosting these assets. As this multi-chain landscape evolves rapidly, other blockchain networks must proactively assess their own quantum attack vulnerabilities to protect their respective distributed ecosystems. The landscape of RWAs is evolving rapidly, and so these estimates should be viewed as a point-in-time assessment of the risk profile for various asset classes that will likely become outdated very quickly. Data gathered from:~\cite{CastleLabs2026realworld}. } \label{fig:eth_admin_vulnerability_rwa}
\end{figure*}

Smart contracts are a source of two quantum vulnerabilities. The first one, which we term the Admin Vulnerability, arises from Account Vulnerability of accounts for which a smart contract reserves enhanced privileges or administrative control, such as the ability to pause execution, upgrade code or extract funds~\cite{Palladino2019erc1967}. Because these administrative keys are rarely rotated and often publicly utilized for governance voting or contract upgrades, they represent a high-value at-rest vulnerability. At the time of writing, among the top 500 contract accounts by ETH balance, at least 70 accounts, with total holdings of about 2.5 million ETH, see \fig{eth_admin_vulnerability_contracts}, are at risk of account takeover by a quantum attacker due to Admin Vulnerability. We expect that a private key derivation attack on these 70 Ethereum accounts using a fast-clock CRQC will take less than 15 hours.

A strategic adversary could prioritize ``high-leverage'' admin accounts that exhibit Admin Vulnerability. These accounts often hold negligible amounts of ETH to minimize exposure, but confer various administrative privileges. A successful quantum attack on an admin account could grant the attacker control over the following (non-exhaustive) list of functions and systems:

\begin{enumerate}

\item Real World Asset (RWA) token issuance, see \fig{eth_admin_vulnerability_rwa}: RWA administrative keys used to authorize the minting and burning of tokens that represent offchain collateral; compromising them allows an attacker to mint fraudulent tokens to collapse the peg between a digital token and the real-world asset backing it.
\item Bridges: Bridge administrators manage the multi-signature schemes that lock assets on one blockchain to mint ``wrapped'' equivalents on another; compromising them allows the attacker to drain the entire liquidity pool backing these cross-chain transfers (This risk supersedes the Code Risk we analyze later, see \fig{eth_code_vulnerability}. For L2 scaling solutions, admin failure is more catastrophic than an ordinary protocol failure). 
\item Oracle Nodes: These accounts broadcast authoritative data feeds (e.g., ETH/USD price) that smart contracts rely on; by hijacking these keys, an attacker could broadcast false price data to trigger catastrophic price action potentially leading to automated liquidations across decentralized lending protocols.
\item Guardians: These privileged accounts act as the emergency safety mechanisms for DeFi applications, holding the power to instantaneously pause execution, freeze user funds, or in some cases, bypass security time-locks to inject code updates.
\end{enumerate}

Quantifying the aggregate financial exposure from Admin Vulnerability as a whole is difficult due to the complex web of dependencies and composability in the Ethereum ecosystem. While the first order 2.5 million ETH (ETH from admin-vulnerable contracts, see \fig{eth_admin_vulnerability_contracts}) and $\sim$200 billion USD in stablecoins and tokenized RWAs (assets tied up in admin-vulnerable contracts, see \fig{eth_admin_vulnerability_rwa}) is large, the second order risk from other ``high-leverage low ETH accounts'' (Bridges, Oracles and Guardians etc.) is arguably much larger due to their compromise potentially leading to mass liquidations across lending markets, peg collapse of stablecoins, freezing of cross-chain liquidity etc. This makes the true value at risk very hard to calculate using standard asset-balance models (especially with growth projections of RWAs to 16.1 trillion USD in 2030~\cite{Gretz2025tokenization}) , as the effective exposure almost certainly encompasses substantial fractions of the entire ecosystem's TVS.

\subsection{Code Vulnerability}
\label{sec:ethereum_code}

The second smart contract vulnerability concerns the code running on EVM. In order to prevent abuse and runaway algorithms from exhausting system resources, the EVM meters the runtime cost of executing smart contracts~\cite{Wood2025ethereum}. The account sending a transaction is charged the so-called ``gas fee'' for the cost of executing the smart contracts it triggers. This provides economic incentives for computational efficiency and discourages implementation of complex algorithms, such as cryptographic primitives, directly in EVM byte code (in line with the famous dictum ``do not roll your own crypto''). Instead, basic cryptographic primitives, such as hash functions and digital signature verification, are provided as standard precompiled contracts~\cite{Wood2025ethereum, Contributors2026precompiled} (also known as ``precompiles'') that a user's contract can interact with. However, at present, none of the precompiled contracts implement any of the modern PQC primitives. This situation causes the second quantum vulnerability in smart contracts (also an at-rest vulnerability), which we call Code Vulnerability and which lies in the fact that no precompiles for post-quantum cryptographic protocols, such as zero-knowledge (ZK) arguments, are currently available on Ethereum. Inevitably, the primitives currently utilized by smart contracts carry their own quantum vulnerabilities.

The severity of the Code Vulnerabilities and the challenges of addressing them are magnified by the extensive ecosystem built on top of Ethereum. In the process of becoming the world's second largest blockchain by market capitalization, Ethereum ran into scalability limits of a single blockchain. Further growth was enabled by development of effective L2 networks, such as state channels and rollups, which process transactions quickly and cheaply offchain and periodically commit transaction batches to an L1 blockchain, such as Ethereum. Assets are moved between the L1 and L2 networks using bridges which rely on event logs, smart contracts, and EOAs to interact with the L1 chain. Therefore, bridges and L2 networks inherit all the vulnerabilities of smart contracts and the Ethereum account model~\cite{Zamyatin2021sok}. In addition, some L2 solutions use quantum-vulnerable zkSNARKs~\cite{Bitansky2012from, BenSasson2019scalable}. The Ethereum community is discussing various approaches for introducing PQC at different layers of the Ethereum stack. In particular, EIP-7932~\cite{Kempton2025eip7932} is a proposal for the introduction of precompiles implementing post-quantum signature schemes.

\begin{figure*}[b]
    \centering
        \includegraphics[width=\linewidth]{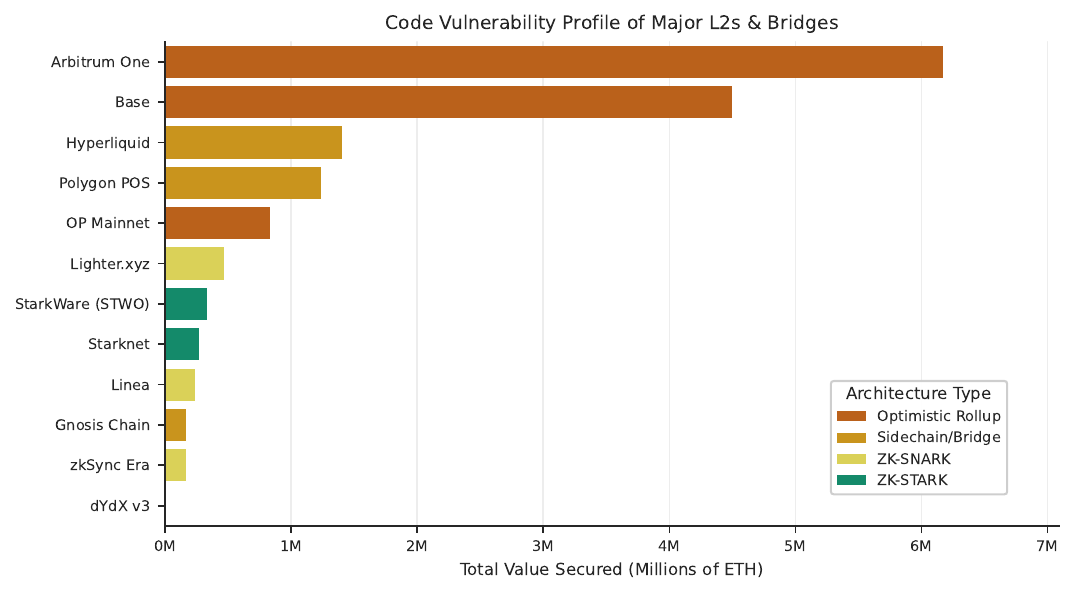}
    \caption{Breakdown of Total Value Secured (TVS) across major scaling protocols, categorized by their underlying security model. Optimistic Rollups (red), Bridges (orange), and zkSNARKs (yellow) currently rely on digital signatures or elliptic curve pairings that are vulnerable to quantum key derivation. Only protocols utilizing zkSTARKs (green) employ hash-based proofs that are currently believed to be resistant to quantum attacks. Across the three examples in each of the four protocol types we chose, the sum of all TVSs is $\sim$15 million ETH. Note: the Layer 2 landscape evolves very rapidly; TVS figures fluctuate daily, and protocols frequently upgrade and change their proving architectures (e.g., migrating from pairing-based to hash-based proofs or hybrid systems that can drastically alter the quantum risk profile). These estimates should be viewed as a point-in-time assessment of the ecosystem's risk profile that will likely become outdated (or incorrect) very quickly.}
    \label{fig:eth_code_vulnerability}
\end{figure*}

We estimate the Code Vulnerability risk as at least 15 million ETH by summing the TVS of a few major quantum-vulnerable L2 protocols and cross-chain bridges, such as Arbitrum, Base, and Optimism obtained from the L2BEAT dashboard~\cite{l2beat2026}, see \fig{eth_code_vulnerability}. However, this risk is partially mitigated by the fact that many L2 solutions operate in a ``Stage 0'' or ``Stage 1'' maturity phase, retaining centralized administrative safeguards. These security councils typically possess the authority to pause the bridge or override proof verification in the event of an anomaly. Paradoxically, this centralization (while an Admin Vulnerability in its own right) provides a temporary defense against Code Vulnerability; if an attacker were to exploit a cryptographic flaw (e.g., forging a validity proof), human administrators could theoretically intervene to freeze the protocol before the theft is finalized. The eventual upgrade of the Ethereum base layer to quantum-secure consensus does not retroactively fix this Code Vulnerability embedded in existing smart contracts (the EVM cannot automatically recompile deployed smart contracts). To completely remove the Code Vulnerability, the onus falls upon the protocol stakeholders and governance councils to coordinate the migration of their specific bridge contracts and administrative multi-signature schemes. Without this distributed effort, a quantum adversary could derive the private keys for these distinct protocols from public chain data, effectively bypassing the security of the base layer to drain these specific protocols.

\subsection{Consensus Vulnerability}
\label{sec:ethereum_consensus}

The last major difference between Bitcoin and Ethereum concerns the consensus mechanism. In 2022, seven years after its launch as a Proof-of-Work blockchain, Ethereum upgraded to the energy-efficient Proof-of-Stake consensus mechanism by merging its execution layer with the new consensus layer, called the Beacon Chain, and effectively ceasing all mining activity~\cite{Kalinin2021eip3675}. In the new protocol, consensus is driven by a one half majority vote among ``validators'', with finality established by a two thirds majority~\cite{Buterin2017casper, Buterin2020combining}. Any Ethereum node can become a validator by staking 32 ETH as collateral. The stake enables the network to incentivize the validator's honest behavior by a combination of financial penalties and rewards~\cite{Buterin2020combining}.

Penalties suffered by a malicious or misbehaving validator include leaking for inactivity and slashing for provably malicious behavior. Rewards for an actively participating honest validator consist of transaction fees from proposing blocks and issuance from attesting. For most validators, part of the transaction fees come from block builders~\cite{DAmato2024eip7732} that extract Maximal Extractable Value (MEV - analogous to the same concept in Bitcoin where it stands for Miner Extractable Value), from their ability to reorder, insert and remove transactions as they build new blocks~\cite{Daian2020flash, Chone2025maximal}. In particular, validators' control over which transactions are executed and when creates opportunities to front-run and sandwich transactions which allows them to benefit financially at the expense of other users of the network~\cite{Daian2020flash, Chone2025maximal}.

Efficient processing of thousands of digital signatures per second from validators signing attestations presents a cryptographic challenge due to the large volume of cryptographic information involved. Ethereum substantially reduces verification overhead by compressing tens of thousands of digital signatures into a few dozen using the quantum-vulnerable BLS signature aggregation protocol~\cite{Boneh2001short} on the BLS12-381 curve. As discussed earlier, Shor's algorithm on BLS12-381 uses a greater amount of quantum memory than that required by our resource estimates for curves like secp256k1. However, we estimate that the additional resource cost is modest. Consequently, the consensus layer of Ethereum should be considered at-rest vulnerable to the same first-generation CRQCs as the execution-layer account model and smart contracts.

The severity of this Consensus Vulnerability depends on the fraction of validators compromised:
\begin{enumerate}
\item If less than a third of validators are compromised, an attacker can force them to ``equivocate'', i.e., sign conflicting blocks for the same slot causing the system to slash their stake and eject them from the network~\cite{Buterin2020combining}.
\item If more than a third of validators are compromised, an attacker can prevent finalisation which requires two thirds supermajority. The attacker-controlled stake would dilute via the activity leak until enough of the honest stake regains the supermajority~\cite{Buterin2020combining}. This constitutes a denial of service attack on finality.
\item If more than half of validators are compromised, an attacker can control the fork choice rule to prevent chain growth or execute deep blockchain reorganizations.
\item If more than two thirds of validators are compromised, the attacker can finalize inconsistent chains.
\end{enumerate}

\begin{figure*}[t]
    \centering
        \includegraphics[width=\linewidth]{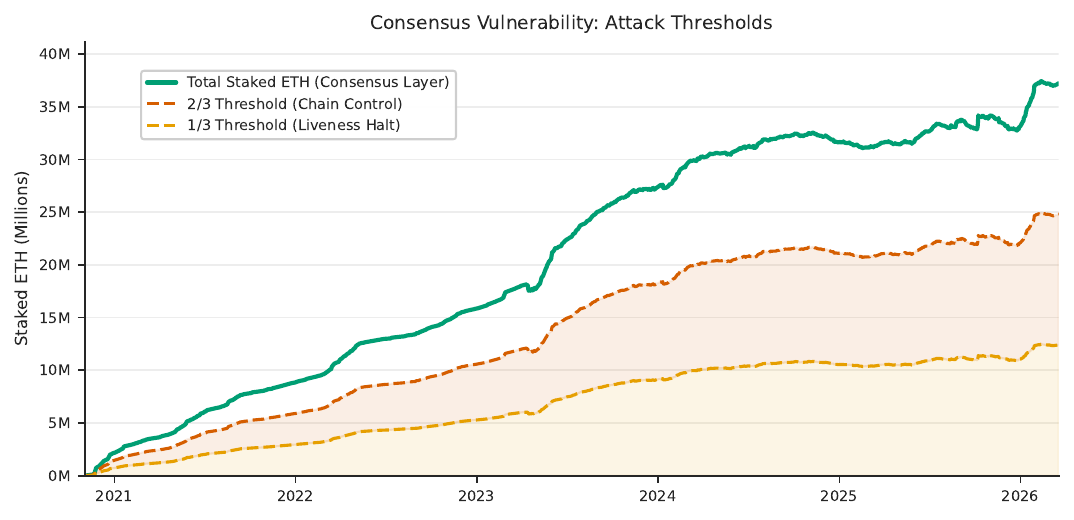}
    \caption{Historical accumulation of ETH in the Beacon Chain deposit contract (0x0...05fa). The green line represents the total stake securing the network (which relies on the vulnerable BLS12-381 curve for signature aggregation). The dashed lines identify the thresholds for quantum attacks:  $>1/3$ of the stake (yellow) allows an attacker to halt finality (liveness failure), $>2/3$ (red) grants the ability to rewrite history and censor transactions (safety failure). Plot generated using data from \texttt{bigquery-public-data.crypto\_ethereum}~\cite{Day2018ethereum}.}
    \label{fig:eth_consensus_vulnerability}
\end{figure*}

As of February 2026, about 37 million ETH is staked and exposed to potential slashing and ejection due to quantum attacks exploiting the Consensus Vulnerability, see \fig{eth_consensus_vulnerability}. The second-order risks of Consensus Vulnerability are also very large. First, the deep integration of Liquid Staking Tokens (LSTs) means that large slashing events could trigger a peg collapse for derivative assets widely used as collateral, precipitating cascading liquidations across the DeFi ecosystem analogous to a traditional bank run. Second, a breach of finality introduces large desynchronization risks for cross-chain infrastructure, where an attacker could exploit deep chain reorganizations to double-spend assets across bridges and Layer 2 networks that relied on the immutability of the base layer. Lastly, absolute control over the blockchain and all recent history would likely remove the network's credible neutrality and possibly the institutional trust required for Ethereum to function as a global settlement layer.

Recovery from a supermajority attack, where more than two thirds of validators are compromised, is impossible within the protocol rules and would necessitate a ``social consensus'' intervention. The community would be forced to coordinate some emergency measures (possibly a hard fork), manually identifying the last valid block and migrating to a new protocol state; a catastrophic event for trust in the system. Consequences of strategically straightforward quantum attacks on Ethereum consensus layer are primarily destructive in nature: loss of stake inflicted on compromised validators' accounts, halt of transaction processing. Strategically more sophisticated attacks are possible and potentially involve manipulation of the selection and ordering of transactions in newly proposed blocks of the same character as the one used by validators to obtain the MEV revenue stream~\cite{Daian2020flash}. Quantum attackers with a fast-clock CRQCs could also attempt to profit from transaction manipulation by attacking the mempool directly rather than by compromising validators~\cite{Daian2020flash}.

Attacks on consensus are made easier by the exposure of validators' public keys in deposit transactions and the validator registry~\cite{Wood2025ethereum, Kalinin2021eip3675}. Another weakness is the absence of expedient mechanisms for key rotation: a validator must go through a lengthy process of withdrawal and re-staking~\cite{Buterin2020combining}. At the same time, the very large number (about one million) of validators together with the stake-weighted random selection process makes attacks on consensus hard and expensive~\cite{Buterin2020combining}. Indeed, there is safety in numbers: even assuming that an early fast-clock CRQC can solve ECDLP on BLS12-381 in the same amount of time as on secp256k1, a quantum attacker with 20 such machines would need more than nine months to derive private keys of a two thirds supermajority of validators. It should be noted that this safety margin assumes a uniformly decentralized validator set, if stake remains concentrated in a few large pools (such as Lido accounting for approx. 20\% of staked ETH~\cite{hildobby2026ethereum}) with admin vulnerabilities, an attacker could target the specific key-management infrastructure of those providers to acquire a supermajority much faster.

Long-term remedy to the Consensus Vulnerability requires a phased transition to new validator credentials based on a suitable post-quantum multi-signature scheme. In order to help identify such a scheme, Ethereum Foundation has conducted research on candidate hash-based replacements for BLS12-381~\cite{Drake2025hashbased}. In the meantime, a fast and effective mechanism for rotating validator keys could serve as a relatively simple stopgap measure for fortifying the Beacon Chain against earlier-than-expected quantum attacks on consensus. Following implementation and deployment of a post-quantum replacement and once a sufficient supermajority of stake has rotated to PQC validator credentials, the system may need to begin ejecting or slashing non-migrated validators to eliminate the lingering attack surface of the legacy cryptosystem.

\subsection{Data Availability Vulnerability}
\label{sec:ethereum_data}

The Layer 2 ecosystem depends on the ability to commit bulk transaction data to the Layer 1 Ethereum blockchain. In order to check integrity of this data, validators used to have to download it via smart contract calldata which was a source of inefficiency. In order to address this throughput bottleneck, Ethereum introduced inexpensive, transient onchain data storage via 128 KiB blobs~\cite{Buterin2022eip4844} together with the Data Availability Sampling (DAS) mechanism~\cite{Ryan2024eip7594} which allows validators to probabilistically verify blob integrity by downloading only small, randomized data chunks called cells. Blobs and DAS led to massive speedups and lower costs for Layer 2 bulk transaction commitments.

DAS employs the KZG polynomial commitment scheme on the BLS12-381 elliptic curve. The binding property of the scheme is vulnerable to quantum attacks, allowing a quantum-capable adversary to forge cell authentication proofs and to deceive validators into believing that a blob is available despite it not being reconstructable from the cells with forged proofs. This can be used to stall rollups, creating opportunities for ransom attacks.

Furthermore, an attacker does not need to run Shor's algorithm for each individual blob. The KZG scheme relies on the so-called trusted setup  ---  a one-time cryptographic ceremony~\cite{Foundation2022kzg} to generate a secret random scalar encrypted in a collection of points on the BLS12-381 elliptic curve. The points comprise the scheme's public parameters, known as the Structured Reference String (SRS), while the random scalar, sometimes referred to as ``toxic waste'', must be destroyed  ---  a failure to erase it creates a permanent backdoor into the protocol. Mathematically, the toxic waste can be recovered as the discrete logarithm of two elliptic curve points in the SRS, so a CRQC can compute it from publicly available parameters. Once this secret is extracted, the adversary obtains a persistent, universal backdoor allowing them to forge data availability proofs at minimal computational cost and with no further need for a CRQC. In effect, a single successful quantum attack on the SRS creates a potentially tradable exploit that gives any adversary without a CRQC a persistent ability to attack DAS. The exploit remains functional until a new trusted setup is created. This consistitutes an example of what we refer to as on-setup attacks.

Attacks against this single point of failure would have severe consequences for the L2 ecosystem. L2 rollups rely on L1 data availability to reconstruct their global state, so by withholding transaction batch data, an attacker effectively halts the L2 sequencers and validators. Throughput constraints and lower cost incentives have driven the majority of the ecosystem to this lower cost quantum-vulnerable architecture since the Dencun network upgrade in March 2024. Hence, the majority of modern Layer 2 data volume is now fundamentally dependent on the KZG scheme. As of February 2026, we estimate Data Availability Vulnerability risk is roughly 15 million ETH in TVS (\fig{eth_code_vulnerability}) which along with second order effects, such as complete liveness failure halting the affected L2 networks, would provide adversaries with leverage for ransom attacks against bridged L1 assets. As with Code Vulnerability, Stage 0 and Stage 1 Admins could partially mitigate impacts by pausing the Layer 2 activity until a workaround is found.

Mitigating the Data Availability Vulnerability requires upgrading the DAS protocol to a PQC commitment scheme before CRQCs come online. Anticipating this need, the Ethereum research community is exploring quantum-resistant alternatives, primarily hash-based polynomial commitments, such as Fast Reed-Solomon Interactive Oracle Proofs of Proximity~\cite{BenSasson2018fast}, and Merkle-tree architectures~\cite{HallAndersen2023foundations}. While transitioning to these post-quantum alternatives will likely introduce higher bandwidth and storage overheads for validators compared to the highly efficient KZG commitments, replacing them is a mandatory step to ensure the long-term survivability of Ethereum's L2 scaling infrastructure.

In summary, Ethereum has a broader overall quantum attack surface than Bitcoin. However, this is compensated by stronger community leadership in the Ethereum Foundation. Operating as a non-profit out of Switzerland since 2014, the Ethereum Foundation supports the ecosystem through funding, research and coordination~\cite{Foundation2026official}. In the spirit of decentralization, the Ethereum foundation does not ``control the decision making process'', but can exert significant influence in advocating for the long-term success of the ecosystem by helping to organize development work, coordinate decision making, and define protocol strategy and roadmap. This influence was demonstrated when it successfully implemented a bailout of The DAO by carrying out a one-off irregular change of blockchain state to reverse the July 2016 hack~\cite{Antonopoulos2018mastering, Detrio2017eip779, Securities2017report}. With time, acceptance of emergency measures appears to have declined as a similar intervention in April 2018 was rejected~\cite{Schoedon2018eip999}. Nevertheless, stronger leadership will likely make it easier and faster for Ethereum to transition to PQC~\cite{Drake2026PQCannouncement} and to adopt any mitigation measures deemed necessary. Moreover, the proven willingness to reach for extraordinary means to preserve the integrity of the system provides a high degree of assurance. By contrast, Bitcoin's decentralized community and the lack of a singular center of power may require a potentially drawn-out process of consensus building. Indeed, the increased storage and compute requirements of PQC may place the Bitcoin community in a position similar to August 2017 when divisions over block size change led to a hard fork that created Bitcoin Cash.

\section{Quantum Vulnerabilities of Other Blockchains}
\label{sec:blockchains}

Bitcoin and Ethereum dominate the cryptocurrency world in terms of economic activity, but a lot of financial and cryptographic innovation and experimentation takes place in the broader ecosystem and on other blockchains. Some of these developments seek to address quantum vulnerabilities while others continue to introduce new ones by building innovative features based on the ECDLP. The peril of the latter developments has been recognized in the community for example by Luke Parker, a prominent developer in the Monero ecosystem, who has called for a moratorium on research and development of quantum-vulnerable protocols~\cite{Parker2024stepping, SyntheticBird2024discussion}. In this section, we briefly examine a broad sample of blockchains and cryptocurrencies from the standpoint of quantum security.

\subsection{Bitcoin and Ethereum Derivatives}
\label{sec:blockchains_derivative}

Among the many early cryptocurrencies based on Bitcoin, one can identify two distinct types. Codebase forks, such as Litecoin (LTC), Zcash (ZEC), and Dogecoin (DOGE), employ modified Bitcoin software and produce their own blockchain starting with a new genesis block. Chain forks, such as Bitcoin Cash (BCH) and eCash (XEC), run modified Bitcoin software and share a part of transaction history with their parent chain up to the block after which they split off as a hard fork. Many elements of our Bitcoin analysis apply to both types of forks with suitable adaptations to account for software customizations, new features, and blockchain history.

For example, Litecoin implemented both SegWit and Taproot upgrades and our earlier discussion of these upgrades applies to it. In particular, litecoin secured by a P2TR locking script exposes the public key and is therefore vulnerable to at-rest attacks while litecoin secured by a P2WPKH script hides the public key behind a hash, so if the key is not reused or otherwise exposed elsewhere, then it is only vulnerable to on-spend attacks. At the same time, as we remarked earlier (see \fig{btc_block_and_network_time}), Litecoin's 2.5 minute average block interval significantly reduces the probability of a successful on-spend attack using early fast-clock CRQCs. Nevertheless, theoretical advances and increases in the number of qubits are very likely to bring on-spend vulnerability to Litecoin eventually.

Dogecoin eschewed the SegWit and Taproot upgrades and therefore does not support Bitcoin's P2WPKH, P2WSH, and P2TR script types. Further, its one minute average block interval makes on-spend attacks using early CRQCs practically impossible under our current assumptions (see \fig{btc_block_and_network_time}). A recent draft Dogecoin Improvement Proposal~\cite{Jefferson2025dip} seeks to add native support for verification logic of quantum-vulnerable zkSNARKs with the objective of enabling L2 scaling solutions. Depending on the chosen ZK protocols, this may introduce a vulnerability to on-setup attacks to Dogecoin.

Bitcoin Cash launched in 2017 as a hard fork from Bitcoin due to disagreement about the solution to scalability issues. Bitcoin Cash never introduced SegWit or Taproot addresses and, like Zcash and Dogecoin, lacks the P2TR vulnerability. At the same time, as a hard fork of Bitcoin, it shares Bitcoin's early transaction history and thus, inherits the problem of P2PK coins, including Satoshi era mining rewards.

Pegged sidechains~\cite{Back2014enabling, Singh2020sidechain} are blockchains built in order to add new functionality to an existing network without creating a new store of value. They rely on bridges to transfer assets between the parent blockchain and the sidechain. For example, Rootstock (RSK)~\cite{Lerner2022rsk, Lerner2019rsk} is an Ethereum-based Bitcoin sidechain with EVM-compatible stateful smart contracts and two types of accounts (EOAs and contract accounts) whose native token (RBTC) is pegged 1:1 to bitcoin. Users can send bitcoin to a special multi-signature Bitcoin address where it is received by an automated mechanism called PowPeg causing release of the corresponding amount of RBTC on the Rootstock sidechain where users can take advantage of a rich ecosystem of DeFi applications enabled by the EVM-compatible smart contracts. Later, they can send their RBTC to a precompiled smart contract called Bridge to trigger the release of the corresponding amount of bitcoin on the Bitcoin blockchain. In addition to the quantum vulnerabilities inherited from Ethereum, such as the Account and Code vulnerabilities, Rootstock introduces new quantum dangers. The system protects the private keys that control the multi-signature address using Hardware Security Modules (HSM). However, an attacker equipped with a CRQC can derive these keys without compromising the HSMs and steal any bitcoin residing at the address. This creates the risk of sidechain bankruptcy by making it unable to satisfy redemptions. Besides technical quantum vulnerabilities, Rootstock enables practices that may expose users to quantum risk. For increased convenience it allows users to omit the explicit Rootstock destination address from the deposit transaction on the Bitcoin side. In this case, the address is automatically derived as a cryptographic hash of the user's public key. Consequently, the destination account on Rootstock ends up controlled by the same private key that signed the first input of the deposit transaction. This is convenient, but gives rise to Offchain Exposure vulnerability by exposing the Bitcoin public key through the user's activities on the Rootstock sidechain.

\subsection{Privacy-Preserving Blockchains}
\label{sec:blockchains_privacy}

In 2022, Litecoin implemented a private transactions protocol called Mimblewimble~\cite{Jedusor2016mimblewimble, Yang2019lip} as a sidechain with blocks embedded in the canonical Litecoin blockchain. Inputs, outputs and amounts of Mimblewimble transactions are hidden or erased by a combination of Pedersen commitments, transaction aggregation, and pruning of spent outputs. In order to eliminate the need for interaction between the sender and receiver transacting on Mimblewimble, Litecoin introduced stealth addresses~\cite{Burkett2020lip} and modified the protocol to use ECDH key exchange for offline derivation of secrets between the sender and receiver.  Pedersen commitments and ECDH key exchange protocol are both vulnerable to quantum attacks. Moreover, the two elliptic curve points that Mimblewimble uses in the Pedersen commitments are fixed public parameters, creating a vulnerability to on-setup attacks. Indeed, an adversary with a CRQC can solve the corresponding ECDLP once to manufacture a tradable exploit that can be used repeatedly to break the binding property of the Pedersen commitments and execute undetected inflation attacks without further need for a CRQC.

The Litecoin community considered a variety of protocols for private transactions, including post-quantum zkSTARKs. However, they ultimately rejected this option due to its high resource cost~\cite{Yang2019lip}. Instead, they adopted a design which provides the option to switch from Pedersen commitments to ElGamal commitments to protect the monetary base at the expense of privacy~\cite{Yang2019lip, Ruffing2018switch}. ElGamal commitments are quantum vulnerable, but in a different way than Pedersen commitments. Before the switch, a quantum adversary can steal and create new coins, because the binding property of Pedersen commitments is vulnerable to quantum attacks. In addition, a quantum adversary can learn transaction amounts, despite quantum resistance of the hiding property of Pedersen commitments, by exploiting quantum vulnerability in ECDH key exchange to learn blinding factors. After the switch, a quantum adversary can learn transaction amounts due to quantum vulnerability in the hiding property of ElGamal commitments, but cannot steal or create new coins, because the scheme's binding property is quantum-resistant.

Zcash (ZEC) is a Bitcoin-based blockchain that employs advanced cryptography to provide confidential transactions. Zcash wallets use zkSNARKs to prove to the network that a transaction is valid without revealing the sender, recipient, or amount sent. Zcash has evolved over three generations of protocols for shielded transactions~\cite{Hopwood2026zcash}. Sprout, the original shielded protocol based on the BCTV14 proof system~\cite{BenSasson2013succinct}, was launched in 2016 and phased out in 2020~\cite{Hopwood2019zip, Lai2020pull}. Sapling, launched in 2018, implemented the Groth16 protocol~\cite{Groth2016size} on the BLS12-381 curve, substantially improving performance. The launch was preceded by a trusted setup ceremony~\cite{Miller2017announcing} which produced public protocol parameters called the Structured Reference String (SRS). As in the case of Ethereum's DAS, the Sapling ceremony also generated a secret scalar, known as the ``toxic waste'', which had to be destroyed in order to avoid leaving behind a permanent backdoor into the protocol. The toxic waste can be recovered from the SRS using a CRQC to create a persistent tradable exploit enabling non-quantum attackers to stealthily inflate monetary base in the Sapling shielded pool. The need for trusted setup and the vulnerability to these on-setup attacks were both eliminated in the newest shielded protocol called Orchard that launched in 2021. The protocol introduced the Halo 2 scheme~\cite{Co2026halo2} using Pasta elliptic curves~\cite{Team2026pasta} and increased scalability by introducing recursive proof composition. Nevertheless, Orchard's improved quantum security posture does not eliminate all quantum risks.

Indeed, many Zcash innovations, including those in the Orchard protocol, use ECDLP-based cryptographic primitives, such as zkSNARKs, Pedersen commitments, and ECDH key exchange, and are therefore vulnerable to quantum attacks. For example, Zcash enables users to publish an effectively unlimited number of public addresses, known as diversified addresses, associated with a single private incoming viewing key. A classical adversary cannot determine whether different diversified addresses belong to the same user --- a privacy feature known as unlinkability. However, the incoming viewing key is the solution to the ECDLP associated with the diversified address, so a quantum attacker can defeat unlinkability by deriving the incoming viewing key. Practical quantum attacks on unlinkability would likely be done offline. Potential online attacks on diversified addresses would not benefit from the speed-up achieved by ``priming'' the quantum computer which we described earlier, because priming depends on the knowledge of the generator which is part of a diversified address, not a fixed constant publicly known before the attack. Consequences of quantum attacks on diversified addresses are limited to the recovery of the incoming view key by Zcash's robust key hierarchy which denies quantum attackers access to the user's spending key.

Another quantum vulnerability concerns encryption. Even though Zcash notes are encrypted onchain using a quantum-resistant authenticated symmetric encryption scheme~\cite{Bradner1997key}, the encryption key is derived using ECDH from the recipient's incoming viewing key and an ephemeral private key. The ephemeral public key is recorded on the blockchain alongside the encrypted note. Consequently, a quantum attacker who knows the target diversified address can recover the encryption key and decrypt the note, including transaction amount and memo. Thus, the most pressing quantum danger for Zcash is the eventual retroactive degradation of privacy by future quantum attacks on known addresses. Beyond privacy, CRQCs will be able to compromise soundness of zkSNARK protocols raising the risk of theft and creation of counterfeit notes. The Zcash community is discussing proposals for addressing quantum vulnerabilities, such as recently proposed foundational features to enable recoverability from quantum attacks~\cite{Hopwood2025zip}. This feature is a part of Zcash's broader plans for post-quantum transition~\cite{Bowe2025zcash}.

Each of Zcash's shielded transaction protocols  ---  Sprout, Sapling and Orchard  ---  has its own distinct value pool with the fourth pool for transparent transactions inherited from Bitcoin. Assets moving between different shielded pools must pass through the transparent pool which allows Zcash to track total monetary supply within each shielded pool using a mechanism called Turnstile~\cite{Company2019turnstile, BoweHopwood2019ZIP209}. This provides the last line of defense against supply inflating attacks~\cite{Bowe2025zcash}. Zcash never implemented Bitcoin's SegWit and Taproot upgrades, so transparent transactions use P2PKH and P2SH script types. On-spend attacks against Zcash are made extremely challenging under our current assumptions by its 75 seconds average block interval (see \fig{btc_block_and_network_time}).

\subsection{Post-Quantum Blockchains}
\label{sec:blockchains_pq}

A few blockchains have made progress in real-world deployment of PQC. In particular, the QRL~\cite{Foundation2026quantum, Waterland2016quantum} launched in 2018 stands out as post-quantum from inception. Its original design was based on the stateful post-quantum signature scheme known as XMSS~\cite{Cooper2020recommendation} and it is currently adding support for the stateless post-quantum signature scheme called CRYSTALS-Dilithium~\cite{Ducas2017crystals} and recently standardized by NIST under the name ML-DSA~\cite{NISTUS2024modulelatticebased_dsa204}. Other examples of post-quantum blockchains include Mochimo (MCM), which uses a variant of hash-based post-quantum Winternitz One-Time Signatures (WOTS)~\cite{Merkle1990certified, Dods2005hash}, and the post-quantum privacy-preserving blockchain Abelian (ABEL), which makes extensive use of lattice-based PQC.

Algorand (ALGO) provides an example of real-world deployment of PQC on an otherwise quantum-vulnerable blockchain. It launched in 2019 as a Pure-Proof-of-Stake blockchain for smart contracts and fast transactions. Smart contracts on Algorand are written in popular high-level programming languages like Python and TypeScript and are compiled to assembly-like Transaction Execution Approval Language (TEAL)~\cite{Foundation2026teal} which executes on the Algorand Virtual Machine (AVM)~\cite{Foundation2026avm}. In addition to builtin single- and multi-signature transactions, Algorand supports stateless smart signatures and stateful smart contracts for DeFi applications. Algorand's consensus and builtin transactions are based on quantum-vulnerable Ed25519 digital signature scheme. However, it has recently deployed post-quantum Falcon digital signatures~\cite{Perlner2025fips, Fouque2020falcon} for smart transactions and state proofs (cryptographic attestations of blockchain state for cross-chain integrations). Algorand has also made Falcon signature verification available as a TEAL primitive~\cite{Foundation2026falcon_verify} to enable development of quantum-safe smart contracts for AVM. These PQC technologies are now publicly available: Algorand executed its first PQC-secured transaction in 2025~\cite{Young2025technical}. Moreover, Algorand enables users to change the private keys associated with their accounts~\cite{Foundation2026rekeying}. While this mechanism does not provide full quantum security at present, it facilitates future PQC migration.

There are also experimental and test deployments of PQC on quantum-vulnerable blockchains. For example, Solana (SOL)~\cite{Yakovenko2017solana} deployed an experimental feature called Solana Winternitz Vault~\cite{Little2026solana, Swayne2025solana} which uses WOTS to protect digital assets. More recently, the XRP Ledger (XRP)~\cite{Chase2018analysis, Developers2026rippled}, deployed post-quantum ML-DSA signatures on its AlphaNet test instance~\cite{ForkLog2025xrp}. The blockchain provides extensive support for RWA tokenization including compliance controls, issuer permissions and asset metadata~\cite{RippleX2025future} and is increasingly utilized by global financial institutions. It currently holds about two thirds of all TBILL tokens backed 1:1 by short-dated U.S. Treasury bills~\cite{CastleLabs2026openeden} with most of the remaining one third held on Ethereum.

\subsection{Stablecoins and Real-World Asset Tokenization}
\label{sec:blockchains_rwa}

Involvement of global financial institutions and RWA tokenization reflect the significant growth and diversification in the digital economy and led to the introduction of new types of digital assets. A prominent new asset class is known as stablecoins which are digital tokens that seek to maintain a stable exchange rate to a fiat currency, commodity or another cryptocurrency, either by means of algorithmic trading or maintenance of liquid reserves. The latter type of stablecoins are a prime example of tokenization where the RWA backing the digital token is a fiat currency.

Unlike Bitcoin and Ethereum, many new cryptocurrencies, called crypto tokens, do not have their own dedicated blockchain infrastructure and are instead implemented as smart contracts on other blockchains, primarily Ethereum. In particular, the two most popular stablecoins, Tether (USDT) and USD Coin (USDC), are both available as ERC-20 tokens on Ethereum. The smart contracts that govern these cryptocurrencies give special administrative privileges to a handful of accounts. These include the right to mint and burn coins, freeze accounts and, most importantly, to upgrade the contract's logic. In particular, minting of unbacked tokens could potentially cause a market-driven peg collapse. It may be possible to contain some of the damage caused by such an attack by early detection and halt in trading. However, the mechanisms involved may need to involve offchain elements, because the logic upgrade privilege confers essentially unlimited control over the onchain contract. Access to this privilege is typically secured by a multi-signature that requires a certain threshold number of valid digital signatures from a group of accounts controlled by important stakeholders. Due to past upgrades, these public keys are exposed on the blockchain which creates the existential risk of a quantum attacker taking complete control over the smart contract. This illustrates the fact that crypto tokens, such as USDT and USDC, are exposed to highly concentrated quantum risk arising from the Admin Vulnerability in the host blockchain.

Cross-chain assets, such as USDT and USDC, may have quantum vulnerabilities on many blockchains and may inherit quantum weaknesses from interoperability mechanisms, such as bridges and Interblockchain Communication Protocol~\cite{Goes2006Interblockchain}. However, their presence on multiple blockchains also introduces a potential new element for the mitigation strategy. As some blockchains adopt PQC and others lag, multi-chain cryptocurrencies will have the option to withdraw from quantum-vulnerable systems while relying on post-quantum blockchains to continue their economic activity securely. There is precedent for a tenant cryptocurrency withdrawing from a host blockchain. For example, in February 2024, Circle announced~\cite{Financial2024circle} that they ceased minting new USDC on the TRON blockchain and allowed users to transfer their coin to other chains until February 2025. The one-year gap between the cessation of minting and termination of redemptions highlights the potential risks and liabilities associated with the long-tail of lingering assets and indicates the need for forward-looking assessment of a host blockchain's vulnerabilities. The possibility of cross-chain migration of high profile tenant cryptocurrencies to post-quantum systems may exert competitive pressure on blockchains such as Ethereum to accelerate PQC adoption. In fact, early developments in this direction have already taken place, for example USDC is available on the Algorand blockchain which supports post-quantum digital signatures. The competitive pressure on the Ethereum ecosystem may further increase with the emergence of EVM-compatible post-quantum blockchains. In fact, the post-quantum Abelian blockchain already supports EVM-compatible smart contracts via a zk-rollup called QDay. Furthermore, QRL's development roadmap~\cite{Foundation2026qrl} lists the support for EVM smart contracts as one of its key objectives.

\subsection{A Taxonomy of Quantum Risk Profiles for Distributed Ledgers}
\label{sec:blockchains_taxonomy}

The variety of quantum risk profiles of popular blockchains can be summarized using four coarse categories. The first one includes post-quantum blockchains, such as the QRL, Mochimo, Abelian and QDay (Abelian's L2 network).

The second one consists of protocols in which it is possible for individual users to avoid long-term exposure of quantum-vulnerable public keys. This category includes UTXO-based ledgers, such as Bitcoin, Litecoin, Dogecoin as well as Cardano~\cite{Hoskinson2017Why} which uses UTXOs extended with logic and data to support smart contracts~\cite{Chakravarty2020extended}. These protocols do not tie wealth to static account identities and allow users to secure assets with ephemeral public keys hidden behind a cryptographic hash. This makes it possible for cautious users to evade at-rest attacks leaving only the possibility of on-spend attacks that are more challenging for a quantum attacker. In fact, in our Scenario 2, early CRQCs are too slow to launch these on-spend attacks. However, the potential resistance of these blockchains to at-rest quantum attacks is often lost in practice due at-rest vulnerabilities arising outside the basic transaction protocol. For Bitcoin and its derivatives, they primarily stem from public key exposure driven by user practices such as address reuse, whereas in Cardano, at-rest vulnerabilities mainly affect staking activity in its Proof-of-Stake consensus protocol as well as voting in its decentralized governance system and originate from onchain exposure of staking and voting keys (distinct from spending keys) in registration certificates~\cite{sro2023understanding}. Thus, an individual user's ability to avoid public key exposure does not preclude systemic risks, such as those arising from vulnerabilities in consensus and governance.

The third category consists of protocols which make long-term exposure of quantum-vulnerable public keys inevitable. It includes blockchains with persistent accounts, such as Ethereum, Solana, Rootstock, Algorand, TRON and the XRP Ledger. These blockchains utilize an account model and either use public keys directly as account addresses (Solana) or expose them in the first transaction. Simultaneously, their ecosystems increasingly leverage onchain account histories to assess financial risk and confer privileges making account migrations costly for users. This design makes active public keys easier for quantum attackers to find compared to UTXO-based blockchains. Moreover, for legacy accounts on Ethereum and Rootstock, and for standard keypair accounts in Solana, this is exacerbated by lack of support for key rotation which effectively locks users into public key exposure. Modern Ethereum, Solana and Rootstock accounts are controlled by smart wallets and support key rotation, but legacy accounts remain a lingering vulnerability. By contrast, Algorand~\cite{Foundation2026rekeying}, TRON~\cite{DAO2026account} and the XRP Ledger~\cite{Ripple2026documentation} support native, protocol-level key rotation. Public blockchains in this category generally support smart contracts. Consequently, the scale of funds at stake in quantum attacks on these blockchains is greatly increased by smart contract vulnerabilities and is expected to grow further due to rising popularity of stablecoins and tokenization. Indeed, potential compromise of administrative keys for a high-value bridge may represent a systemic existential threat.

Finally, the fourth category comprises the quantum vulnerable privacy-preserving blockchains, such as Zcash, Monero and Litecoin's Mimblewimble sidechain. In addition to the forward-looking threats, such as asset theft and supply inflation, the privacy-preserving blockchains face retroactive degradation of privacy. An adversary with a CRQC could potentially deanonymize years of historical confidential transactions for known addresses, making PQC migration of these blockchains not only a way of preventing future quantum attacks, but also a means of reestablishing privacy today and thus a critical imperative.

\section{Risks and Challenges in Migrating to Post-Quantum Cryptography}
\label{sec:pqc}

The root cause of the numerous quantum vulnerabilities discussed in this whitepaper lies in the widespread use of cryptographic protocols, especially digital signatures, based on ECDLP  ---  a computational problem believed to be hard for classical computers that is known to be efficiently solvable on quantum machines~\cite{Shor1997polynomialtime}. Therefore, the only durable long-term solution to these vulnerabilities lies in upgrading the underlying schemes to post-quantum alternatives~\cite{Alagic2024transition}.

The task of upgrading blockchains to PQC faces significant headwinds in multiple areas. With few exceptions~\cite{McEliece1978Public,Lamport1979constructing,Merkle1979Secrecy}, PQC protocols are relatively new and have faced less scrutiny and seen less real-world usage than protocols based on ECDLP. They are divided into five categories~\cite{Chen2016report} depending on the choice of the computational problem suspected to be hard for classical and quantum computers, that they use to replace ECDLP as a hardness assumption. The categories include cryptosystems based on lattices~\cite{Peikert2015decade}, hash functions~\cite{Fathalla2024beyond}, codes~\cite{Weger2022survey}, multivariate polynomials~\cite{Ding2006multivariate}, and isogenies~\cite{DeFeo2017mathematics}.

The relative novelty of these schemes raises justifiable concerns about the possibility of undiscovered weaknesses against both classical and quantum attacks. Indeed, the Supersingular Isogeny Diffie-Hellman protocol~\cite{Jao2011towards} has already succumbed to a classical attack~\cite{Castryck2022an}. Security of other post-quantum cryptosystems is an active area of research. In particular, research on quantum algorithms based on Regev's reduction~\cite{Regev2005lattices} has recently led to the discovery of a novel quantum algorithm called Decoded Quantum Interferometry (DQI)~\cite{Jordan2025optimization}. Even though the algorithm tackles an approximate optimization problem rather than a cryptographic task, algorithms based on Regev's reduction are intimately connected with the Short Integer Solution (SIS)~\cite{Ajtai1999generating} and Learning With Errors (LWE)~\cite{Regev2005lattices} problems which feature prominently in hardness assumptions of some of the lattice-based post-quantum cryptosystems. A different line of research, investigating quantum algorithms based on the Kikuchi method~\cite{Wein2025kikuchi, Hastings2020classical}, yielded a new quantum algorithm~\cite{Schmidhuber2025quartic} that achieves a quartic (square of square) speedup over the best known classical algorithm for sparse Learning Parity with Noise (LPN), a problem related to LWE. It is conceivable that further progress in these active areas of research yields unexpected breakthroughs regarding quantum attacks on lattice-based cryptosystems. On the other hand, if these lines of research yield no efficient algorithms for SIS or LWE, then the extra scrutiny will further boost confidence in lattice-based cryptography. Currently, lattice- and hash-based cryptosystems appear to be the most promising of the PQC schemes and have been standardized by the U.S. National Institute for Standards and Technology (NIST)~\cite{NISTUS2024modulelatticebased_kem203, NISTUS2024modulelatticebased_dsa204, NISTUS2024stateless, Cooper2020recommendation}.

A further complication arises from additional requirements of blockchain applications, such as signature aggregation and recursive proof aggregation. Such schemes are needed to aggregate large numbers of signatures in Ethereum's Proof-of-Stake consensus mechanism and would enable blockchains to maintain high transaction rates. Ethereum Foundation has supported and conducted research on analyzing suitable PQC protocols, such as certain variants of the hash-based XMSS signature scheme~\cite{Drake2025hashbased}.

Software implementations of PQC schemes are another source of uncertainty. They are inevitably newer and less battle-hardened than older libraries for ECDLP-based schemes, so any upgrade that introduces them into a blockchain software stack brings with it a chance of new software bugs and subtle security weaknesses. For example, some LWE-based schemes, such as Falcon~\cite{Perlner2025fips, Fouque2020falcon}, involve sampling from discrete Gaussian distribution which has in the past created side-channel vulnerabilities~\cite{Marzougui2023feasibility}.

Use of post-quantum signature schemes also increases resource requirements. Indeed, a key reason for the choice of ECDLP-based signature schemes in blockchain networks is their excellent performance: the popular variants of these signature schemes use small public keys and signatures, which reduces storage and bandwidth requirements, and are efficient to sign and verify, which reduces compute requirements. Unfortunately, none of the PQC signature schemes matches these excellent performance characteristics. Indeed, the storage and compute requirements of PQC schemes are typically one or more orders of magnitude greater than those of the legacy ECDLP-based protocols~\cite{PQShield2026postquantum, Milton2025bitcoin}. Indeed, the high cost of verifying SQISign signatures~\cite{Feo2020sqisign} raised concerns about the possibility of denial of service attacks~\cite{Beast2025p2qrh}. Storage requirements of post-quantum Falcon signatures used on Algorand blockchain are about 1280 bytes in size while quantum-vulnerable ECDSA signatures used in Bitcoin are 64-73 bytes. These increases also affect usability as they may lead to much larger payment addresses~\cite{vanOorschot1999parallel}.

Moreover, a common approach to fortifying digital signatures against possible weaknesses in a novel post-quantum scheme is to use composite signatures~\cite{Bradner1997key, Bindel2023note} that combine a proven scheme, e.g., based on ECDLP, to protect against forgery by classical computers and a novel PQC scheme, e.g., based on LWE, to protect against forgery by quantum computers. This choice enhances security at an even greater performance cost than the PQC scheme alone.

If Bitcoin keeps the block size unchanged while introducing post-quantum signatures, then fewer transactions will fit into a block, reducing future transaction rate. However, increasing the block size proved controversial in the past. In particular, the August 2017 hard fork that created Bitcoin Cash highlights the divisiveness of issues surrounding design choices that affect compute resources needed to run a Bitcoin full node. These choices raise fears that increased resource requirements portend network centralization, which is at odds with some of the foundational ethos of Bitcoin.

The resource costs of post-quantum digital signatures could be absorbed by L2 scaling solutions, amortized using signature aggregation~\cite{Aardal2024aggregating, Drake2025hashbased} or reduced via succinct proofs~\cite{Quan2022improving}, such as post-quantum hash-based SNARKs, that allow recursive proof composition~\cite{Bitansky2012recursive} and promise high efficiency.

Once PQC is adopted, digital asset holders will generally need to initiate transactions to send their assets from old quantum-vulnerable addresses to new post-quantum ones. On some blockchains, this process is expected to run into technical bottlenecks. On the Bitcoin blockchain, at the current transaction rate, the migration would take several months even if the network processed only asset migrations~\cite{Pont2024downtime}. Thus, ideally, the process should start years before the quantum threat truly emerges in order to avoid causing problematic network congestion. We expect this asset migration to take less time on younger blockchains and on blockchains that implement the account model, such as Ethereum, due to the resulting asset consolidation. The dangers associated with the long tail of asset migration can be mitigated using commit-reveal schemes~\cite{Stewart2018committing} which can be used to protect existing script types against on-spend attacks at the cost of increased complexity and duration of transactions involving non-migrated assets.

A related issue concerns the need to alert and educate a very large number of cryptocurrency holders about the need to migrate to post-quantum addresses. This calls for updates to User Interfaces (UIs), ensuring quantum-safe default settings, and possibly even coordinated messaging campaigns.

The technical and social complexities of switching blockchains to post-quantum signature schemes indicate that the process will take years and cannot be delayed until the exact timeline and feasibility of constructing CRQCs become completely clear (and again, those details might not be broadcast publicly). At the same time, these complexities and challenges are feasible to overcome as demonstrated by Algorand, Solana and the XRP Ledger which have made notable progress in real-world adoption of PQC, as well as by QRL, Abelian and Mochimo which have used PQC protocols from inception.

Nevertheless, even when blockchains implement post-quantum signature schemes and asset owners are able to transact out of quantum-vulnerable to quantum-safe addresses, many blockchains will still contain quantum-vulnerable assets, such as Bitcoin's P2PK coins. The P2PK locking scripts have seen little use since 2009 and a majority of the 1.7 millions bitcoin have not been transacted since 2009 despite the fact that such coins are known to be vulnerable to at-rest quantum attacks. This leads to the widespread assumption that (in most cases), the private keys of these Bitcoin UTXOs have been lost. For example, accounts believed to have been owned by Satoshi Nakamoto hold about 1 million bitcoin~\cite{Sklavos2005implementation, Lerner2013well} behind these P2PK scripts and the most common theories are that Satoshi has either died or deliberately burned his private keys (or never recorded them in the first place) either out of some sort of ideological conviction or because they were almost worthless at the time they were initially mined~\cite{Humayun2018``satoshi, FinlowBates2023in, McCormack2018bitcoins, Notes2024new}. If true, such assets can never be safely migrated to a more secure protocol and remain fixed on the ledger without those private keys. Thus, there is no obvious way to update Bitcoin in a fashion that would keep lost P2PK coins and other dormant assets spendable by their owners and also protected against quantum computers. This makes them an especially enticing target for quantum computers and pose a policy challenge for the Bitcoin community and governments.

\section{Dormant Digital Assets}
\label{sec:dormant}

\subsection{The Challenge Posed by Dormant Assets}
\label{sec:dormant_challenge}

Inevitably, some vulnerable assets will not migrate to post-quantum protocols in time or possibly ever, perhaps because their owners do not learn of the threat until it is too late or perhaps because they have lost their private keys. The Ethereum blockchain's contract accounts present similar long-tail migration issues. Thus, in addition to planning and executing upgrades to cryptographic protocols, each cryptocurrency community also faces challenges regarding quantum-vulnerable assets and smart contracts that may linger on public blockchains for an extended or indefinite period of time.

\begin{figure*}[t]
    \centering
        \includegraphics[width=\linewidth]{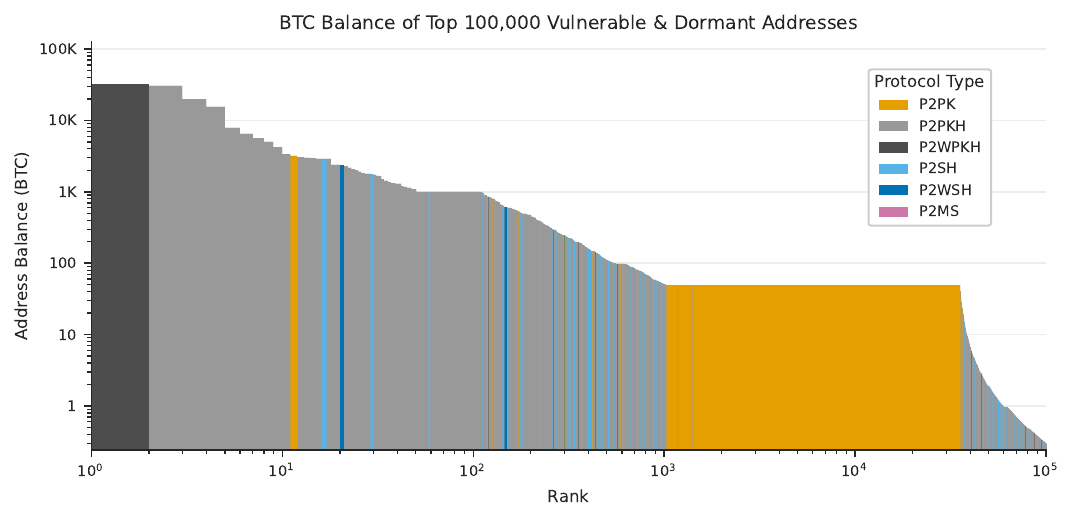}
    \caption{Similar to \fig{btc_vulnerable_addresses}, this plot displays the BTC balance of the top 100,000 Bitcoin addresses ranked by value that are vulnerable to at-rest attacks from exposed or re-used keys and have not initiated a spend in the last 5 years (This ignores things like receiving dust~\cite{Bradbury2013problem}). The sum of all BTC in vulnerable addresses in this list is $\sim$2.3 million BTC. Plot generated using data from \texttt{bigquery-public-data.crypto\_bitcoin}~\cite{Day2018bitcoin}.}
    \label{fig:btc_vulnerable_dormant_addresses}
\end{figure*}

Despite lack of unambiguous precedent, many jurisdictions could classify accessing abandoned cryptographic assets, such as the P2PK coins, without authorization as theft. However, we maintain that if protocol changes are not made, vulnerable assets will eventually be cracked by quantum computers and taken irrespective of the law. In the absence of a clear resolution, these assets are likely to become a lucrative target for bad actors. We quantify the scale of some of the dormant assets at stake in \fig{btc_vulnerable_dormant_addresses}.

Quantum computing providers could put in place systems to detect and deny workloads involving Shor's algorithm, but sufficiently sophisticated actors may find ways to circumvent such safeguards, e.g., by using circuit obfuscation schemes~\cite{Bartusek2024obfuscation, Huang2025obfuscation}. Dormant digital assets, such as P2PK coins, are a sufficiently high value target that the asymmetry of cyberdefense suggests an ``assume breach'' mindset. If taking the coins is clearly deemed illegal, then in the absence of further policy changes, possible outcomes include:
A technologically sophisticated adversary builds their own CRQC and uses it to steal vulnerable bitcoins.
An adversary hacks into a CRQC and uses it to steal vulnerable bitcoins.
An adversary coerces (e.g., through extortion, bribery or violence) people known to have access to a CRQC into stealing or facilitating the theft of vulnerable bitcoins.
A customer of a company cleared to use the device misappropriates their access to steal vulnerable bitcoins or their access is commandeered by a hacker.
This list is not exhaustive. We cannot predict exactly how these coins would be taken but we maintain that if Bitcoin protocol changes are not made, they will eventually be taken (perhaps using devices based in the U.S., perhaps not). Without protocol changes, laws criminalizing the taking of these coins merely ensure they are eventually taken by nefarious agents. This possibility, combined with the enormous value of the assets in question, calls for a resolute policy response.

In the rest of this section we discuss two complementary sets of policy options available to address the challenge of dormant assets. First, we consider the three main policy options discussed in the Bitcoin community, called Do Nothing, Burn, and Hourglass. We also propose a solution based on a new dedicated sidechain that employs offchain cryptographic proofs of ownership to maximize the amount of dormant assets that return to their rightful owners. We call our approach ``bad sidechain'' since it is inspired by the notion of ``bad bank''~\cite{Martini2009bad} used for orderly resolution of distressed assets in traditional finance. Next, we consider public policy responses available to governments. We begin by explaining the practical reasons why legally mandating the destruction of the dormant assets is not viable. Finally, we describe three realistic public policy options: regulated digital salvage, national security response, and government engagement with the Bitcoin community.

\subsection{Bitcoin Community's Options for the Challenge of Dormant Assets}
\label{sec:dormant_bitcoin}

\subsubsection{Do Nothing, Burn, and Hourglass}
\label{sec:dormant_bitcoin_three}

\begin{figure*}[t]
    \centering
        \includegraphics[width=\linewidth]{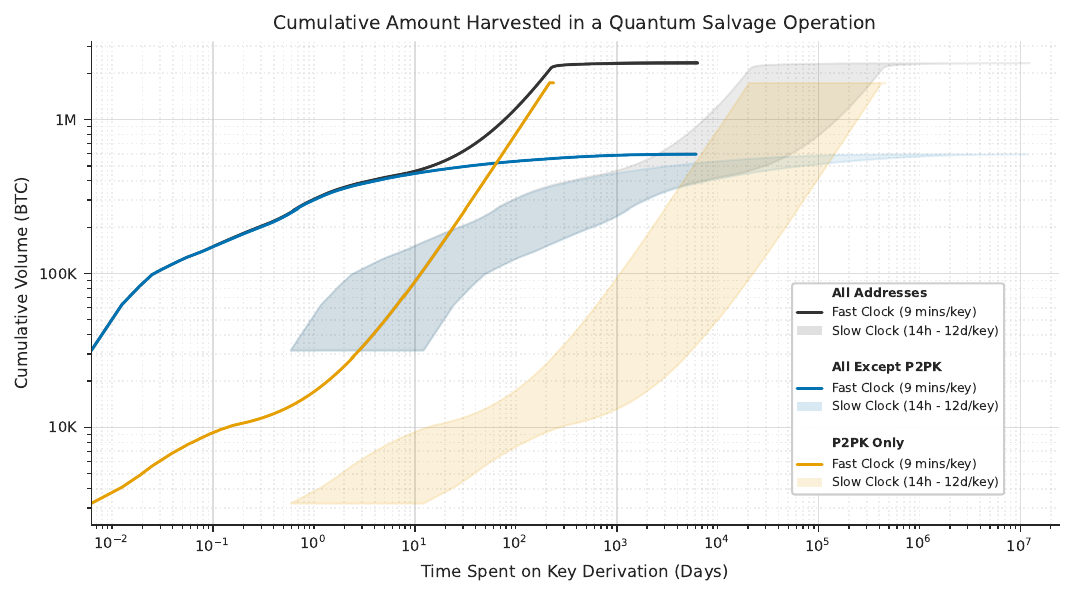}
    \caption{Cumulative amount of money harvested in a quantum salvage operation. We consider a single fast-clock quantum computer or a single slow-clock quantum computer deriving private keys for public keys exposed in P2PK addresses in decreasing order of amount locked. We assume 9 minutes per key derivation on the high clock speed architecture, where we are guided by our rigorous resource estimates, and between $\sim$14 hours and $\sim$12 days per derivation on the slow-clock speed architecture, where we employ a simplified model dominated by magic state cultivation with 10,000 to 200,000 qubits dedicated to cultivation (yielding 200 and 4000 magic T states per second, respectively). The first wallet holds approximately 2 thousand BTC while the total amount harvested is approximately 2.3 million BTC. Only public keys inactive for at least five years are included. Plot generated using data from \texttt{bigquery-public-data.crypto\_bitcoin}~\cite{Day2018bitcoin} combined with chainstate database and raw blockchain data to account for the number of private keys an attacker must derive for each multi-signature address.}
    \label{fig:btc_salvage_rate}
\end{figure*}

The Bitcoin community (especially those with large Bitcoin holdings) might not be enthusiastic to have quantum computers salvage the lost Satoshi era coins. Any sudden increase of total Bitcoin supply would likely lead to a decrease in the value of Bitcoin. We note that the extent of this depends strongly on how, and how quickly, the coins are reintroduced and how many new buyers step forward. Because large quantum salvage operations would themselves quickly become significant holders of the cryptocurrency, they would have incentive to take steps to safeguard its overall value, for example by reintroducing the coins slowly into circulation. We expect the total trove of P2PK coins to take months (in Scenario 1) or years (in Scenario 2) to unlock, see \fig{btc_salvage_rate}.

The Bitcoin community is considering three general categories of approaches to dealing with dormant assets~\cite{Li2025quantum}:
Do Nothing: the protocol is kept unchanged allowing quantum attackers to acquire the dormant assets.
Burn: the protocol is modified to render dormant assets unspendable after a certain date, effectively confiscating all assets locked by P2PK scripts.
Hourglass~\cite{Beast2025bip}: the protocol is modified to limit the rate at which dormant assets may be spent.
These proposals apply a one-size-fits-all policy blind to potential offchain proofs of ownership and are therefore typically only proposed as a means of dealing with the narrow problem of P2PK coins. Later, we contribute a fourth type of solution that takes offchain proofs of ownership into account, applicable to other classes of dormant assets.

The Do Nothing approach is motivated by the desire to protect property rights, avoid setting a confiscatory precedent and preserve Bitcoin's fixed monetary supply, but risks price volatility as the dormant assets are reintroduced into circulation. By contrast, Burn prevents potential supply shock, but expropriates dormant assets. Indeed, any bitcoin owner who does not transact into a post-quantum protocol in time would permanently lose their assets. Burn could also potentially hurt the price of Bitcoin due the damaging precedent it sets. Accordingly, the Burn proposal has proven controversial. Hourglass seeks to strike a balance between these competing concerns by avoiding confiscation while stabilizing Bitcoin price. It also sets up a bottleneck for spending dormant coins which may incentivize parties with CRQCs to enter into bidding wars that transfer some of the abandoned wealth to Bitcoin miners.

In their most straightforward incarnation, Burn and Hourglass proposals entail modifications to the Bitcoin protocol that impose restrictions on transactions spending dormant coins: either by banning them outright after a certain date (in the case of Burn) or limiting their number to for example one per block (in the case of Hourglass). The key challenge for these proposals arises from the need for a broad consensus necessary to implement protocol changes.

The Bitcoin protocol is maintained by a decentralized community of developers and the core network consists of miners (which validate transactions and bundle them into new blocks) and full nodes (which relay blocks and transactions and maintain a full and independent record of the entire blockchain). Software developers, businesses operating mining pools, bitcoin holders, and individual users running full nodes, who together form the Bitcoin community, need to work together to implement and adopt protocol changes, such as Burn and Hourglass.

There are two ways that technical changes can be made to the Bitcoin protocol. The most desirable type of update is known as a ``soft fork''. A soft fork is possible when changes are ``backwards compatible'' and a decisive majority of miners and node operators agree to implement a proposal. The basic idea of a soft fork is that as long as the proposed change is narrowing the protocol rules (the new rules are a subset of the prior rules, e.g. banning certain UTXOs) then as long as a majority only acknowledge blocks that follow these rules, the longest blockchains (and thus, ultimately, the blocks that become permanent) will also follow these rules. However, if the change does not constitute a narrowing of the consensus rules or a clear majority cannot be established, then the only way to push a change is via hard fork. The community generally tries to avoid hard forks, because they result in multiple versions of the cryptocurrency, as happened for example in the 2017 hard fork that created Bitcoin Cash. From a technical perspective, both Burn and Hourglass proposals could be implemented as a soft fork since banning or limiting transactions that spend dormant coins is a narrowing of the protocol rules. However, this requires broad support in the community.

Consensus remains elusive. Indeed, different members of the Bitcoin community have different philosophical beliefs and financial incentives. Those who consider digital property rights fundamental tend to have strong objections to the Burn proposal. Large Bitcoin holders are likely concerned about a potential supply shock and its effect on Bitcoin price. Miners may welcome Do Nothing and Hourglass proposals due to potential increase in transaction fees and volumes. The diversity and complexity of the Bitcoin community makes the ultimate outcome of these ongoing debates hard to predict. Indeed, an informal poll at 2025 Presidio Bitcoin Quantum Summit in San Francisco saw roughly equal support for each of the three categories of solutions ~\cite{Li2025quantum}. In the event of a hard fork, every owner of the original digital asset becomes an owner of an equal amount of each of the two cryptocurrencies emerging from the fork. Relative value of the two assets depends on the ensuing market dynamics. Example developments that may affect relative price of the two cryptocurrencies include significant sales of one of the assets by a major cryptocurrency holder, especially when combined with simultaneous purchases of the other as well as new buyers stepping forward to build up positions in reaction to the fork's effects on predictions of future returns. By contrast, miners generally follow the changing valuations of digital assets.

Irrespective of the outcome of the debate, governments will need to decide whether or not to legalize the salvage of dormant assets for the reasons outlined in the prior section. Indeed, it might be reasonable policy for governments to both legalize the salvage of dormant assets while also recommending that the Bitcoin community burn them. This is not contradictory because it would be taking a position that it is best for the Bitcoin community to protect itself from any adverse economic consequences of quantum salvage, while also recognizing that it cannot control how the protocol is updated.

\subsubsection{Bad Sidechain}
\label{sec:dormant_bitcoin_sidechain}

Even though the Bitcoin blockchain does not provide any mechanism to distinguish the true owner of dormant assets from someone who used a CRQC to derive the private key, for some assets there are offchain mechanisms to do this. Indeed, many public keys are derived from a mnemonic code  ---  a group of easy to remember words  --- 
\cite{Palatinus2013bip} which a quantum attacker
cannot recover, but the original owner could present as proof of ownership. Similarly, users can create offchain cryptographic proofs of ownership in systems such as Project 11's yellowpages registry~\cite{Deegan2025hello}.

None of these offchain proofs of ownership is accepted on the Bitcoin blockchain, but the community could launch a special-purpose pegged sidechain~\cite{Back2014enabling, Singh2020sidechain} where CRQC operators could send the recovered dormant assets, such as P2PK coins, for resolution. The sidechain would implement manual or automated processes for identifying true owners of the original coins and would transfer the assets back to the Bitcoin network once the ownership question is resolved. The recovery sidechain could be derived from a post-quantum variant of Ethereum to enable the use of smart contracts to automate some or all of the resolution processes and its bridge to Bitcoin could be based on the Rootstock blockchain~\cite{Lerner2022rsk} which brings EVM-compatible smart contracts to Bitcoin. CRQC operators may be incentivized to send dormant coins to the recovery sidechain by contingent fees or forced to do so by a protocol upgrade, law, or both.

To expedite recovery, the setup could allow greater centralization than what is acceptable on the Bitcoin blockchain. For example, it could be governed by a consortium of stakeholders, perhaps organized in analogy to Liquid Federation that runs the Liquid Network sidechain~\cite{Nick2020liquid}. The consortium could approve offchain ownership proof mechanisms, set recovery fees, decide burn schedules, adjust the peg and manage the pace of return of recovered coins to control inflationary effects. At a high level, the sidechain would serve a function bearing some analogy to the ``bad bank'' used in the resolution of distressed assets on the books of a failing bank in traditional finance~\cite{Martini2009bad}.

In the presence of a broad agreement on what types of proofs of ownership are acceptable, one can imagine a decentralized market ecosystem of multiple recovery sidechains competing for CRQC-recovered dormant assets. Each sidechain would accept financial risks arising from the legal obligation to return funds to the original owners and would bear the costs of acquiring dormant assets from CRQC operators while attempting to turn a profit by charging owners a deposit fee for the amount of time the recovered assets were held on the sidechain and eventually on the portion of assets for which the return obligation ultimately expires. Compared to Do Nothing, Burn, and Hourglass, the Bad Sidechain approach maximizes the value returned to the rightful owners as well as the value retained in the cryptocurrency community (assuming this community develops, deploys and operates the recovery sidechains) at the cost of greater technical, political and legal complexity.

\subsection{Public Policy Options for the Challenge of Dormant Assets}
\label{sec:dormant_govt}

The most important objective of the transition to quantum-secure blockchain technology is to protect ongoing onchain economic activities. It is possible, even likely, that concerted efforts will fully attain this goal before the first CRQCs come online. By contrast, the objective of preserving transaction confidentiality on privacy-preserving blockchains, such as Zcash and Monero, cannot be fully achieved due to retroactive degradation of ECDLP-protected privacy of known addresses by quantum-capable adversaries.

Another objective is to prevent rogue actors from acquiring substantial quantum-vulnerable assets. This objective is unlikely to be fully achieved by means available to the cryptocurrency communities alone. Indeed, the Bitcoin community in particular remains divided about the handling of dormant assets~\cite{Li2025quantum} and some options being considered, such as Do Nothing and Hourglass discussed earlier, explicitly allow dormant assets to be acquired by actors with CRQCs. Even if the Burn policy ultimately wins on all major blockchains, its practical application may achieve only partial success due to complexities of dealing with the inevitable long tail of unmigrated assets with lost keys and in abandoned wallets. These complexities are addressed by our Bad Sidechain proposal, but its political and technical complexities might be prohibitive. Moreover, policy disagreements may potentially lead to hard forks. Some of the forked blockchains may then retain dormant assets whose fiat value, while reduced compared to the pre-fork amount, can still be substantial. Furthermore, even if disagreements and hard forks are avoided, the long time required to migrate assets to post-quantum protocols~\cite{Pont2024downtime} may leave a window of opportunity for actors with CRQCs to acquire a portion of dormant assets. 

These challenges suggest that the problem of dormant assets is unlikely to be solved completely before the first CRQCs are built. Therefore, governments may need to consider their own options for dealing with offchain consequences. Next, we explain why legally mandated ``destruction'' of dormant assets would be an extremely challenging and likely ineffective policy option for governments before outlining three more realistic policy responses.

\subsubsection{Regulated Destruction of Dormant Assets is Not Viable Policy}
\label{sec:dormant_govt_destruction}

Many legal systems around the world place constraints on the power of governments to take over or destroy private property, such as the Fourth and Fifth Amendments to the United States Constitution. However, even governments that lack or manage to overcome such constraints cannot unilaterally ``destroy'' dormant cryptocurrency assets. Due to the cryptographic relationships between the blocks on the chain, rewriting the contents of a block requires changing all the blocks that follow it. However, some dormant assets, in particular an overwhelming majority of the P2PK coins, are recorded in some of the oldest blocks of the chain (see \fig{btc_transaction_output_fractions}), so erasing them would entail rewriting essentially the entire blockchain history. The computational cost of doing so is equivalent to the computational cost of re-mining more than a decade of Bitcoin's history.

However, a government could require Bitcoin miners in their jurisdiction to reject transactions that spend dormant assets. The cryptocurrency community refers to this as transaction censorship and regards it as extremely controversial. Indeed, removing the need for third party intermediaries together with the associated costs and risks, including transaction censorship, are among Bitcoin's central design objectives~\cite{Nakamoto2008bitcoin, Voskuil2020cryptoeconomics, Li2025quantum}. Blockchain networks exhibit a significant degree of censorship resistance due to the decentralized nature of their architecture, transaction settlement mechanism and governance. If a government banned mining transactions spending dormant assets, then foreign miners outside cooperating jurisdictions could continue to include them in their blocks. Thus, this policy action can delay and increase costs for offending transactions, but does not prevent them from ultimately being settled on the blockchain. Indeed, this is what happened after the U.S. Treasury's Office of Foreign Assets Control (OFAC) sanctioned Tornado Cash~\cite{Nadler2023tornado, Pertsev2019Tornado}, a smart contract protocol, in August 2022~\cite{Wahrstatter2024blockchain, Brownworth2024regulating}. The sanctions were successful in substantially reducing user participation in the protocol and delayed its transactions as many block builders and validators chose to cooperate. However, they ultimately failed to prevent Tornado Cash transactions from eventually being settled by uncooperating network participants.

Therefore, a government would in fact have to require domestic miners not only to reject transactions that spend dormant assets but also valid blocks mined elsewhere and containing banned transactions. Such an action would likely result in Bitcoin mining moving abroad, as it did in 2021 following the ban on Bitcoin mining in China~\cite{Tovanich2022evolution, Shen2021chinas, Reuters2021us}. Alternatively, if a sufficiently large number of miners stayed, the blockchain would undergo a hard fork splitting into two  ---  a domestically mined censored chain and a foreign-mined uncensored one  ---  while leaving every owner with an equal amount of digital assets on both forks. Ultimately, the action would solve the challenge of dormant assets on the censored domestic blockchain while leaving it unsolved on the uncensored one. The degree of success would depend on the relative value of the two cryptocurrencies which would in turn depend on the behavior of market participants, especially large cryptocurrency holders. Given the importance of censorship resistance to cryptocurrency communities, it is reasonable to expect the value of the censored cryptocurrency to suffer significantly due to loss of trust and abandonment by users. In addition, offchain, the action could trigger litigation against expropriation and destruction of billions of USD of private property. At the end of the day, despite the high costs, legally-mandated transaction censorship would likely fail to secure dormant assets against quantum attacks as they would remain vulnerable on the uncensored foreign-mined blockchain.

In order for this type of forceful policy response to have any hope of success, a large fraction of all miners globally would have to acquiesce to transaction censorship. Rough estimates put the joint mining power of the United States, China, Kazakhstan, Canada and the European Union at about 85\% of global mining power~\cite{Finance2026cambridge} (these estimates must be treated with caution since reliable information about miner location is not easily available). This falls short of majorities generally associated with consensus-based protocol upgrades, such as Taproot which reached 99.85\% of miner support~\cite{Walker2026taproot}. Moreover, despite the estimates that miners are concentrated in the above countries, the cryptocurrency mining industry is mobile and responsive to financial incentives~\cite{Tovanich2022evolution}, so in order to deny rogue actors equipped with CRQCs access to dormant assets, a broad, multi-lateral government response would likely be needed to coordinate a long-lasting impact. However, each individual government would be incentivized to allow the proceeds from liquidation of these assets to be realized in their jurisdiction, creating the dynamics of a multi-party Prisoner's Dilemma~\cite{Gokhale2010evolutionary, Poundstone1992prisoners}. This would constrain and complicate international cooperation~\cite{Snyder1971``prisoners} while traditional foreign policy tools for overcoming these limits and complications via institution building~\cite{KEOHANE1984hegemony} would likely prove inadequate due to the possibility of state support for some CRQC-equipped rogue actors. Given the inherent challenges of tightly enforced and sustained international regulation of a globally dispersed and mobile industry, this policy action would face immense practical barriers to success, in addition to forceful opposition from the cryptocurrency community.

\subsubsection{Digital Salvage}
\label{sec:dormant_govt_salvage}

Bitcoin's ownership model (as well as that of many other cryptocurrencies) is based on the principle that the owner of any given asset is anyone possessing the knowledge of the private key corresponding to the public key to which the asset is locked on the ledger. This principle rests on the assumption that the private key cannot be computed by anyone else. However, a CRQC can quickly derive a private key from any public key, thus breaking the foundation of Bitcoin's model of ownership. As soon as actors with access to CRQCs enter the scene, confidence that parties transacting in Bitcoin once had in the ownership of their respective digital assets is irrevocably lost. At this point, none of Bitcoin's onchain mechanisms provide a reliable means of identifying asset owners.

One public policy option treats dormant assets as effectively abandoned property subject to regulated digital salvage (akin to ``sunken treasure''). Existing legal frameworks and property laws provide general guidance for dealing with various types of lost, abandoned and unclaimed assets and help imagine what digital salvage might look like. Furthermore, principles such as escheatment~\cite{Kenton2025escheat}, treasure trove, adverse possession, and maritime salvage offer general precedent for regulated recovery of abandoned property. A legal argument that dormant assets are lost or unowned can be made based on the fact that none of the information on the Bitcoin blockchain enables one to distinguish the true owner from someone with access to a CRQC (though in some cases offchain information may fill that gap). Another general legal principle that may apply is the doctrine of laches~\cite{MerriamWebsterLaches}; if an owner of dormant coins has known for years that their assets are at risk and has failed to transact them to a post-quantum address, then they may be deemed to have failed to assert their rights through inaction.

In the United States, a specific law that might provide a concrete policy blueprint is the Revised Uniform Unclaimed Property Act (2016)~\cite{Commission2016revised} which governs the handling of abandoned property, both tangible and intangible, including ``virtual currency''. In particular, its Article 10 regulates examinations of records for the purpose of identification of unclaimed assets and the transfer of their custody to state authorities. RUUPA implementations specify inactivity time necessary for an asset to be deemed abandoned and regulate fees for record examinations. Following the logic of Article 10, a contract auditor might assist a government's administrator in examination of transaction records on the blockchain to identify digital assets whose public keys have been exposed and which therefore lack reliably identifiable owners. Subsequently, the auditor could use a CRQC to derive the private keys of the identified assets to facilitate the transfer of their custody to state authorities. This type of policy option would introduce a powerful, market-based incentive for quantum computer development and would allow for the resulting financial gains to be transparently integrated into the formal, taxable economy, before they are recovered by bad actors.

However, even though these laws and principles provide a potential basis to design a digital salvage policy, none of them meets the challenge in their present form. In particular, RUUPA deals with assets in possession and control of a business entity called a ``holder'' and distinct from the assets' owner. However, no party involved in the operation of the Bitcoin blockchain clearly meets the legal requirements to be the ``holder'' of the dormant coins. Indeed, none of them possess or control the coins since none of them know the private key. Moreover, RUUPA is a model law of which individual US states enact their own variants while the Bitcoin blockchain is a single world-wide distributed ledger. In the absence of legal precedent, it is ambiguous as to whether the Bitcoin blockchain falls under the purview of RUUPA and it is unclear how one would resolve inevitable conflicts in its application in different states.

\subsubsection{National Security Response}
\label{sec:dormant_govt_sec}

The transnational character of the Bitcoin blockchain suggests a second type of policy option that recognizes the spectre of dormant assets falling to rogue actors as a national security risk. Indeed, some governments will have the option of using CRQCs (or paying a bounty to companies) to acquire these assets (possibly to burn them by sending them to the unspendable \texttt{OP\_RETURN} address~\cite{Walker2025opreturn}) as a national security matter. As before, blockchain's loss of the ability to reliably identify asset owners combined with the laches doctrine~\cite{MerriamWebsterLaches} enables governments to argue that the original owners, through years of inaction, have failed to assert their property rights.

\subsubsection{Engagement with Bitcoin Community}
\label{sec:dormant_govt_engagement}

Finally, governments have the option to engage in a dialog with the cryptocurrency community to advocate for solutions based on their broader offchain consequences. Bitcoin has its origins in the cypherpunk movement espousing the ``code is law'' philosophy~\cite{Lessig1999code} and traditionally characterized by a techno-libertarian ethos~\cite{Russell2006rough}. The community fiercely guards its independence and has a strong preference for reasoned technical solutions arrived at through consensus-seeking discussion over forceful legal interventions. Its general skepticism of government intervention has also been expressed by Bitcoin's pseudonymous creator Satoshi Nakamoto in the genesis block (block zero) which immortalizes a news headline: ``The Times 03/Jan/2009 Chancellor on brink of second bailout for banks''. All of this suggests that the Bitcoin community may prove a potentially challenging partner for governments. However, in recent years, as the Bitcoin community has come to resemble the wider society, the original techno-libertarian ethos has lost some of its former centrality potentially making them easier for governments to engage with it. Moreover, the community has always had a strong sense of the ecosystem's own interests, including its financial stability and continued success, suggesting potentially shared interests with governments, especially for large cryptocurrency holders. This creates the possibility of a productive engagement and an opportunity for governments to shape the community's ongoing debates about the challenge of P2PK coins and other dormant assets.

\section{Outlook}
\label{sec:outlook}

The emergence of CRQCs represents a serious threat to cryptocurrencies that demands a close examination of possible developments at the intersection of quantum computing and digital finance. While the quantum computing and cryptocurrency communities have largely operated in isolation, the significant reduction in resource requirements detailed here necessitates a convergence of these two worlds. Our finding that a superconducting architecture could break 256-bit ECDLP with fewer than half a million physical qubits challenges conventional wisdom regarding the timeline of the threat. This is not merely a distant danger to dormant keys; the potential for early fast-clock CRQCs to launch on-spend attacks within Bitcoin's ten minute average block time places active transactions at immediate risk, dismantling the assumption that fee-based prioritization alone can outrun quantum adversaries.

This rapidly closing window forces the Bitcoin community to face urgent and difficult decisions regarding legacy assets, such as the 1.7 million bitcoin locked in P2PK scripts and an even greater amount of assets vulnerable due to address reuse. Because dormant assets cannot be migrated via standard software updates, they represent a fixed, multibillion-dollar target that will inevitably become accessible to a quantum actor. One way to address this would be a bifurcated approach that respects the distinct authorities of the state and the decentralized network. Governments have a range of policy options to prevent these assets from falling to rogue actors, such as regulated salvage, national security intervention, and engagement with cryptocurrency communities to shape technical solutions. These policy actions help enable the funds resulting from the recovery of P2PK coins and other dormant assets to flow into the formal, taxable economy rather than empowering political adversaries, rogue states, or criminal syndicates. A clear policy framework could act as a complement to, rather than a substitute for, technical intervention. The Bitcoin community retains the autonomy to decide whether to ``burn'' these vulnerable coins to preserve the chain's financial stability, allowing these strategies to function in parallel: governments create legal containers to manage the potential chaos of inevitable salvage, while the community simultaneously weighs the difficult choice of breaking immutability to destroy the vulnerable supply.

Ethereum also faces substantial at-rest vulnerabilities that must be dealt with before CRQCs arrive. The account model uses vulnerable elliptic curves as a core component of onchain identity, putting all accounts that have carried out transactions at risk including high value accounts, such as exchange hot wallets. Smart contracts with exposed admin keys that cannot be easily rotated (without draining and replacing the contracts themselves) create a logistical bottleneck for security upgrades that puts ``low ether, high leverage'' accounts and contracts responsible for tokenized real-world assets, oracles, bridges, guardians, etc. at risk. Moreover, the potential compromise of validators threatens the integrity of the Proof-of-Stake consensus mechanism itself, creating an existential risk to the chain's continued operation. Finally, the vulnerability of Data Availability Sampling mechanism opens it up to on-setup attacks that can be launched without a quantum computer using a reusable exploit created once on a CRQC. Ethereum is more exposed than Bitcoin due to the prevalence of at-rest vulnerabilities, but its recent active steps towards PQC migration promise a potentially more expedient transition to quantum-safe protocols.

Ongoing developments in other blockchain technologies and financial services also add to the urgency. In particular, tokenization of real-world assets, actively taking place on blockchains such as Ethereum and Solana, is expected to open up markets projected to exceed 16 trillion USD by 2030~\cite{Gretz2025tokenization}. At the same time, the cryptographic protocols underlying tokenization remain vulnerable to quantum attacks. In fact, upgrades and new features continue to be built on quantum-vulnerable cryptography and in some cases even weaken existing protections. Given that tokenization is projected to push the potential value of quantum-vulnerable assets above the ``too-big-to-fail'' economic stability thresholds, the situation may justify policy interventions, such as a mandate to deploy post-quantum cryptography.

We contend that the amount of time remaining before the arrival of CRQCs still exceeds the amount of time needed to migrate public blockchains to PQC, though the margin for error is increasingly narrow. Therefore, we have offered updated resource estimates for quantum attacks on blockchain cryptography together with an analysis of vulnerabilities and mitigations in order to urge all vulnerable cryptocurrency communities to begin PQC transition immediately while its timely completion is still the likely prospect. We have also discussed policy options to address offchain consequences of some of the possible migration paths. Finally, we have drawn attention to the numerous ongoing efforts towards PQC transition, such as the development of post-quantum blockchains like the QRL and Abelian, the integration of post-quantum protocols on Algorand, post-quantum experimentation on Solana and the XRP Ledger, as well as the active research and development initiatives spearheaded by the Ethereum Foundation. Above all, these trailblazing projects demonstrate that transition to post-quantum cryptography is realistic and instill hope that it will have been completed before the first CRQCs come online.

\subsection*{Acknowledgements}

We thank Greg Kahanamoku-Meyer and Martin Eker\r{a} for discussions related to the cryptanalysis resource estimates presented in this work. We thank Alan Ho, Alex Pruden, Ansgar Dietrichs, Conor Deegan, Pierre-Luc Dallaire-Demers and Liam Horne for discussions about the quantum vulnerabilities of cryptocurrencies. We thank Mark Zhandry and Alex Meiburg for discussions about our approach to zero-knowledge proofs. We thank James Manyika, Kate Weber, Sergio Boixo, Sophie Schmieg and Thomas Plumb-Reyes for careful review and feedback on this whitepaper. We thank Keegan Ryan from Trail of Bits for catching a software bug that admitted an exploit against the soundness of an earlier version of our zero-knowledge proof.

\subsection*{Financial Conflict of Interest Statement}

We the authors attest that at the time of the initial arXiv and IACR ePrint publication of this article, none of us hold any short positions against any cryptocurrency assets. Some of us hold long positions in cryptocurrencies, including some that involve the use of post-quantum cryptography. The authors reserve the right to initiate any positions in these assets in the future.

\bibliography{refs}

@misc{112026bitcoin,
 author = {{Project 11}},
 howpublished = {Technical Database},
 note = {A comprehensive audit of {Bitcoin} addresses with exposed public keys.},
 title = {{Bitcoin Risq List}},
 url = {https://www.projecteleven.com/bitcoin-risq-list},
 urldate = {2026-03-23},
 year = {2026}
}

@misc{Aardal2024aggregating,
 author = {Marius A. Aardal and Diego F. Aranha and Katharina Boudgoust and Sebastian Kolby and Akira Takahashi},
 howpublished = {Cryptology {ePrint} Archive, Paper 2024/311},
 title = {Aggregating Falcon Signatures with {LaBRADOR}},
 url = {https://eprint.iacr.org/2024/311},
 year = {2024}
}

@misc{Aaronson2025more,
 author = {Aaronson, Scott},
 day = {21},
 howpublished = {\url{https://scottaaronson.blog/?p=9425}},
 month = {dec},
 note = {Accessed: 2026-03-22},
 title = {{More on whether useful quantum computing is ``imminent'' --- Shtetl-Optimized}},
 year = {2025}
}

@misc{Aaronson2026reducing,
 author = {Aaronson, Scott},
 day = {15},
 howpublished = {\url{https://scottaaronson.blog/?p=9564}},
 month = {feb},
 note = {Accessed: 2026-03-22},
 title = {{On reducing the cost of breaking {{RSA-2048}} to 100,000 physical qubits --- Shtetl-Optimized}},
 year = {2026}
}

@article{Acharya2023suppressing,
 author = {Acharya, Rajeev and Aleiner, Igor and Allen, Richard and Andersen, Trond I. and Ansmann, Markus and Arute, Frank and Arya, Kunal and Asfaw, Abraham and Atalaya, Juan and Babbush, Ryan and Bacon, Dave and Bardin, Joseph C. and Basso, Joao and Bengtsson, Andreas and Boixo, Sergio and Bortoli, Gina and Bourassa, Alexandre and Bovaird, Jenna and Brill, Leon and Broughton, Michael and Buckley, Bob B. and Buell, David A. and Burger, Tim and Burkett, Brian and Bushnell, Nicholas and Chen, Yu and Chen, Zijun and Chiaro, Ben and Cogan, Josh and Collins, Roberto and Conner, Paul and Courtney, William and Crook, Alexander L. and Curtin, Ben and Debroy, Dripto M. and Del Toro Barba, Alexander and Demura, Sean and Dunsworth, Andrew and Eppens, Daniel and Erickson, Catherine and Faoro, Lara and Farhi, Edward and Fatemi, Reza and Flores Burgos, Leslie and Forati, Ebrahim and Fowler, Austin G. and Foxen, Brooks and Giang, William and Gidney, Craig and Gilboa, Dar and Giustina, Marissa and Grajales Dau, Alejandro and Gross, Jonathan A. and Habegger, Steve and Hamilton, Michael C. and Harrigan, Matthew P. and Harrington, Sean D. and Higgott, Oscar and Hilton, Jeremy and Hoffmann, Markus and Hong, Sabrina and Huang, Trent and Huff, Ashley and Huggins, William J. and Ioffe, Lev B. and Isakov, Sergei V. and Iveland, Justin and Jeffrey, Evan and Jiang, Zhang and Jones, Cody and Juhas, Pavol and Kafri, Dvir and Kechedzhi, Kostyantyn and Kelly, Julian and Khattar, Tanuj and Khezri, Mostafa and Kieferov{\'a}, M{\'a}ria and Kim, Seon and Kitaev, Alexei and Klimov, Paul V. and Klots, Andrey R. and Korotkov, Alexander N. and Kostritsa, Fedor and Kreikebaum, John Mark and Landhuis, David and Laptev, Pavel and Lau, Kim-Ming and Laws, Lily and Lee, Joonho and Lee, Kenny and Lester, Brian J. and Lill, Alexander and Liu, Wayne and Locharla, Aditya and Lucero, Erik and Malone, Fionn D. and Marshall, Jeffrey and Martin, Orion and McClean, Jarrod R. and McCourt, Trevor and McEwen, Matt and Megrant, Anthony and Meurer Costa, Bernardo and Mi, Xiao and Miao, Kevin C. and Mohseni, Masoud and Montazeri, Shirin and Morvan, Alexis and Mount, Emily and Mruczkiewicz, Wojciech and Naaman, Ofer and Neeley, Matthew and Neill, Charles and Nersisyan, Ani and Neven, Hartmut and Newman, Michael and Ng, Jiun How and Nguyen, Anthony and Nguyen, Murray and Niu, Murphy Yuezhen and O'Brien, Thomas E. and Opremcak, Alex and Platt, John and Petukhov, Andre and Potter, Rebecca and Pryadko, Leonid P. and Quintana, Chris and Roushan, Pedram and Rubin, Nicholas C. and Saei, Negar and Sank, Daniel and Sankaragomathi, Kannan and Satzinger, Kevin J. and Schurkus, Henry F. and Schuster, Christopher and Shearn, Michael J. and Shorter, Aaron and Shvarts, Vladimir and Skruzny, Jindra and Smelyanskiy, Vadim and Smith, W. Clarke and Sterling, George and Strain, Doug and Szalay, Marco and Torres, Alfredo and Vidal, Guifre and Villalonga, Benjamin and Vollgraff Heidweiller, Catherine and White, Theodore and Xing, Cheng and Yao, Z. Jamie and Yeh, Ping and Yoo, Juhwan and Young, Grayson and Zalcman, Adam and Zhang, Yaxing and Zhu, Ningfeng},
 doi = {10.1038/s41586-022-05434-1},
 issn = {1476-4687},
 journal = {Nature},
 month = {February},
 number = {7949},
 pages = {676--681},
 publisher = {Springer Science and Business Media LLC},
 title = {Suppressing quantum errors by scaling a surface code logical qubit},
 url = {http://dx.doi.org/10.1038/s41586-022-05434-1},
 volume = {614},
 year = {2023}
}

@article{Acharya2024quantum,
 author = {Acharya, Rajeev and Abanin, Dmitry A. and Aghababaie-Beni, Laleh and Aleiner, Igor and Andersen, Trond I. and Ansmann, Markus and Arute, Frank and Arya, Kunal and Asfaw, Abraham and Astrakhantsev, Nikita and Atalaya, Juan and Babbush, Ryan and Bacon, Dave and Ballard, Brian and Bardin, Joseph C. and Bausch, Johannes and Bengtsson, Andreas and Bilmes, Alexander and Blackwell, Sam and Boixo, Sergio and Bortoli, Gina and Bourassa, Alexandre and Bovaird, Jenna and Brill, Leon and Broughton, Michael and Browne, David A. and Buchea, Brett and Buckley, Bob B. and Buell, David A. and Burger, Tim and Burkett, Brian and Bushnell, Nicholas and Cabrera, Anthony and Campero, Juan and Chang, Hung-Shen and Chen, Yu and Chen, Zijun and Chiaro, Ben and Chik, Desmond and Chou, Charina and Claes, Jahan and Cleland, Agnetta Y. and Cogan, Josh and Collins, Roberto and Conner, Paul and Courtney, William and Crook, Alexander L. and Curtin, Ben and Das, Sayan and Davies, Alex and De Lorenzo, Laura and Debroy, Dripto M. and Demura, Sean and Devoret, Michel and Di Paolo, Agustin and Donohoe, Paul and Drozdov, Ilya and Dunsworth, Andrew and Earle, Clint and Edlich, Thomas and Eickbusch, Alec and Elbag, Aviv Moshe and Elzouka, Mahmoud and Erickson, Catherine and Faoro, Lara and Farhi, Edward and Ferreira, Vinicius S. and Burgos, Leslie Flores and Forati, Ebrahim and Fowler, Austin G. and Foxen, Brooks and Ganjam, Suhas and Garcia, Gonzalo and Gasca, Robert and Genois, {\'E}lie and Giang, William and Gidney, Craig and Gilboa, Dar and Gosula, Raja and Dau, Alejandro Grajales and Graumann, Dietrich and Greene, Alex and Gross, Jonathan A. and Habegger, Steve and Hall, John and Hamilton, Michael C. and Hansen, Monica and Harrigan, Matthew P. and Harrington, Sean D. and Heras, Francisco J. H. and Heslin, Stephen and Heu, Paula and Higgott, Oscar and Hill, Gordon and Hilton, Jeremy and Holland, George and Hong, Sabrina and Huang, Hsin-Yuan and Huff, Ashley and Huggins, William J. and Ioffe, Lev B. and Isakov, Sergei V. and Iveland, Justin and Jeffrey, Evan and Jiang, Zhang and Jones, Cody and Jordan, Stephen and Joshi, Chaitali and Juhas, Pavol and Kafri, Dvir and Kang, Hui and Karamlou, Amir H. and Kechedzhi, Kostyantyn and Kelly, Julian and Khaire, Trupti and Khattar, Tanuj and Khezri, Mostafa and Kim, Seon and Klimov, Paul V. and Klots, Andrey R. and Kobrin, Bryce and Kohli, Pushmeet and Korotkov, Alexander N. and Kostritsa, Fedor and Kothari, Robin and Kozlovskii, Borislav and Kreikebaum, John Mark and Kurilovich, Vladislav D. and Lacroix, Nathan and Landhuis, David and Lange-Dei, Tiano and Langley, Brandon W. and Laptev, Pavel and Lau, Kim-Ming and Le Guevel, Lo{\"\i}ck and Ledford, Justin and Lee, Joonho and Lee, Kenny and Lensky, Yuri D. and Leon, Shannon and Lester, Brian J. and Li, Wing Yan and Li, Yin and Lill, Alexander T. and Liu, Wayne and Livingston, William P. and Locharla, Aditya and Lucero, Erik and Lundahl, Daniel and Lunt, Aaron and Madhuk, Sid and Malone, Fionn D. and Maloney, Ashley and Mandr\'{a}, Salvatore and Manyika, James and Martin, Leigh S. and Martin, Orion and Martin, Steven and Maxfield, Cameron and McClean, Jarrod R. and McEwen, Matt and Meeks, Seneca and Megrant, Anthony and Mi, Xiao and Miao, Kevin C. and Mieszala, Amanda and Molavi, Reza and Molina, Sebastian and Montazeri, Shirin and Morvan, Alexis and Movassagh, Ramis and Mruczkiewicz, Wojciech and Naaman, Ofer and Neeley, Matthew and Neill, Charles and Nersisyan, Ani and Neven, Hartmut and Newman, Michael and Ng, Jiun How and Nguyen, Anthony and Nguyen, Murray and Ni, Chia-Hung and Niu, Murphy Yuezhen and O'Brien, Thomas E. and Oliver, William D. and Opremcak, Alex and Ottosson, Kristoffer and Petukhov, Andre and Pizzuto, Alex and Platt, John and Potter, Rebecca and Pritchard, Orion and Pryadko, Leonid P. and Quintana, Chris and Ramachandran, Ganesh and Reagor, Matthew J. and Redding, John and Rhodes, David M. and Roberts, Gabrielle and Rosenberg, Eliott and Rosenfeld, Emma and Roushan, Pedram and Rubin, Nicholas C. and Saei, Negar and Sank, Daniel and Sankaragomathi, Kannan and Satzinger, Kevin J. and Schurkus, Henry F. and Schuster, Christopher and Senior, Andrew W. and Shearn, Michael J. and Shorter, Aaron and Shutty, Noah and Shvarts, Vladimir and Singh, Shraddha and Sivak, Volodymyr and Skruzny, Jindra and Small, Spencer and Smelyanskiy, Vadim and Smith, W. Clarke and Somma, Rolando D. and Springer, Sofia and Sterling, George and Strain, Doug and Suchard, Jordan and Szasz, Aaron and Sztein, Alex and Thor, Douglas and Torres, Alfredo and Torunbalci, M. Mert and Vaishnav, Abeer and Vargas, Justin and Vdovichev, Sergey and Vidal, Guifre and Villalonga, Benjamin and Heidweiller, Catherine Vollgraff and Waltman, Steven and Wang, Shannon X. and Ware, Brayden and Weber, Kate and Weidel, Travis and White, Theodore and Wong, Kristi and Woo, Bryan W. K. and Xing, Cheng and Yao, Z. Jamie and Yeh, Ping and Ying, Bicheng and Yoo, Juhwan and Yosri, Noureldin and Young, Grayson and Zalcman, Adam and Zhang, Yaxing and Zhu, Ningfeng and Zobrist, Nicholas},
 doi = {10.1038/s41586-024-08449-y},
 issn = {1476-4687},
 journal = {Nature},
 month = {December},
 number = {8052},
 pages = {920--926},
 publisher = {Springer Science and Business Media LLC},
 title = {Quantum error correction below the surface code threshold},
 url = {http://dx.doi.org/10.1038/s41586-024-08449-y},
 volume = {638},
 year = {2024}
}

@article{Aggarwal2018quantum,
 author = {Aggarwal, Divesh and Brennen, Gavin and Lee, Troy and Santha, Miklos and Tomamichel, Marco},
 doi = {10.5195/ledger.2018.127},
 issn = {2379-5980},
 journal = {Ledger},
 month = {October},
 publisher = {University Library System, University of Pittsburgh},
 title = {Quantum Attacks on Bitcoin, and How to Protect Against Them},
 url = {http://dx.doi.org/10.5195/ledger.2018.127},
 volume = {3},
 year = {2018}
}

@inproceedings{Ajtai1999generating,
 address = {Berlin, Heidelberg},
 author = {Ajtai, Mikl{'o}s},
 booktitle = {Automata, Languages and Programming},
 editor = {Wiedermann, Jir{'i}
and van Emde Boas, Peter
and Nielsen, Mogens},
 isbn = {978-3-540-48523-0},
 pages = {1--9},
 publisher = {Springer Berlin Heidelberg},
 title = {Generating Hard Instances of the Short Basis Problem},
 year = {1999}
}

@techreport{Alagic2024transition,
 author = {Alagic, Gorjan and Chen, Lily and Jordan, Stephen and Liu, Yi-Kai and Moody, Dustin and Peralta, Rene and Perlner, Ray and Sonneman, Daniel},
 day = {12},
 doi = {10.6028/NIST.IR.8547.ipd},
 institution = {National Institute of Standards and Technology},
 month = {nov},
 number = {NIST IR 8547},
 title = {{Transition to Post-Quantum Cryptography Standards}},
 type = {NIST Internal Report (Initial Public Draft)},
 url = {https://csrc.nist.gov/pubs/ir/8547/ipd},
 year = {2024}
}

@techreport{Alice2022abelian,
 author = {Alice and Bob and Eve and Lambda},
 day = {18},
 institution = {The {Abelian} Foundation},
 month = {feb},
 title = {{Abelian ({ABEL}) -- A Quantum-Resistant Cryptocurrency Balancing Privacy and Accountability}},
 type = {Cryptographic White Paper},
 url = {https://download.pqabelian.io/release/docs/whitepaper.pdf},
 year = {2022}
}

@misc{Amy2016estimating,
 author = {Matthew Amy and Olivia Di Matteo and Vlad Gheorghiu and Michele Mosca and Alex Parent and John Schanck},
 howpublished = {Cryptology {ePrint} Archive, Paper 2016/992},
 title = {Estimating the cost of generic quantum pre-image attacks on {SHA}-2 and {SHA}-3},
 url = {https://eprint.iacr.org/2016/992},
 year = {2016}
}

@article{Anderson1990technological,
 author = {Philip Anderson and Michael L. Tushman},
 issn = {00018392},
 journal = {Administrative Science Quarterly},
 number = {4},
 pages = {604--633},
 publisher = {[Sage Publications, Inc., Johnson Graduate School of Management, Cornell University]},
 title = {Technological Discontinuities and Dominant Designs: A Cyclical Model of Technological Change},
 url = {http://www.jstor.org/stable/2393511},
 urldate = {2026-03-22},
 volume = {35},
 year = {1990}
}

@book{antonopoulos2014mastering,
 author = {Antonopoulos, A.M.},
 isbn = {9781491902646},
 lccn = {2015413040},
 publisher = {O'Reilly Media},
 title = {Mastering Bitcoin: Unlocking Digital Cryptocurrencies},
 url = {https://books.google.com/books?id=IXmrBQAAQBAJ},
 year = {2014}
}

@book{Antonopoulos2018mastering,
 author = {Antonopoulos, A.M. and D, G.W.P.},
 edition = {1st},
 isbn = {9781491971918},
 publisher = {O'Reilly Media},
 title = {Mastering Ethereum: Building Smart Contracts and DApps},
 url = {https://books.google.com/books?id=nJJ5DwAAQBAJ},
 year = {2018}
}

@article{Awschalom2021development,
 author = {Awschalom, David and Berggren, Karl K. and Bernien, Hannes and Bhave, Sunil and Carr, Lincoln D. and Davids, Paul and Economou, Sophia E. and Englund, Dirk and Faraon, Andrei and Fejer, Martin and Guha, Saikat and Gustafsson, Martin V. and Hu, Evelyn and Jiang, Liang and Kim, Jungsang and Korzh, Boris and Kumar, Prem and Kwiat, Paul G. and Lon\v{c}ar, Marko and Lukin, Mikhail D. and Miller, David A.B. and Monroe, Christopher and Nam, Sae Woo and Narang, Prineha and Orcutt, Jason S. and Raymer, Michael G. and Safavi-Naeini, Amir H. and Spiropulu, Maria and Srinivasan, Kartik and Sun, Shuo and Vu\v{c}kovi{\'c}, Jelena and Waks, Edo and Walsworth, Ronald and Weiner, Andrew M. and Zhang, Zheshen},
 doi = {10.1103/prxquantum.2.017002},
 issn = {2691-3399},
 journal = {PRX Quantum},
 month = {February},
 number = {1},
 publisher = {American Physical Society (APS)},
 title = {Development of Quantum Interconnects (QuICs) for Next-Generation Information Technologies},
 url = {http://dx.doi.org/10.1103/PRXQuantum.2.017002},
 volume = {2},
 year = {2021}
}

@article{Babbush2021focus,
 author = {Babbush, Ryan and McClean, Jarrod R. and Newman, Michael and Gidney, Craig and Boixo, Sergio and Neven, Hartmut},
 doi = {10.1103/prxquantum.2.010103},
 issn = {2691-3399},
 journal = {PRX Quantum},
 month = {March},
 number = {1},
 publisher = {American Physical Society (APS)},
 title = {Focus beyond Quadratic Speedups for Error-Corrected Quantum Advantage},
 url = {http://dx.doi.org/10.1103/PRXQuantum.2.010103},
 volume = {2},
 year = {2021}
}

@article{Babbush2025,
 author = {Ryan Babbush and Robbie King and Sergio Boixo and William Huggins and Tanuj Khattar and Guang Hao Low and Jarrod R. McClean and Thomas O'Brien and Nicholas Rubin},
 journal = {arXiv:2511.09124},
 title = {The Grand Challenge of Quantum Applications},
 url = {https://arxiv.org/abs/2511.09124},
 year = {2025}
}

@article{Babbush2025grand,
 author = {Babbush, Ryan and King, Robbie and Boixo, Sergio and Huggins, William and Khattar, Tanuj and Low, Guang Hao and McClean, Jarrod R. and O'Brien, Thomas and Rubin, Nicholas C.},
 journal = {arXiv:2511.09124},
 title = {The Grand Challenge of Quantum Applications},
 url = {https://arxiv.org/abs/2511.09124},
 year = {2025}
}

@techreport{Back2014enabling,
 author = {Back, Adam and Corallo, Matt and Dashjr, Luke and Friedenbach, Mark and Maxwell, Gregory and Miller, Andrew and Poelstra, Andrew and Tim{'o}n, Jorge and Wuille, Pieter},
 day = {22},
 institution = {Blockstream},
 month = {oct},
 number = {Commit: {{5620e43}}},
 title = {{Enabling Blockchain Innovations with Pegged Sidechains}},
 type = {Protocol White Paper},
 url = {https://blockstream.com/sidechains.pdf},
 year = {2014}
}

@techreport{Barker2020NIST,
  author = {Barker, Elaine},
  title = {{NIST SP 800-57 Part 1 Rev. 5}: Recommendation for Key Management: Part 1 -- {G}eneral},
  institution = {National Institute of Standards and Technology},
  year = {2020},
  month = may,
  type = {NIST Special Publication},
  number = {800-57 Part 1 Rev. 5},
  address = {Gaithersburg, MD, USA},
  doi = {10.6028/NIST.SP.800-57pt1r5},
  url = {https://csrc.nist.gov/pubs/sp/800/57/pt1/r5/final}
}

@misc{Barreto2002constructing,
 author = {Paulo S.  L.  M.  Barreto and Ben Lynn and Michael Scott},
 howpublished = {Cryptology {ePrint} Archive, Paper 2002/088},
 title = {Constructing Elliptic Curves with Prescribed Embedding Degrees},
 url = {https://eprint.iacr.org/2002/088},
 year = {2002}
}

@misc{Barreto2005pairingfriendly,
 author = {Paulo S.  L.  M.  Barreto and Michael Naehrig},
 howpublished = {Cryptology {ePrint} Archive, Paper 2005/133},
 title = {Pairing-Friendly Elliptic Curves of Prime Order},
 url = {https://eprint.iacr.org/2005/133},
 year = {2005}
}

@phdthesis{Bartusek2024obfuscation,
 author = {Bartusek, James},
 month = {Aug},
 number = {UCB/EECS-2024-169},
 school = {EECS Department, University of California, Berkeley},
 title = {Obfuscation of Quantum Computation},
 url = {http://www2.eecs.berkeley.edu/Pubs/TechRpts/2024/EECS-2024-169.html},
 year = {2024}
}

@techreport{Beast2024bip,
 author = {Beast, Hunter and Heilman, Ethan and Duke, Isabel Foxen},
 day = {18},
 institution = {Bitcoin Core},
 month = {dec},
 note = {Source: \url{https://github.com/bitcoin/bips/blob/master/bip-0360.mediawiki}. Accessed: 2026-03-22},
 number = {360},
 title = {{BIP 360: {Pay-to-Merkle-Root} ({P2MR})}},
 type = {Bitcoin Improvement Proposal},
 url = {https://bips.dev/360/},
 year = {2024}
}

@misc{Beast2025bip,
 author = {Beast, Hunter and Casey, Michael},
 day = {29},
 howpublished = {Bitcoin Improvement Proposal (Draft), {GitHub} Repository},
 month = {apr},
 note = {Proposes a soft fork consensus rule restricting {P2PK} throughput to mitigate inflationary quantum attacks.},
 title = {{BIP Draft}: {Hourglass} Spending Rules},
 url = {https://github.com/cryptoquick/bips/blob/hourglass/bip-hourglass.mediawiki},
 urldate = {2026-03-23},
 year = {2025}
}

@misc{Beast2025p2qrh,
 author = {Beast, Hunter},
 day = {19},
 howpublished = {Bitcoin Development Mailing List ({Google Groups})},
 month = {feb},
 note = {Discussion of {BIP-360} featuring contributions from {Matt Corallo}, {Jonas Nick}, and others. Highlights the {DDoS} risk of {Post-Quantum} transition scripts in {Bitcoin}.},
 title = {{P2QRH} / {BIP-360} Update},
 url = {https://groups.google.com/g/bitcoindev/c/oQKezDOc4us},
 urldate = {2026-03-23},
 year = {2025}
}

@misc{BenSasson2013succinct,
 author = {Eli Ben-Sasson and Alessandro Chiesa and Eran Tromer and Madars Virza},
 howpublished = {Cryptology {ePrint} Archive, Paper 2013/879},
 title = {Succinct Non-Interactive Zero Knowledge for a von Neumann Architecture},
 url = {https://eprint.iacr.org/2013/879},
 year = {2013}
}

@inproceedings{BenSasson2018fast,
 author = {Ben-Sasson, Eli and Bentov, Iddo and Horesh, Yinon and Riabzev, Michael},
 copyright = {Creative Commons Attribution 3.0 Unported license},
 doi = {10.4230/LIPICS.ICALP.2018.14},
 journal = {LIPIcs, Volume 107, ICALP 2018},
 language = {English},
 pages = {14:1-14:17},
 publisher = {Schloss Dagstuhl--Leibniz-Zentrum f\"ur Informatik},
 title = {Fast Reed-Solomon Interactive Oracle Proofs of Proximity},
 url = {https://drops.dagstuhl.de/entities/document/10.4230/LIPIcs.ICALP.2018.14},
 volume = {107},
 year = {2018}
}

@misc{BenSasson2018scalable,
 author = {Eli Ben-Sasson and Iddo Bentov and Yinon Horesh and Michael Riabzev},
 howpublished = {Cryptology {ePrint} Archive, Paper 2018/046},
 title = {Scalable, transparent, and post-quantum secure computational integrity},
 url = {https://eprint.iacr.org/2018/046},
 year = {2018}
}

@inproceedings{BenSasson2019scalable,
 address = {Berlin, Heidelberg},
 author = {Ben-Sasson, Eli and Bentov, Iddo and Horesh, Yinon and Riabzev, Michael},
 booktitle = {Advances in Cryptology--CRYPTO 2019: 39th Annual International Cryptology Conference, Santa Barbara, CA, USA, August 18--22, 2019, Proceedings, Part III},
 doi = {10.1007/978-3-030-26954-8_23},
 isbn = {978-3-030-26953-1},
 location = {Santa Barbara, CA, USA},
 numpages = {32},
 pages = {701--732},
 publisher = {Springer-Verlag},
 title = {Scalable Zero Knowledge with No Trusted Setup},
 url = {https://doi.org/10.1007/978-3-030-26954-8_23},
 year = {2019}
}

@inbook{Bernstein2009introduction,
 address = {Berlin, Heidelberg},
 author = {Bernstein, Daniel J.},
 booktitle = {Post-Quantum Cryptography},
 doi = {10.1007/978-3-540-88702-7_1},
 editor = {Bernstein, Daniel J.
and Buchmann, Johannes
and Dahmen, Erik},
 isbn = {978-3-540-88702-7},
 pages = {1--14},
 publisher = {Springer Berlin Heidelberg},
 title = {Introduction to post-quantum cryptography},
 url = {https://doi.org/10.1007/978-3-540-88702-7_1},
 year = {2009}
}

@article{Bernstein2017,
 author = {Bernstein, Daniel J. and Lange, Tanja},
 day = {01},
 doi = {10.1038/nature23461},
 issn = {1476-4687},
 journal = {Nature},
 month = {Sep},
 number = {7671},
 pages = {188-194},
 title = {Post-quantum cryptography},
 url = {https://doi.org/10.1038/nature23461},
 volume = {549},
 year = {2017}
}

@book{Bernstein2017postquantum,
 author = {D.J. Bernstein and T. Lange},
 language = {English},
 publisher = {International Association for Cryptologic Research},
 series = {Cryptology ePrint Archive},
 title = {Post-quantum cryptography: dealing with the fallout of physics success},
 year = {2017}
}

@article{Berry2019qubitization,
 author = {Berry, Dominic W. and Gidney, Craig and Motta, Mario and McClean, Jarrod R. and Babbush, Ryan},
 doi = {10.22331/q-2019-12-02-208},
 issn = {2521-327X},
 journal = {Quantum},
 month = {December},
 pages = {208},
 publisher = {Verein zur Forderung des Open Access Publizierens in den Quantenwissenschaften},
 title = {Qubitization of Arbitrary Basis Quantum Chemistry Leveraging Sparsity and Low Rank Factorization},
 url = {http://dx.doi.org/10.22331/q-2019-12-02-208},
 volume = {3},
 year = {2019}
}

@misc{Bindel2023note,
 author = {Nina Bindel and Britta Hale},
 howpublished = {Cryptology {ePrint} Archive, Paper 2023/423},
 title = {A Note on Hybrid Signature Schemes},
 url = {https://eprint.iacr.org/2023/423},
 year = {2023}
}

@inproceedings{Bitansky2012from,
 address = {New York, NY, USA},
 author = {Bitansky, Nir and Canetti, Ran and Chiesa, Alessandro and Tromer, Eran},
 booktitle = {Proceedings of the 3rd Innovations in Theoretical Computer Science Conference},
 doi = {10.1145/2090236.2090263},
 isbn = {9781450311151},
 location = {Cambridge, Massachusetts},
 numpages = {24},
 pages = {326--349},
 publisher = {Association for Computing Machinery},
 series = {ITCS '12},
 title = {From extractable collision resistance to succinct non-interactive arguments of knowledge, and back again},
 url = {https://doi.org/10.1145/2090236.2090263},
 year = {2012}
}

@misc{Bitansky2012recursive,
 author = {Nir Bitansky and Ran Canetti and Alessandro Chiesa and Eran Tromer},
 howpublished = {Cryptology {ePrint} Archive, Paper 2012/095},
 title = {Recursive Composition and Bootstrapping for {SNARKs} and Proof-Carrying Data},
 url = {https://eprint.iacr.org/2012/095},
 year = {2012}
}

@manual{Bitmain2021S19Pro,
 author = {{Bitmain}},
 title = {Antminer {S19 Pro} Specifications},
 organization = {Bitmain Technologies Ltd.},
 year = {2021},
 month = may,
 day = {27},
 note = {Created Feb 27, 2020. Updated May 27, 2021.},
 url = {https://support.bitmain.com/hc/en-us/articles/900000261726-S19-Pro-Specifications},
}

@article{Bluvstein2025faulttolerant,
 author = {Bluvstein, Dolev and Geim, Alexandra A. and Li, Sophie H. and Evered, Simon J. and Bonilla Ataides, J. Pablo and Baranes, Gefen and Gu, Andi and Manovitz, Tom and Xu, Muqing and Kalinowski, Marcin and Majidy, Shayan and Kokail, Christian and Maskara, Nishad and Trapp, Elias C. and Stewart, Luke M. and Hollerith, Simon and Zhou, Hengyun and Gullans, Michael J. and Yelin, Susanne F. and Greiner, Markus and Vuleti{\'c}, Vladan and Cain, Madelyn and Lukin, Mikhail D.},
 doi = {10.1038/s41586-025-09848-5},
 issn = {1476-4687},
 journal = {Nature},
 month = {November},
 number = {8095},
 pages = {39--46},
 publisher = {Springer Science and Business Media LLC},
 title = {A fault-tolerant neutral-atom architecture for universal quantum computation},
 url = {http://dx.doi.org/10.1038/s41586-025-09848-5},
 volume = {649},
 year = {2025}
}

@inbook{Boneh2001short,
 author = {Boneh, Dan and Lynn, Ben and Shacham, Hovav},
 booktitle = {Advances in Cryptology -- ASIACRYPT 2001},
 doi = {10.1007/3-540-45682-1_30},
 isbn = {9783540456827},
 issn = {0302-9743},
 pages = {514--532},
 publisher = {Springer Berlin Heidelberg},
 title = {Short Signatures from the Weil Pairing},
 url = {http://dx.doi.org/10.1007/3-540-45682-1_30},
 year = {2001}
}

@misc{Bos2013elliptic,
 author = {Joppe W.  Bos and J.  Alex Halderman and Nadia Heninger and Jonathan Moore and Michael Naehrig and Eric Wustrow},
 howpublished = {Cryptology {ePrint} Archive, Paper 2013/734},
 title = {Elliptic Curve Cryptography in Practice},
 url = {https://eprint.iacr.org/2013/734},
 year = {2013}
}

@misc{Bowe2025zcash,
 author = {Bowe, Sean},
 day = {16},
 howpublished = {\url{https://seanbowe.com/blog/zcash-and-quantum-computers/}},
 month = {oct},
 note = {Accessed: 2026-03-22},
 title = {{Zcash and Quantum Computers}},
 year = {2025}
}

@misc{BoweHopwood2019ZIP209,
 author = {Bowe, Sean and Hopwood, Daira-Emma},
 day = {25},
 howpublished = {Zcash Improvement Proposal ({ZIP}) 209},
 month = {feb},
 title = {{ZIP 209}: Prohibit Negative Shielded Chain Value Pool Balances},
 url = {https://github.com/zcash/zips/blob/master/zips/zip-0209.rst},
 urldate = {2026-03-23},
 year = {2019}
}

@article{Bradbury2013problem,
 author = {Danny Bradbury},
 doi = {https://doi.org/10.1016/S1361-3723(13)70101-5},
 issn = {1361-3723},
 journal = {Computer Fraud \& Security},
 number = {11},
 pages = {5-8},
 title = {The problem with Bitcoin},
 url = {https://www.sciencedirect.com/science/article/pii/S1361372313701015},
 volume = {2013},
 year = {2013}
}

@book{Bradner1997key,
 author = {Bradner, S.},
 doi = {10.17487/rfc2119},
 institution = {RFC Editor},
 month = {March},
 title = {Key words for use in RFCs to Indicate Requirement Levels},
 url = {http://dx.doi.org/10.17487/RFC2119},
 year = {1997}
}

@article{Bravyi2024highthreshold,
 author = {Bravyi, Sergey and Cross, Andrew W. and Gambetta, Jay M. and Maslov, Dmitri and Rall, Patrick and Yoder, Theodore J.},
 doi = {10.1038/s41586-024-07107-7},
 issn = {1476-4687},
 journal = {Nature},
 month = {March},
 number = {8005},
 pages = {778--782},
 publisher = {Springer Science and Business Media LLC},
 title = {High-threshold and low-overhead fault-tolerant quantum memory},
 url = {http://dx.doi.org/10.1038/s41586-024-07107-7},
 volume = {627},
 year = {2024}
}

@techreport{Brown2010sec,
 author = {Brown, Daniel R. L.},
 day = {27},
 institution = {Certicom Research},
 month = {jan},
 note = {Standards for Efficient Cryptography (SEC).},
 number = {Version 2.0},
 title = {{SEC 2: Recommended Elliptic Curve Domain Parameters}},
 type = {Technical Specification},
 url = {https://www.secg.org/sec2-v2.pdf},
 year = {2010}
}

@misc{Brown2020rolling,
 author = {Daniel R.  L.  Brown},
 title = {Rolling up sleeves when subversion's in a field?},
 howpublished = {Cryptology {ePrint} Archive, Paper 2020/074},
 year = {2020},
 url = {https://eprint.iacr.org/2020/074}
}

@techreport{Brownworth2024regulating,
 author = {Anders Brownworth and Jon Durfee and Michael Junho Lee and Antoine Martin},
 doi = {10.59576/sr.1112},
 institution = {Federal Reserve Bank of New York},
 month = {Aug},
 number = {1112},
 title = {Regulating Decentralized Systems: Evidence from Sanctions on Tornado Cash},
 type = {Staff Reports},
 url = {https://www.newyorkfed.org/research/staff_reports/sr1112.html},
 year = {2024}
}

@techreport{BTSE2023FUD,
 author = {{BTSE}},
 title = {Crypto Trading Psychology: Dealing with {FUD} and {FOMO} in the Cryptocurrency Market},
 institution = {{BTSE}},
 year = {2023},
 month = apr,
 day = {21},
 type = {Special Report},
 url = {https://www.btse.com/blog/wp-content/uploads/2023/04/Crypto-Trading-Psychology.pdf},
 note = {{BTSE} Trading Psychology Series. Accessed March 28, 2026.}
}

@misc{Bunz2017Bulletproofs,
 author = {Benedikt B{\"u}nz and Jonathan Bootle and Dan Boneh and Andrew Poelstra and Pieter Wuille and Greg Maxwell},
 title = {Bulletproofs: Short Proofs for Confidential Transactions and More},
 howpublished = {Cryptology {ePrint} Archive, Paper 2017/1066},
 year = {2017},
 url = {https://eprint.iacr.org/2017/1066}
}

@techreport{Burkett2020lip,
 author = {Burkett, David},
 day = {28},
 institution = {Litecoin Project},
 month = {feb},
 note = {Accessed: 2026-03-22},
 number = {4},
 title = {{LIP 004: One-Sided Transactions in {MimbleWimble} ({Consensus layer})}},
 type = {Litecoin Improvement Proposal},
 url = {https://github.com/litecoin-project/lips/blob/master/lip-0004.mediawiki},
 year = {2020}
}

@techreport{Buterin2014ethereum,
 author = {Buterin, Vitalik},
 institution = {Ethereum Foundation},
 month = {jan},
 title = {{Ethereum: A Next-Generation Smart Contract \& Decentralized Application Platform}},
 type = {Protocol White Paper},
 url = {https://ethereum.org/en/whitepaper/},
 year = {2014}
}

@article{Buterin2017casper,
 author = {Buterin, Vitalik and Griffith, Virgil},
 journal = {arXiv:1710.09437},
 title = {Casper the Friendly Finality Gadget},
 url = {https://arxiv.org/abs/1710.09437},
 year = {2017}
}

@article{Buterin2020combining,
 author = {Buterin, Vitalik and Hernandez, Diego and Kamphefner, Thor and Pham, Khiem and Qiao, Zhi and Ryan, Danny and Sin, Juhyeok and Wang, Ying and Zhang, Yan X},
 journal = {arXiv:2003.03052},
 title = {Combining GHOST and Casper},
 url = {https://arxiv.org/abs/2003.03052},
 year = {2020}
}

@techreport{Buterin2021erc4337,
 author = {Buterin, Vitalik and Weiss, Yoav and Tirosh, Dror and Nacson, Shahaf and Forshtat, Alex and Gazso, Kristof and Hess, Tjaden},
 day = {29},
 institution = {Ethereum Foundation},
 month = {sep},
 note = {Accessed: 2026-03-23},
 number = {4337},
 title = {{ERC-4337: Account Abstraction Using Alt Mempool [DRAFT]}},
 type = {Ethereum Improvement Proposal},
 url = {https://eips.ethereum.org/EIPS/eip-4337},
 year = {2021}
}

@techreport{Buterin2022eip4844,
 author = {Buterin, Vitalik and Feist, Dankrad and Loerakker, Diederik and Kadianakis, George and Garnett, Matt and Taiwo, Mofi and Dietrichs, Ansgar},
 day = {25},
 institution = {Ethereum Foundation},
 month = {feb},
 note = {Accessed: 2026-03-23},
 number = {4844},
 title = {{EIP-4844: Shard Blob Transactions}},
 type = {Ethereum Improvement Proposal},
 url = {https://eips.ethereum.org/EIPS/eip-4844},
 year = {2022}
}

@techreport{Buterin2024eip7702,
 author = {Buterin, Vitalik and Wilson, Sam and Dietrichs, Ansgar and lightclient},
 day = {7},
 institution = {Ethereum Foundation},
 month = {may},
 note = {Accessed: 2026-03-23},
 number = {7702},
 title = {{EIP-7702: Set Code for EOAs}},
 type = {Ethereum Improvement Proposal},
 url = {https://eips.ethereum.org/EIPS/eip-7702},
 year = {2024}
}

@misc{Carter2026Bitcoin,
 author = {Carter, Nic},
 day = {74},
 howpublished = {\url{https://murmurationstwo.substack.com/p/bitcoin-developers-are-mostly-not}},
 month = {feb},
 note = {Accessed: 2026-03-22},
 title = {{Bitcoin developers are mostly not concerned about quantum risk -- Murmurations II}},
 year = {2026}
}

@misc{CastleLabs2026openeden,
 author = {{Castle Labs, Inc.}},
 howpublished = {\url{https://app.rwa.xyz/assets/TBILL}},
 note = {Accessed on 2026-02-24 and observed ~61.8 million USD on the XRP Ledger out of total supply of ~92.5 million USD.},
 title = {{OpenEden T-Bill (TBILL) Vault --- Real-World Asset Analytics Dashboard (RWA.xyz)}},
 year = {2026}
}

@misc{CastleLabs2026realworld,
 author = {{Castle Labs, Inc.}},
 howpublished = {\url{https://app.rwa.xyz}},
 note = {Accessed: 2026-03-21},
 title = {{Real-World Asset Analytics Dashboard (RWA.xyz)}},
 year = {2026}
}

@misc{Castryck2022an,
 author = {Wouter Castryck and Thomas Decru},
 howpublished = {Cryptology {ePrint} Archive, Paper 2022/975},
 title = {An efficient key recovery attack on {SIDH}},
 url = {https://eprint.iacr.org/2022/975},
 year = {2022}
}

@misc{Center2026certcc,
 author = {{CERT Coordination Center}},
 howpublished = {Official Policy Document},
 institution = {Software Engineering Institute, Carnegie Mellon University},
 title = {{CERT/CC} Vulnerability Disclosure Policy},
 url = {https://certcc.github.io/certcc_disclosure_policy/},
 urldate = {2026-03-23},
 year = {2026}
}

@inproceedings{Chakravarty2020extended,
 address = {Cham},
 author = {Chakravarty, Manuel M. T.
and Chapman, James
and MacKenzie, Kenneth
and Melkonian, Orestis
and Peyton Jones, Michael
and Wadler, Philip},
 booktitle = {Financial Cryptography and Data Security},
 editor = {Bernhard, Matthew
and Bracciali, Andrea
and Camp, L. Jean
and Matsuo, Shin'ichiro
and Maurushat, Alana
and R{\o}nne, Peter B.
and Sala, Massimiliano},
 isbn = {978-3-030-54455-3},
 pages = {525--539},
 publisher = {Springer International Publishing},
 title = {The Extended UTXO Model},
 year = {2020}
}

@article{Chan2025practical,
 author = {Chan, Ming Lai and Capatos, Aliki Anna and Lodahl, Peter and S{\o}rensen, Anders S{\o}ndberg and Paesani, Stefano},
 journal = {arXiv:2507.16152},
 title = {Practical blueprint for low-depth photonic quantum computing with quantum dots},
 url = {https://arxiv.org/abs/2507.16152},
 year = {2025}
}

@article{Chase2018analysis,
 author = {Chase, Brad and MacBrough, Ethan},
 journal = {arXiv:1802.07242},
 title = {{Analysis of the XRP Ledger Consensus Protocol}},
 url = {https://arxiv.org/abs/1802.07242},
 year = {2018}
}

@article{Chen2016Algorand,
 author = {Chen, Jing and Micali, Silvio},
 title = {Algorand},
 url = {https://arxiv.org/abs/1607.01341},
 journal = {arXiv:1607.01341},
 year = {2016},
}

@book{Chen2016report,
 author = {Chen, Lily and Jordan, Stephen and Liu, Yi-Kai and Moody, Dustin and Peralta, Rene and Perlner, Ray and Smith-Tone, Daniel},
 doi = {10.6028/nist.ir.8105},
 institution = {National Institute of Standards and Technology},
 month = {April},
 title = {Report on Post-Quantum Cryptography},
 url = {http://dx.doi.org/10.6028/NIST.IR.8105},
 year = {2016}
}

@misc{Chevignard2026reducing,
 author = {Cl{\'e}mence Chevignard and Pierre-Alain Fouque and Andr{\'e} Schrottenloher},
 howpublished = {Cryptology {ePrint} Archive, Paper 2026/280},
 title = {Reducing the Number of Qubits in Quantum Discrete Logarithms on Elliptic Curves},
 url = {https://eprint.iacr.org/2026/280},
 year = {2026}
}

@techreport{Chone2025maximal,
 author = {Chon{'e}, Anne and Canals, Max Diaz I},
 day = {1},
 institution = {European Securities and Markets Authority},
 month = {jul},
 number = {ESMA50-481369926-29744},
 title = {{Maximal Extractable Value: Implications for crypto markets}},
 type = {Regulatory Risk Analysis},
 url = {https://www.esma.europa.eu/sites/default/files/2025-07/ESMA50-481369926-29744_Maximal_Extractable_Value_Implications_for_crypto_markets.pdf},
 year = {2025}
}

@article{chou2025race,
 author = {Chou, J. and Manyika, J. and Neven, H.},
 journal = {Foreign Affairs},
 title = {The Race to Lead the Quantum Future},
 year = {2025}
}

@misc{CMCGlossaryFUD,
 author = {{CoinMarketCap}},
 title = {{FUD}},
 howpublished = {CoinMarketCap Glossary},
 year = {2021},
 url = {https://coinmarketcap.com/academy/glossary/fud},
 note = {Accessed 2026-03-28}
}

@manual{Co2026halo2,
 author = {{Electric Coin Co.}},
 title = {{The halo2 Book}},
 url = {https://zcash.github.io/halo2/},
 year = {2026}
}

@misc{CoinPhoton2024bitcoin,
 author = {{CoinPhoton}},
 day = {24},
 howpublished = {Binance Square},
 month = {feb},
 note = {Documents network congestion between {December 2023} and {January 2024} with a peak of over 194,000 unconfirmed transactions.},
 title = {Bitcoin Network Clears Congestion After Bullish Rally},
 url = {https://www.binance.com/en/square/post/4577163603186},
 urldate = {2026-03-23},
 year = {2024}
}

@misc{Commission2016revised,
 author = {{Uniform Law Commission}},
 howpublished = {Uniform Law Commission},
 note = {Last updated in 2016; originally promulgated in 1954.},
 shorttitle = {{RUUPA}},
 title = {Revised {Uniform Unclaimed Property Act}},
 type = {Model Act},
 url = {https://www.uniformlaws.org/committees/community-home?CommunityKey=4b7c796a-f158-47bc-b5b1-f3f9a6e404fa},
 year = {2016}
}

@misc{Company2019turnstile,
 author = {{Electric Coin Company}},
 howpublished = {Official Zcash Blog},
 title = {Turnstile Enforcement Against Counterfeiting},
 url = {https://z.cash/turnstile-enforcement-against-counterfeiting/},
 urldate = {2026-03-23},
 year = {2019}
}

@misc{Contributors2026precompiled,
 author = {{evm.codes Contributors}},
 howpublished = {Interactive {EVM} Reference ({Osaka} Hardfork Spec)},
 note = {Catalog of precompiled contracts on Ethereum, including {ECDSA} recovery (0x01) and {BLS12-381} operations (0x0b-0x11). Source code available at: \url{https://github.com/duneanalytics/evm.codes}.},
 title = {Precompiled Contracts},
 url = {https://www.evm.codes/precompiled},
 urldate = {2026-03-22},
 year = {2026}
}

@book{Cooper2020recommendation,
 author = {Cooper, David A. and Apon, Daniel C. and Dang, Quynh H. and Davidson, Michael S. and Dworkin, Morris J. and Miller, Carl A.},
 doi = {10.6028/nist.sp.800-208},
 institution = {National Institute of Standards and Technology},
 month = {October},
 title = {Recommendation for Stateful Hash-Based Signature Schemes},
 url = {http://dx.doi.org/10.6028/NIST.SP.800-208},
 year = {2020}
}

@techreport{Costa2025postquantum,
 author = {Costa, Daniel Bruno Corvelo},
 day = {3},
 institution = {U.S. Securities and Exchange Commission (SEC)},
 month = {sep},
 note = {Prepared for the U.S. Crypto Assets Task Force -- {SEC}},
 title = {{Post-Quantum Financial Infrastructure Framework (PQFIF): A Roadmap for the Quantum-Safe Transition of Global Financial Infrastructure}},
 type = {Regulatory Submission},
 url = {https://www.sec.gov/files/cft-written-input-daniel-bruno-corvelo-costa-090325.pdf},
 year = {2025}
}

@article{CuellarGempeler2025cheesecloth,
 address = {New York, NY, USA},
 articleno = {46},
 author = {Cu'{e}llar Gempeler, Santiago and Harris, Bill and Parker, James and Pernsteiner, Stuart and Sweet, Ian and Tromer, Eran},
 day = {11},
 doi = {10.1145/3747589},
 issn = {2471-2566},
 issue_date = {November 2025},
 journal = {ACM Trans. Priv. Secur.},
 month = {sep},
 number = {4},
 numpages = {35},
 publisher = {Association for Computing Machinery},
 title = {Cheesecloth: Zero-Knowledge Proofs of Real-World Vulnerabilities},
 url = {https://doi.org/10.1145/3747589},
 volume = {28},
 year = {2025}
}

@inproceedings{Daian2020flash,
 author = {Daian, Philip and Goldfeder, Steven and Kell, Tyler and Li, Yunqi and Zhao, Xueyuan and Bentov, Iddo and Breidenbach, Lorenz and Juels, Ari},
 booktitle = {2020 IEEE Symposium on Security and Privacy (SP)},
 doi = {10.1109/SP40000.2020.00040},
 number = {},
 pages = {910-927},
 title = {Flash Boys 2.0: Frontrunning in Decentralized Exchanges, Miner Extractable Value, and Consensus Instability},
 volume = {},
 year = {2020}
}

@article{DallaireDemers2025brace,
 author = {Dallaire-Demers, Pierre-Luc and Doyle, William and Foo, Timothy},
 journal = {arXiv:2508.14011},
 title = {Brace for impact: ECDLP challenges for quantum cryptanalysis},
 url = {https://arxiv.org/abs/2508.14011},
 year = {2025}
}

@techreport{DAmato2024eip7732,
 author = {D'Amato, Francesco and Monnot, Barnab{\'e} and Neuder, Michael and Potuz and Traglia, Justin and Tsao, Terence},
 day = {28},
 institution = {Ethereum Foundation},
 month = {jun},
 note = {Accessed: 2026-03-23},
 number = {7732},
 title = {{EIP-7732: Enshrined Proposer-Builder Separation [DRAFT]}},
 type = {Ethereum Improvement Proposal},
 url = {https://eips.ethereum.org/EIPS/eip-7732},
 year = {2024}
}

@misc{DAO2026account,
 author = {{TRON DAO}},
 howpublished = {TRON Developer Network},
 title = {Account Permission Management},
 url = {https://developers.tron.network/docs/multi-signature},
 urldate = {2026-03-23},
 year = {2026}
}

@misc{Day2018bitcoin,
 author = {Day, Allen and Bookman, Colin},
 day = {8},
 howpublished = {Google Cloud Blog},
 month = {feb},
 note = {Developed in collaboration with {Kaggle}.},
 title = {{Bitcoin} in {BigQuery}: blockchain analytics on public data},
 url = {https://cloud.google.com/blog/topics/public-datasets/bitcoin-in-bigquery-blockchain-analytics-on-public-data},
 urldate = {2026-03-23},
 year = {2018}
}

@misc{Day2018ethereum,
 author = {Day, Allen and Medvedev, Evgeny},
 day = {29},
 howpublished = {Google Cloud Blog},
 month = {aug},
 note = {Developed in collaboration with {Kaggle}, {AMPOS}, and the {Ethereum ETL} project.},
 title = {{Ethereum} in {BigQuery}: a Public Dataset for smart contract analytics},
 url = {https://cloud.google.com/blog/products/data-analytics/ethereum-bigquery-public-dataset-smart-contract-analytics},
 urldate = {2026-03-23},
 year = {2018}
}

@misc{Deegan2025hello,
 author = {Deegan, Conor},
 day = {10},
 howpublished = {Project Eleven Blog},
 month = {jun},
 title = {hello yellowpages},
 url = {https://blog.projecteleven.com/posts/hello-yellowpages},
 urldate = {2026-03-23},
 year = {2025}
}

@misc{Deegan2025look,
 author = {Deegan, Conor},
 day = {9},
 howpublished = {Project Eleven Blog},
 month = {jun},
 title = {A look at post quantum proposals for Bitcoin},
 url = {https://blog.projecteleven.com/posts/a-look-at-post-quantum-proposals-for-bitcoin},
 urldate = {2026-03-23},
 year = {2025}
}

@misc{Deegan2025quantum,
 author = {Deegan, Conor},
 day = {22},
 howpublished = {Project Eleven Blog},
 month = {jul},
 title = {Quantum vulnerability of Bitcoin addresses},
 url = {https://blog.projecteleven.com/posts/quantum-vulnerability-of-bitcoin-addresses},
 urldate = {2026-03-23},
 year = {2025}
}

@article{DeFeo2017mathematics,
 author = {De Feo, Luca},
 journal = {arXiv:1711.04062},
 title = {Mathematics of Isogeny Based Cryptography},
 url = {https://arxiv.org/abs/1711.04062},
 year = {2017}
}

@techreport{Detrio2017eip779,
 author = {Detrio, Casey},
 day = {26},
 institution = {Ethereum Foundation},
 month = {nov},
 note = {Accessed: 2026-03-23},
 number = {779},
 title = {{EIP-779: Hardfork Meta: DAO Fork}},
 type = {Ethereum Improvement Proposal},
 url = {https://eips.ethereum.org/EIPS/eip-779},
 year = {2017}
}

@misc{Developers2026rippled,
 author = {{Ripple Developers}},
 howpublished = {GitHub Repository},
 title = {rippled: {XRP} Ledger Network Server},
 url = {https://github.com/XRPLF/rippled},
 urldate = {2026-03-23},
 year = {2026}
}

@misc{Developers2026segregated,
 author = {{Bitcoin Core Developers}},
 howpublished = {Official Bitcoin Core Documentation},
 title = {Segregated Witness Wallet Development Guide},
 url = {https://bitcoincore.org/en/segwit_wallet_dev/},
 urldate = {2026-03-23},
 year = {2026}
}

@article{Diffie1976new,
 author = {Diffie, W. and Hellman, M.},
 doi = {10.1109/tit.1976.1055638},
 issn = {1557-9654},
 journal = {IEEE Transactions on Information Theory},
 month = {November},
 number = {6},
 pages = {644--654},
 publisher = {Institute of Electrical and Electronics Engineers (IEEE)},
 title = {New directions in cryptography},
 url = {http://dx.doi.org/10.1109/TIT.1976.1055638},
 volume = {22},
 year = {1976}
}

@book{Ding2006multivariate,
 author = {Ding, Jintai and Gower, Jason and Schmidt, Dieter},
 doi = {10.1007/978-0-387-36946-4},
 isbn = {978-0-387-32229-2},
 journal = {Advances in Information Security},
 month = {01},
 pages = {},
 title = {Multivariate Public Key Cryptosystems},
 volume = {25},
 year = {2006}
}

@inproceedings{Dods2005hash,
 address = {Berlin, Heidelberg},
 author = {Dods, C. and Smart, N. P. and Stam, M.},
 booktitle = {Proceedings of the 10th International Conference on Cryptography and Coding},
 doi = {10.1007/11586821_8},
 isbn = {354030276X},
 location = {Cirencester, UK},
 numpages = {20},
 pages = {96--115},
 publisher = {Springer-Verlag},
 series = {IMA'05},
 title = {Hash based digital signature schemes},
 url = {https://doi.org/10.1007/11586821_8},
 year = {2005}
}

@misc{Drake2025hashbased,
 author = {Justin Drake and Dmitry Khovratovich and Mikhail Kudinov and Benedikt Wagner},
 doi = {10.62056/aey7qjp10},
 howpublished = {Cryptology {ePrint} Archive, Paper 2025/055},
 title = {Hash-Based Multi-Signatures for Post-Quantum Ethereum},
 url = {https://eprint.iacr.org/2025/055},
 year = {2025}
}

@misc{Drake2026PQCannouncement,
 author = {Drake, justin},
 day = {23},
 howpublished = {X (formerly Twitter)},
 month = {jan},
 title = {Today marks an inflection in the Ethereum Foundation's long-term quantum strategy.},
 url = {https://x.com/drakefjustin/status/2014791629408784816},
 urldate = {2026-03-22},
 year = {20256}
}

@misc{Ducas2017crystals,
 author = {Leo Ducas and Tancrede Lepoint and Vadim Lyubashevsky and Peter Schwabe and Gregor Seiler and Damien Stehle},
 howpublished = {Cryptology {ePrint} Archive, Paper 2017/633},
 title = {{CRYSTALS} -- Dilithium: Digital Signatures from Module Lattices},
 url = {https://eprint.iacr.org/2017/633},
 year = {2017}
}

@article{Ekera2019revisiting,
 author = {Eker\r{a}, Martin},
 journal = {arXiv:1905.09084},
 title = {Revisiting Shor's quantum algorithm for computing general discrete logarithms},
 url = {https://arxiv.org/abs/1905.09084},
 year = {2019}
}

@phdthesis{Ekera2024factoring,
 address = {Stockholm, Sweden},
 author = {Eker\r{a}, Martin},
 day = {25},
 month = {oct},
 school = {KTH Royal Institute of Technology},
 title = {{On factoring integers, and computing discrete logarithms and orders, quantumly}},
 url = {https://kth.diva-portal.org/smash/get/diva2:1902626/FULLTEXT01.pdf},
 year = {2024}
}

@article{Elgamal1985public,
 author = {Elgamal, T.},
 doi = {10.1109/TIT.1985.1057074},
 journal = {IEEE Transactions on Information Theory},
 number = {4},
 pages = {469-472},
 title = {A public key cryptosystem and a signature scheme based on discrete logarithms},
 volume = {31},
 year = {1985}
}

@techreport{Entriken2018erc721,
 author = {Entriken, William and Shirley, Dieter and Evans, Jacob and Sachs, Nastassia},
 day = {24},
 institution = {Ethereum Foundation},
 month = {jan},
 note = {Accessed: 2026-03-23},
 number = {721},
 title = {{ERC-721: Non-Fungible Token Standard}},
 type = {Ethereum Improvement Proposal},
 url = {https://eips.ethereum.org/EIPS/eip-721},
 year = {2018}
}

@inproceedings{Escofet2023interconnect,
 address = {New York, NY, USA},
 author = {Escofet, Pau and Rached, Sahar Ben and Rodrigo, Santiago and Almudever, Carmen G. and Alarc'{o}n, Eduard and Abadal, Sergi},
 booktitle = {Proceedings of the 16th International Workshop on Network on Chip Architectures},
 doi = {10.1145/3610396.3623267},
 isbn = {9798400703072},
 location = {Toronto, ON, Canada},
 numpages = {6},
 pages = {34--39},
 publisher = {Association for Computing Machinery},
 series = {NoCArc '23},
 title = {Interconnect Fabrics for Multi-Core Quantum Processors: A Context Analysis},
 url = {https://doi.org/10.1145/3610396.3623267},
 year = {2023}
}

@article{Fathalla2024beyond,
 author = {Fathalla, Efat and Azab, Mohamed},
 doi = {10.1109/ACCESS.2024.3485602},
 journal = {IEEE Access},
 number = {},
 pages = {175969-175987},
 title = {Beyond Classical Cryptography: A Systematic Review of Post-Quantum Hash-Based Signature Schemes, Security, and Optimizations},
 volume = {12},
 year = {2024}
}

@misc{Feo2020sqisign,
 author = {Luca De Feo and David Kohel and Antonin Leroux and Christophe Petit and Benjamin Wesolowski},
 howpublished = {Cryptology {ePrint} Archive, Paper 2020/1240},
 title = {{SQISign}: compact post-quantum signatures from quaternions and isogenies},
 url = {https://eprint.iacr.org/2020/1240},
 year = {2020}
}

@article{Feynman1982simulating,
 author = {Feynman, Richard P},
 journal = {International Journal of Theoretical Physics},
 number = {6/7},
 title = {Simulating physics with computers},
 url = {https://link.springer.com/article/10.1007/BF02650179},
 volume = {21},
 year = {1982}
}

@inproceedings{Fiat1987to,
 address = {Berlin, Heidelberg},
 author = {Fiat, Amos
and Shamir, Adi},
 booktitle = {Advances in Cryptology --- CRYPTO' 86},
 editor = {Odlyzko, Andrew M.},
 isbn = {978-3-540-47721-1},
 pages = {186--194},
 publisher = {Springer Berlin Heidelberg},
 title = {How To Prove Yourself: Practical Solutions to Identification and Signature Problems},
 year = {1987}
}

@misc{Finance2026cambridge,
 author = {{Cambridge Centre for Alternative Finance}},
 howpublished = {Interactive Data Dashboard},
 institution = {University of Cambridge Judge Business School},
 note = {Approximate geographic distribution of {Bitcoin} hashrate and estimated total network electricity consumption.},
 title = {{Cambridge Bitcoin Electricity Consumption Index (CBECI)}: Mining Map},
 url = {https://ccaf.io/cbnsi/cbeci/mining_map},
 urldate = {2026-03-23},
 year = {2026}
}

@misc{Financial2024circle,
 author = {{Circle Internet Financial}},
 day = {21},
 howpublished = {Official Blog Announcement},
 month = {feb},
 title = {Circle is Discontinuing Support for {USDC} on the {TRON} Blockchain},
 url = {https://www.circle.com/blog/circle-is-discontinuing-support-for-usdc-on-the-tron-blockchain},
 urldate = {2026-03-23},
 year = {2024}
}

@article{Fink2025larry,
 author = {Fink, Larry and Goldstein, Rob},
 day = {1},
 journal = {The Economist},
 month = {dec},
 note = {Reprint available at: \url{https://www.blackrock.com/corporate/literature/article-reprint/larry-fink-rob-goldstein-economist-op-ed-tokenization.pdf}},
 series = {By Invitation},
 title = {{Larry Fink and Rob Goldstein on how tokenisation could transform finance}},
 url = {https://www.economist.com/by-invitation/2025/12/01/larry-fink-and-rob-goldstein-on-how-tokenisation-could-transform-finance},
 year = {2025}
}

@misc{FinlowBates2023in,
 author = {Finlow-Bates, Keir},
 day = {29},
 howpublished = {Medium},
 month = {dec},
 title = {In Search of Lost Keys},
 url = {https://kf106.medium.com/in-search-of-lost-keys-69f624c0c28e},
 urldate = {2026-03-22},
 year = {2023}
}

@misc{ForkLog2025xrp,
 author = {{ForkLog}},
 day = {26},
 howpublished = {ForkLog Magazine},
 month = {dec},
 title = {{XRP Ledger} Implements Quantum Threat Protection},
 url = {https://forklog.com/en/xrp-ledger-implements-quantum-threat-protection/},
 urldate = {2026-03-23},
 year = {2025}
}

@misc{Foundation2022kzg,
 author = {{Ethereum Foundation}},
 howpublished = {Official Protocol Website},
 note = {A massively multi-party computation to produce the Structured Reference String for {KZG} polynomial commitments used in {EIP-4844}.},
 title = {The {KZG} Trusted Setup Ceremony},
 url = {https://ceremony.ethereum.org/},
 urldate = {2026-03-22},
 year = {2022}
}

@misc{Foundation2026asset,
 author = {{Algorand Foundation}},
 howpublished = {Official Ecosystem Portal},
 title = {Asset tokenization on {Algorand}: Powering a more open financial system},
 url = {https://algorand.co/solutions/tokenization},
 urldate = {2026-03-23},
 year = {2026}
}

@misc{Foundation2026avm,
 author = {{Algorand Foundation}},
 howpublished = {Algorand Developer Documentation},
 title = {{Algorand Virtual Machine}},
 url = {https://dev.algorand.co/concepts/smart-contracts/avm/},
 urldate = {2026-03-23},
 year = {2026}
}

@misc{Foundation2026falcon_verify,
 author = {{Algorand Foundation}},
 howpublished = {Algorand Developer Reference},
 title = {{Algorand} {TEAL} Opcodes: {falcon\_verify}},
 url = {https://dev.algorand.co/reference/algorand-teal/opcodes/#falcon_verify},
 urldate = {2026-03-23},
 year = {2026},
 note = {bytecode \texttt{0x85}, availabile in v12.}
}

@misc{Foundation2026official,
 author = {{Ethereum Foundation}},
 howpublished = {\url{https://ethereum.foundation/}},
 note = {Accessed: 2026-03-22},
 title = {{Official Website of the Ethereum Foundation}},
 year = {2026}
}

@misc{Foundation2026qrl,
 author = {{The QRL Foundation}},
 howpublished = {Official Project Roadmap},
 title = {{QRL 2.0 (Codenamed Project Zond)}},
 url = {https://www.theqrl.org/roadmap/},
 urldate = {2026-03-22},
 year = {2026}
}

@misc{Foundation2026quantum,
 author = {{The QRL Foundation}},
 title = {{Quantum Resistant Ledger}: A visionary, future-proof blockchain with unparalleled security},
 url = {https://www.theqrl.org/a-visionary-future-proof-blockchain-with-unparalleled-security/the-qrl-explained.pdf},
 urldate = {2026-03-22},
 year = {2026}
}

@misc{Foundation2026rekeying,
 author = {{Algorand Foundation}},
 howpublished = {Algorand Developer Documentation},
 title = {Rekeying Accounts},
 url = {https://dev.algorand.co/concepts/accounts/rekeying/},
 urldate = {2026-03-23},
 year = {2026}
}

@misc{Foundation2026rwa,
 author = {{Stellar Development Foundation}},
 howpublished = {Official Ecosystem Portal},
 title = {{RWA} tokenization},
 url = {https://stellar.org/use-cases/tokenization},
 urldate = {2026-03-23},
 year = {2026}
}

@misc{Foundation2026teal,
 author = {{Algorand Foundation}},
 howpublished = {Algorand Developer Documentation},
 title = {{Algorand} {TEAL}: Transaction Execution Approval Language},
 url = {https://dev.algorand.co/concepts/smart-contracts/languages/teal/},
 urldate = {2026-03-23},
 year = {2026}
}

@techreport{Fouque2020falcon,
 author = {Fouque, Pierre-Alain and Hoffstein, Jeffrey and Kirchner, Paul and Lyubashevsky, Vadim and Pornin, Thomas and Prest, Thomas and Ricosset, Thomas and Seiler, Gregor and Whyte, William and Zhang, Zhenfei},
 day = {1},
 institution = {NIST Post-Quantum Cryptography Project},
 month = {oct},
 number = {Specification v1.2},
 title = {{Falcon: Fast-Fourier Lattice-based Compact Signatures over {NTRU}}},
 type = {Technical Specification (NIST PQC Submission)},
 url = {https://falcon-sign.info/falcon.pdf},
 year = {2020}
}

@article{fowler2012surface,
 author = {Fowler, Austin G and Mariantoni, Matteo and Martinis, John M and Cleland, Andrew N},
 journal = {Physical Review A-Atomic, Molecular, and Optical Physics},
 number = {3},
 pages = {032324},
 publisher = {APS},
 title = {Surface codes: Towards practical large-scale quantum computation},
 url = {Surface codes: Towards practical large-scale quantum computation},
 volume = {86},
 year = {2012}
}

@article{Fowler2012timeoptimal,
 author = {Fowler, Austin G.},
 journal = {arXiv:1210.4626},
 title = {Time-optimal quantum computation},
 url = {https://arxiv.org/abs/1210.4626},
 year = {2012}
}

@misc{Friedenbach2021why,
 author = {Friedenbach, Mark},
 day = {15},
 howpublished = {Freicoin Substack},
 month = {mar},
 title = {Why I'm against {Taproot} and recommend against activation on {Bitcoin}},
 url = {https://freicoin.substack.com/p/why-im-against-taproot},
 urldate = {2026-03-23},
 year = {2021}
}

@misc{Fukuda2025grand,
 author = {Kigen Fukuda and Shin'ichiro Matsuo and Yuji Suga and Tadahiko Ito},
 howpublished = {Cryptology {ePrint} Archive, Paper 2025/1626},
 title = {The Grand Challenge of {PQC} Migration: Analysis of Modern Blockchain and Intertwined Human Egoisms},
 url = {https://eprint.iacr.org/2025/1626},
 year = {2025}
}

@article{gheorghiu2019benchmarking,
 author = {Gheorghiu, Vlad and Mosca, Michele},
 journal = {arXiv:1902.02332},
 title = {Benchmarking the quantum cryptanalysis of symmetric, public-key and hash-based cryptographic schemes},
 url = {https://arxiv.org/abs/1902.02332},
 year = {2019}
}

@article{Gidney2018halving,
 author = {Gidney, Craig},
 doi = {10.22331/q-2018-06-18-74},
 issn = {2521-327X},
 journal = {Quantum},
 month = {June},
 pages = {74},
 publisher = {Verein zur Forderung des Open Access Publizierens in den Quantenwissenschaften},
 title = {Halving the cost of quantum addition},
 url = {http://dx.doi.org/10.22331/q-2018-06-18-74},
 volume = {2},
 year = {2018}
}

@article{Gidney2019flexible,
 author = {Gidney, Craig and Fowler, Austin G.},
 journal = {arXiv:1905.08916},
 title = {Flexible layout of surface code computations using AutoCCZ states},
 url = {https://arxiv.org/abs/1905.08916},
 year = {2019}
}

@misc{Gidney2019verifying,
 author = {Gidney, Craig},
 day = {7},
 howpublished = {\url{https://algassert.com/post/1903}},
 month = {aug},
 note = {Accessed: 2026-03-22},
 title = {{Verifying Measurement Based Uncomputation --- Algassert}},
 year = {2019}
}

@article{Gidney2021to,
 author = {Gidney, Craig and Eker\r{a}, Martin},
 doi = {10.22331/q-2021-04-15-433},
 issn = {2521-327X},
 journal = {{Quantum}},
 month = {April},
 pages = {433},
 publisher = {{Verein zur F{"{o}}rderung des Open Access Publizierens in den Quantenwissenschaften}},
 title = {How to factor 2048 bit {RSA} integers in 8 hours using 20 million noisy qubits},
 url = {https://doi.org/10.22331/q-2021-04-15-433},
 volume = {5},
 year = {2021}
}

@article{Gidney2023yoked,
 author = {Gidney, Craig and Newman, Michael and Brooks, Peter and Jones, Cody},
 journal = {arXiv:2312.04522},
 title = {Yoked surface codes},
 url = {https://arxiv.org/abs/2312.04522},
 year = {2023}
}

@article{Gidney2024magic,
 author = {Gidney, Craig and Shutty, Noah and Jones, Cody},
 journal = {arXiv:2409.17595},
 title = {Magic state cultivation: growing T states as cheap as CNOT gates},
 url = {https://arxiv.org/abs/2409.17595},
 year = {2024}
}

@article{Gidney2025Factoring,
 author = {Gidney, Craig},
 journal = {arXiv:2505.15917},
 title = {How to factor 2048 bit {RSA} integers with less than a million noisy qubits},
 url = {https://arxiv.org/abs/2505.15917},
 year = {2025}
}

@article{Gidney2025to,
 author = {Gidney, Craig},
 journal = {arXiv:2505.15917},
 title = {How to factor 2048 bit {RSA} integers with less than a million noisy qubits},
 url = {https://arxiv.org/abs/2505.15917},
 year = {2025}
}

@misc{Gidney2025yoked,
 author = {Gidney, Craig and Newman, Michael},
 day = {11},
 howpublished = {Talk at the 7th International Conference on Quantum Error Correction ({QEC25}), New Haven, Connecticut, USA},
 month = {aug},
 title = {Yoked surface codes},
 url = {https://yale.hosted.panopto.com/Panopto/Pages/Viewer.aspx?id=423a1d0e-42cd-4fa4-9ae8-b33000f2acf0&start=3740},
 urldate = {2026-03-23},
 year = {2025}
}

@article{Goes2006Interblockchain,
 author = {Goes, Christopher},
 journal = {arXiv:2006.15918},
 title = {The Interblockchain Communication Protocol: An Overview},
 url = {https://arxiv.org/abs/2006.15918},
 year = {2006}
}

@article{Gokhale2010evolutionary,
 author = {Chaitanya S. Gokhale  and Arne Traulsen },
 doi = {10.1073/pnas.0912214107},
 eprint = {https://www.pnas.org/doi/pdf/10.1073/pnas.0912214107},
 journal = {Proceedings of the National Academy of Sciences},
 number = {12},
 pages = {5500-5504},
 title = {Evolutionary games in the multiverse},
 url = {https://www.pnas.org/doi/abs/10.1073/pnas.0912214107},
 volume = {107},
 year = {2010}
}

@misc{Goldberg2021cairo,
 author = {Lior Goldberg and Shahar Papini and Michael Riabzev},
 howpublished = {Cryptology {ePrint} Archive, Paper 2021/1063},
 title = {Cairo--a Turing-complete {STARK}-friendly {CPU} architecture},
 url = {https://eprint.iacr.org/2021/1063},
 year = {2021}
}

@inproceedings{Goldwasser1985knowledge,
 address = {New York, NY, USA},
 author = {Goldwasser, S and Micali, S and Rackoff, C},
 booktitle = {Proceedings of the Seventeenth Annual ACM Symposium on Theory of Computing},
 doi = {10.1145/22145.22178},
 isbn = {0897911512},
 location = {Providence, Rhode Island, USA},
 numpages = {14},
 pages = {291--304},
 publisher = {Association for Computing Machinery},
 series = {STOC '85},
 title = {The knowledge complexity of interactive proof-systems},
 url = {https://doi.org/10.1145/22145.22178},
 year = {1985}
}

@article{Gouzien2023performance,
 author = {Gouzien, {\'E}lie and Ruiz, Diego and Le R{\'e}gent, Francois-Marie and Guillaud, J{\'e}r{\'e}mie and Sangouard, Nicolas},
 doi = {10.1103/physrevlett.131.040602},
 issn = {1079-7114},
 journal = {Physical Review Letters},
 month = {July},
 number = {4},
 publisher = {American Physical Society (APS)},
 title = {Performance Analysis of a Repetition Cat Code Architecture: Computing 256-bit Elliptic Curve Logarithm in 9 Hours with 126,133 Cat Qubits},
 url = {http://dx.doi.org/10.1103/PhysRevLett.131.040602},
 volume = {131},
 year = {2023}
}

@techreport{Gretz2025tokenization,
 author = {Gretz, Philipp and Gross, Dimitri and Haynes, Eric Marti and Tarek, Yehia and Toma, Andrei and Chitgupi, Aneesha and Kurowski, Tomasz},
 institution = {Nethermind and {PwC Germany}},
 month = {feb},
 note = {Distributed in collaboration with Global Finance \& Technology Network (GFTN).},
 title = {{Tokenization Standards: Taming the Regulatory Menagerie}},
 type = {Industry Research Report},
 url = {https://gftn.co/hubfs/Tokenization_Standards_Nethermind_PwC_GFTN.pdf},
 year = {2025}
}

@article{Griffiths1996semiclassical,
 author = {Griffiths, Robert B. and Niu, Chi-Sheng},
 doi = {10.1103/physrevlett.76.3228},
 issn = {1079-7114},
 journal = {Physical Review Letters},
 month = {April},
 number = {17},
 pages = {3228--3231},
 publisher = {American Physical Society (APS)},
 title = {Semiclassical {Fourier Transform} for Quantum Computation},
 url = {http://dx.doi.org/10.1103/PhysRevLett.76.3228},
 volume = {76},
 year = {1996}
}

@misc{Groth2016size,
 author = {Jens Groth},
 howpublished = {Cryptology {ePrint} Archive, Paper 2016/260},
 title = {On the Size of Pairing-based Non-interactive Arguments},
 url = {https://eprint.iacr.org/2016/260},
 year = {2016}
}

@misc{Gudgeon2019sok,
 author = {Lewis Gudgeon and Pedro Moreno-Sanchez and Stefanie Roos and Patrick McCorry and Arthur Gervais},
 howpublished = {Cryptology {ePrint} Archive, Paper 2019/360},
 title = {{SoK}: Layer-Two Blockchain Protocols},
 url = {https://eprint.iacr.org/2019/360},
 year = {2019}
}

@misc{Habovstiak2025hashed,
 author = {Habov
{s}tiak, Martin},
 dat = {16},
 howpublished = {Bitcoin Development Mailing List ({Google Groups})},
 month = {mar},
 note = {Mailing list thread featuring contributions from {Lloyd Fournier}, {Agustin Cruz}, {Antoine Poinsot}, and others. Proposes a commit-reveal scheme to protect hashed pubkeys from mempool-snatching by quantum adversaries.},
 title = {Hashed keys are actually fully quantum secure},
 url = {https://groups.google.com/g/bitcoindev/c/jr1QO95k6Uc},
 urldate = {2026-03-23},
 year = {2025}
}

@misc{HallAndersen2023foundations,
 author = {Mathias Hall-Andersen and Mark Simkin and Benedikt Wagner},
 doi = {10.62056/a09qudhdj},
 howpublished = {Cryptology {ePrint} Archive, Paper 2023/1079},
 title = {Foundations of Data Availability Sampling},
 url = {https://eprint.iacr.org/2023/1079},
 year = {2023}
}

@misc{Haner2020improved,
 author = {Thomas H{\"a}ner and Samuel Jaques and Michael Naehrig and Martin Roetteler and Mathias Soeken},
 howpublished = {Cryptology {ePrint} Archive, Paper 2020/077},
 title = {Improved Quantum Circuits for Elliptic Curve Discrete Logarithms},
 url = {https://eprint.iacr.org/2020/077},
 year = {2020}
}

@book{Harris1998complete,
 title={The Complete Sales Letter Book: Model Letters for Every Selling Situation},
 author={Harris, R. and McIntyre, A.},
 isbn={9780765600837},
 lccn={97033176},
 series={Sharpe Professional},
 url={https://books.google.com/books?id=bSbenQEACAAJ},
 year={1998},
 publisher={Sharpe Professional}
}

@article{Hastings2020classical,
 author = {Hastings, Matthew B.},
 doi = {10.22331/q-2020-02-27-237},
 issn = {2521-327X},
 journal = {Quantum},
 month = {February},
 pages = {237},
 publisher = {Verein zur Forderung des Open Access Publizierens in den Quantenwissenschaften},
 title = {Classical and Quantum Algorithms for Tensor Principal Component Analysis},
 url = {http://dx.doi.org/10.22331/q-2020-02-27-237},
 volume = {4},
 year = {2020}
}

@article{He2025discovering,
 author = {He, Austin Yubo and Liu, Zi-Wen},
 journal = {arXiv:2502.14372},
 title = {Discovering highly efficient low-weight quantum error-correcting codes with reinforcement learning},
 url = {https://arxiv.org/abs/2502.14372},
 year = {2025}
}

@misc{Heilman2025changes,
 author = {Heilman, Ethan},
 day = {1},
 howpublished = {Delving Bitcoin},
 month = {jul},
 note = {Post \#29 identifies the two cases arising from consideration of CRQC clock speed.},
 title = {Changes to {BIP-360} --- {Pay to Quantum Resistant Hash (P2QRH)}},
 url = {https://delvingbitcoin.org/t/changes-to-bip-360-pay-to-quantum-resistant-hash-p2qrh/1811/29},
 urldate = {2026-03-23},
 year = {2025}
}

@misc{hildobby2026ethereum,
 author = {hildobby},
 howpublished = {Dune Analytics},
 note = {Industry-standard real-time tracker for {Ethereum} staking metrics.},
 title = {Ethereum {ETH} Staking Dashboard},
 url = {https://dune.com/hildobby/eth2-staking},
 urldate = {2026-03-23},
 year = {2026}
}

@misc{Holmes2021assessment,
 author = {Stephen Holmes and Liqun Chen},
 howpublished = {Cryptology {ePrint} Archive, Paper 2021/967},
 title = {Assessment of Quantum Threat To Bitcoin and Derived Cryptocurrencies},
 url = {https://eprint.iacr.org/2021/967},
 year = {2021}
}

@techreport{Hopwood2019zip,
 author = {Hopwood, Daira-Emma},
 day = {29},
 institution = {Electric Coin Co.},
 month = {mar},
 note = {Accessed: 2026-03-21},
 number = {211},
 title = {{ZIP 211: Disabling Addition of New Value to the Sprout Chain Value Pool}},
 type = {Zcash Improvement Proposal},
 url = {https://zips.z.cash/zip-0211},
 year = {2019}
}

@misc{Hopwood2025zip,
 author = {Hopwood, Daira-Emma and Grigg, Jack},
 day = {31},
 howpublished = {Zcash Improvement Proposal (Draft)},
 month = {mar},
 note = {Credits: Sean Bowe. Discussion available at {GitHub} Issue \#1135: \url{https://github.com/zcash/zips/issues/1135}},
 title = {{ZIP 2005}: Quantum Recoverability},
 url = {https://zips.z.cash/draft-ecc-quantum-recoverability},
 urldate = {2026-03-23},
 year = {2025}
}

@techreport{Hopwood2026zcash,
 author = {Hopwood, Daira-Emma and Bowe, Sean and Hornby, Taylor and Wilcox, Nathan},
 day = {19},
 institution = {Electric Coin Co.},
 month = {mar},
 note = {Accessed: 2026-03-21},
 title = {{Zcash Protocol Specification, Version 2025.6.3-84-gf58f5c [NU6]}},
 type = {Protocol Specification},
 url = {https://zips.z.cash/protocol/nu6.pdf},
 year = {2026}
}

@techreport{Hoskinson2017Why,
 author = {Hoskinson, Charles},
 title = {Why We Are Building {C}ardano: {A} Subjective Approach},
 institution = {IOHK},
 year = {2017},
 month = {jun},
 day = {28},
 type = {Foundational Whitepaper},
 url = {https://whitepaper.io/document/581/cardano-whitepaper},
 note = {Accessed March 26, 2026.}
}

@article{Huang2025obfuscation,
 author = {Huang, Mi-Ying and Tang, Er-Cheng},
 journal = {arXiv:2507.11970},
 title = {Obfuscation of Unitary Quantum Programs},
 url = {https://arxiv.org/abs/2507.11970},
 year = {2025}
}

@inbook{Humayun2018``satoshi,
 author = {Humayun, Mariam and Belk, Russell},
 doi = {10.1108/S0885-211120180000019002},
 isbn = {978-1-78743-907-8},
 journal = {Research in Consumer Behavior},
 month = {02},
 pages = {19-35},
 title = {``{Satoshi is Dead. Long Live Satoshi}'': {The Curious Case of Bitcoin's Creator}},
 volume = {19},
 year = {2018}
}

@article{Ilie2020unstable,
 author = {Ilie, Dragos I. and Werner, Sam M. and Stewart, Iain and Knottenbelt, William J.},
 journal = {arXiv:2006.03044},
 title = {Unstable Throughput: When the Difficulty Algorithm Breaks},
 url = {https://arxiv.org/abs/2006.03044},
 year = {2020}
}

@techreport{ISOIEC2018isoiec,
 address = {Geneva, Switzerland},
 author = {{ISO/IEC}},
 institution = {International Organization for Standardization and International Electrotechnical Commission},
 title = {{ISO/IEC 29147:2018 Information technology --- Security techniques --- Vulnerability disclosure}},
 type = {Standard},
 url = {https://www.iso.org/standard/72311.html},
 year = {2018}
}

@misc{Jancar2026bls12381,
 author = {Jancar, Jan and Sedlacek, Vladimir},
 howpublished = {\url{https://std.neuromancer.sk/bls/BLS12-381}},
 note = {Technical specifications for the BLS12-381 pairing-friendly elliptic curve. Accessed: 2026-03-21},
 title = {{Parameters for {BLS12-381} --- Standard Curves Database}},
 year = {2026}
}

@misc{Jancar2026secp256k1,
 author = {Jancar, Jan and Sedlacek, Vladimir},
 howpublished = {\url{https://std.neuromancer.sk/secg/secp256k1}},
 note = {Technical specifications for the Koblitz curve used by Bitcoin and Ethereum. Accessed: 2026-03-21},
 title = {{Parameters for {secp256k1} --- Standard Curves Database}},
 year = {2026}
}

@inbook{Jao2011towards,
 author = {Jao, David and De Feo, Luca},
 booktitle = {Post-Quantum Cryptography},
 doi = {10.1007/978-3-642-25405-5_2},
 isbn = {9783642254055},
 issn = {1611-3349},
 pages = {19--34},
 publisher = {Springer Berlin Heidelberg},
 title = {Towards Quantum-Resistant Cryptosystems from Supersingular Elliptic Curve Isogenies},
 url = {http://dx.doi.org/10.1007/978-3-642-25405-5_2},
 year = {2011}
}

@misc{Jedusor2016mimblewimble,
 author = {{Tom Elvis Jedusor}},
 day = {19},
 howpublished = {Original text file distributed via IRC},
 month = {jul},
 title = {{MIMBLEWIMBLE}},
 url = {https://scalingbitcoin.org/papers/mimblewimble.txt},
 urldate = {2026-03-23},
 year = {2016}
}

@misc{Jefferson2025dip,
 author = {Jefferson, Jordan},
 day = {22},
 howpublished = {Dogecoin Improvement Proposal (Draft), {GitHub} Issue \#3869},
 month = {jul},
 note = {Proposes a new opcode for onchain verification of {SNARKs} to enable {zk-rollups} on {Dogecoin}.},
 title = {{DIP}: {\texttt{OP\_CHECKZKP}} Zero-Knowledge Upgrade},
 url = {https://github.com/dogecoin/dogecoin/discussions/3869},
 urldate = {2026-03-22},
 year = {2025}
}

@article{Johnson2001elliptic,
 author = {Johnson, Don and Menezes, Alfred and Vanstone, Scott},
 doi = {10.1007/s102070100002},
 issn = {1615-5262},
 journal = {International Journal of Information Security},
 month = {August},
 number = {1},
 pages = {36--63},
 publisher = {Springer Science and Business Media LLC},
 title = {The Elliptic Curve Digital Signature Algorithm (ECDSA)},
 url = {http://dx.doi.org/10.1007/s102070100002},
 volume = {1},
 year = {2001}
}

@article{jones2012layered,
 author = {Jones, N Cody and Van Meter, Rodney and Fowler, Austin G and McMahon, Peter L and Kim, Jungsang and Ladd, Thaddeus D and Yamamoto, Yoshihisa},
 journal = {Physical Review X},
 number = {3},
 pages = {031007},
 publisher = {APS},
 title = {Layered architecture for quantum computing},
 url = {https://link.aps.org/doi/10.1103/PhysRevX.2.031007},
 volume = {2},
 year = {2012}
}

@article{Jones2013lowoverhead,
 author = {Jones, Cody},
 doi = {10.1103/physreva.87.022328},
 issn = {1094-1622},
 journal = {Physical Review A},
 month = {February},
 number = {2},
 publisher = {American Physical Society (APS)},
 title = {Low-overhead constructions for the fault-tolerant Toffoli gate},
 url = {http://dx.doi.org/10.1103/PhysRevA.87.022328},
 volume = {87},
 year = {2013}
}

@article{Jordan2025optimization,
 author = {Jordan, Stephen P. and Shutty, Noah and Wootters, Mary and Zalcman, Adam and Schmidhuber, Alexander and King, Robbie and Isakov, Sergei V. and Khattar, Tanuj and Babbush, Ryan},
 doi = {10.1038/s41586-025-09527-5},
 issn = {1476-4687},
 journal = {Nature},
 month = {October},
 number = {8086},
 pages = {831--836},
 publisher = {Springer Science and Business Media LLC},
 title = {Optimization by decoded quantum interferometry},
 url = {http://dx.doi.org/10.1038/s41586-025-09527-5},
 volume = {646},
 year = {2025}
}

@article{Joseph2022,
 author = {Joseph, David
and Misoczki, Rafael
and Manzano, Marc
and Tricot, Joe
and Pinuaga, Fernando Dominguez
and Lacombe, Olivier
and Leichenauer, Stefan
and Hidary, Jack
and Venables, Phil
and Hansen, Royal},
 day = {01},
 doi = {10.1038/s41586-022-04623-2},
 issn = {1476-4687},
 journal = {Nature},
 month = {May},
 number = {7909},
 pages = {237-243},
 title = {Transitioning organizations to post-quantum cryptography},
 url = {https://doi.org/10.1038/s41586-022-04623-2},
 volume = {605},
 year = {2022}
}

@techreport{Kalinin2021eip3675,
 author = {Kalinin, Mikhail and Ryan, Danny and Buterin, Vitalik},
 day = {22},
 institution = {Ethereum Foundation},
 month = {jul},
 note = {Accessed: 2026-03-23},
 number = {3675},
 title = {{EIP-3675: Upgrade consensus to Proof-of-Stake}},
 type = {Ethereum Improvement Proposal},
 url = {https://eips.ethereum.org/EIPS/eip-3675},
 year = {2021}
}

@inproceedings{Kate2010constantsize,
 address = {Berlin, Heidelberg},
 author = {Kate, Aniket
and Zaverucha, Gregory M.
and Goldberg, Ian},
 booktitle = {Advances in Cryptology - ASIACRYPT 2010},
 editor = {Abe, Masayuki},
 isbn = {978-3-642-17373-8},
 pages = {177--194},
 publisher = {Springer Berlin Heidelberg},
 title = {Constant-Size Commitments to Polynomials and Their Applications},
 year = {2010}
}

@techreport{Kempton2025eip7932,
 author = {Kempton, James},
 day = {12},
 institution = {Ethereum Foundation},
 month = {apr},
 note = {Accessed: 2026-03-23},
 number = {7932},
 title = {{EIP-7932: Secondary Signature Algorithms [DRAFT]}},
 type = {Ethereum Improvement Proposal},
 url = {https://eips.ethereum.org/EIPS/eip-7932},
 year = {2025}
}

@misc{Kenton2025escheat,
 author = {Kenton, Will},
 dat = {4},
 howpublished = {Investopedia},
 month = {sep},
 note = {Reviewed by Toby Walters, defines "escheat" as "when a government obtains ownership of unclaimed property or estate assets due to there being no identifiable heirs or beneficiaries"},
 title = {Escheat: Meaning, Process, and Reclaiming Assets},
 url = {https://www.investopedia.com/terms/e/escheat.asp},
 urldate = {2026-03-22},
 year = {2025}
}

@book{KEOHANE1984hegemony,
 author = {Robert O. Keohane},
 edition = {REV - Revised},
 isbn = {9780691076768},
 publisher = {Princeton University Press},
 title = {After Hegemony: Cooperation and Discord in the World Political Economy},
 url = {http://www.jstor.org/stable/j.ctt7sq9s},
 urldate = {2026-03-22},
 year = {1984}
}

@article{Kerckhoffs1883la,
 author = {Kerckhoffs, Auguste},
 journal = {Journal des Sciences Militaires},
 language = {French},
 month = {jan # " and " # feb},
 pages = {5--38, 161--191},
 title = {{La Cryptographie Militaire}},
 volume = {IX},
 year = {1883}
}

@misc{Kim2026new,
 author = {Hyunji Kim and Kyungbae Jang and Siyi Wang and Anubhab Baksi and Gyeongju Song and Hwajeong Seo and Anupam Chattopadhyay},
 howpublished = {Cryptology {ePrint} Archive, Paper 2026/106},
 title = {New Quantum Circuits for {ECDLP}: Breaking Prime Elliptic Curve Cryptography in Minutes},
 url = {https://eprint.iacr.org/2026/106},
 year = {2026}
}

@article{Koblitz1987Ecc,
 author = {Neal Koblitz},
 issn = {00255718, 10886842},
 journal = {Mathematics of Computation},
 number = {177},
 pages = {203--209},
 publisher = {American Mathematical Society},
 title = {Elliptic Curve Cryptosystems},
 url = {http://www.jstor.org/stable/2007884},
 urldate = {2026-03-21},
 volume = {48},
 year = {1987}
}

@book{koe2020zero,
 author = {koe and Alonso, Kurt M. and Noether, Sarang},
 day = {4},
 edition = {2nd},
 month = {apr},
 number = {v2.0.0},
 publisher = {Monero Research Lab},
 title = {{Zero to Monero: Second Edition --- a technical guide to a private digital currency; for beginners, amateurs, and experts}},
 url = {https://www.getmonero.org/library/Zero-to-Monero-2-0-0.pdf},
 year = {2020}
}

@misc{Koshelev2022generation,
 author = {Dmitrii Koshelev},
 title = {Generation of "independent" points on elliptic curves by means of Mordell--Weil lattices},
 howpublished = {Cryptology {ePrint} Archive, Paper 2022/794},
 year = {2022},
 url = {https://eprint.iacr.org/2022/794}
}

@misc{l2beat2026,
  author = {{L2BEAT sp. z o.o.}},
  title = {L2BEAT: The state of the layer two ecosystem},
  year = {2026},
  url = {https://l2beat.com},
  urldate = {2026-03-24}
}

@misc{Labs2026groth16,
 author = {{Succinct Labs}},
 howpublished = {Official Documentation},
 note = {Clarifies the zero-knowledge properties of {SP1} proofs, noting that {Groth16/PLONK} proofs are {ZK} and {STARK} proofs currently are not.},
 title = {{Groth16, PLONK, and the Zero-Knowledgeness of SP1}},
 url = {https://docs.succinct.xyz/docs/sp1/security/security-model#groth16-plonk-and-the-zero-knowledgeness-of-sp1},
 urldate = {2026-03-23},
 year = {2026}
}

@misc{Labs2026sp1,
 author = {{Succinct Labs}},
 howpublished = {GitHub Repository},
 title = {{SP1}: The Zero-Knowledge Virtual Machine ({zkVM})},
 url = {https://github.com/succinctlabs/sp1},
 urldate = {2026-03-23},
 year = {2026}
}

@misc{Lai2020pull,
 author = {Lai, Ying Tong and {Zcash Developers}},
 day = {1},
 howpublished = {Zcash GitHub Repository},
 month = {may},
 note = {Implements {ZIP-211}.},
 title = {Pull Request \#4489: Disable addition of new value to the {Sprout} value pool},
 url = {https://github.com/zcash/zcash/pull/4489},
 urldate = {2026-03-23},
 year = {2020}
}

@techreport{Lamport1979constructing,
 author = {Lamport, Leslie},
 edition = {SRI International},
 month = {October},
 note = {This paper was published by IEEE in the Proceedings of HICSS-43 in January, 2010.},
 number = {CSL-98},
 title = {Constructing Digital Signatures from a One Way Function},
 url = {https://www.microsoft.com/en-us/research/publication/constructing-digital-signatures-one-way-function/},
 year = {1979}
}

@misc{LearnMeABitcoin,
 author = {Walker, Greg},
 howpublished = {\url{https://learnmeabitcoin.com}},
 note = {Accessed: 2026-03-22},
 title = {{Learn Me A Bitcoin}},
 year = {2026}
}

@techreport{Lebrun2021erc3643,
 author = {Lebrun, Joachim and Malghem, Tony and Thizy, Kevin and Falempin, Luc and Boudjemaa, Adam},
 day = {9},
 institution = {Ethereum Foundation},
 month = {jul},
 note = {Accessed: 2026-03-23},
 number = {3643},
 title = {{ERC-3643: T-REX - Token for Regulated EXchanges}},
 type = {Ethereum Improvement Proposal},
 url = {https://eips.ethereum.org/EIPS/eip-3643},
 year = {2021}
}

@article{Lee2020hypercontraction,
 author = {Lee, Joonho and Berry, Dominic W and Gidney, Craig and Huggins, William J and McClean, Jarrod R and Wiebe, Nathan and Babbush, Ryan},
 journal = {PRX quantum},
 number = {3},
 pages = {030305},
 publisher = {APS},
 title = {Even more efficient quantum computations of chemistry through tensor hypercontraction},
 url = {https://journals.aps.org/prxquantum/abstract/10.1103/PRXQuantum.2.030305},
 volume = {2},
 year = {2021}
}

@misc{Lerner2013well,
 author = {Lerner, Sergio Demian},
 day = {17},
 howpublished = {Bitslog},
 month = {apr},
 note = {Seminal research identifying the "Patoshi" mining pattern and estimating the size of {Satoshi Nakamoto}'s early {Bitcoin} holdings.},
 title = {The Well Deserved Fortune of {Satoshi Nakamoto}, {Bitcoin} creator, Visionary and Genius},
 url = {https://bitslog.com/2013/04/17/the-well-deserved-fortune-of-satoshi-nakamoto/},
 urldate = {2026-03-23},
 year = {2013}
}

@techreport{Lerner2019rsk,
 author = {Lerner, Sergio Demian},
 day = {29},
 institution = {RSK Labs},
 month = {jan},
 number = {Revision 11},
 title = {{RSK}: {Bitcoin} Powered Smart Contracts},
 type = {Protocol White Paper},
 url = {https://rootstock.io/rsk-white-paper-updated.pdf},
 year = {2019}
}

@misc{Lerner2022rsk,
 author = {Sergio Demian Lerner and Javier \'Alvarez Cid-Fuentes and Julian Len and Rams\`es Fern\`andez-Val\`encia and Patricio Gallardo and Nicol{\'a}s Vescovo and Ra\`ul Laprida and Shreemoy Mishra and Federico Jinich and Diego Masini},
 howpublished = {Cryptology {ePrint} Archive, Paper 2022/684},
 title = {{RSK}: A Bitcoin sidechain with stateful smart-contracts},
 url = {https://eprint.iacr.org/2022/684},
 year = {2022}
}

@book{Lessig1999code,
 address = {New York, NY},
 author = {Lessig, Lawrence},
 description = {dret'd bibliography},
 isbn = {0465039138},
 publisher = {Basic Books},
 title = {Code and Other Laws of Cyberspace},
 year = {1999}
}

@article{LewisPye2023permissionless,
 author = {Lewis-Pye, Andrew and Roughgarden, Tim},
 journal = {arXiv:2304.14701},
 title = {Permissionless Consensus},
 url = {https://arxiv.org/abs/2304.14701},
 year = {2023}
}

@article{Li2019Electronic,
 author = {Li, Zhendong and Li, Junhao and Dattani, Nikesh S and Umrigar, CJ and Chan, Garnet Kin},
 journal = {The Journal of chemical physics},
 number = {2},
 publisher = {AIP Publishing},
 title = {The electronic complexity of the ground-state of the FeMo cofactor of nitrogenase as relevant to quantum simulations},
 url = {https://pubs.aip.org/aip/jcp/article/150/2/024302/197301/The-electronic-complexity-of-the-ground-state-of},
 volume = {150},
 year = {2019}
}

@techreport{Li2025quantum,
 author = {Li, Alex and others},
 day = {31},
 institution = {Human Rights Foundation},
 month = {oct},
 note = {Informed by technical insights from the 2025 Presidio Bitcoin Quantum Summit.},
 title = {{The Quantum Threat to {Bitcoin}}},
 type = {Human Rights Research Report},
 url = {https://hrf.org/latest/the-quantum-threat-to-Bitcoin/},
 year = {2025}
}

@misc{Linus2023bitvm,
 author = {Linus, Robin},
 day = {12},
 howpublished = {\url{https://bitvm.org/bitvm.pdf}},
 month = {oct},
 note = {Technical White Paper},
 title = {{BitVM: Compute Anything on {Bitcoin}}},
 year = {2023}
}

@article{Litinski2023to,
 author = {Litinski, Daniel},
 journal = {arXiv:2306.08585},
 title = {How to compute a 256-bit elliptic curve private key with only 50 million Toffoli gates},
 url = {https://arxiv.org/abs/2306.08585},
 year = {2023}
}

@misc{Little2026solana,
 author = {Little, Dean and Silur and Ashimine, Ikko Eltociear},
 howpublished = {GitHub Repository},
 title = {{Solana Winternitz Vault}: A Quantum-Resistant Lamport Vault using {WOTS}},
 url = {https://github.com/blueshift-gg/solana-winternitz-vault},
 urldate = {2026-03-23},
 year = {2026}
}

@inproceedings{Lokhava2019stellar,
author = {Lokhava, Marta and Losa, Giuliano and Mazi\`{e}res, David and Hoare, Graydon and Barry, Nicolas and Gafni, Eli and Jove, Jonathan and Malinowsky, Rafa\l{} and McCaleb, Jed},
title = {Fast and secure global payments with Stellar},
year = {2019},
isbn = {9781450368735},
publisher = {Association for Computing Machinery},
address = {New York, NY, USA},
url = {https://doi.org/10.1145/3341301.3359636},
doi = {10.1145/3341301.3359636},
booktitle = {Proceedings of the 27th ACM Symposium on Operating Systems Principles},
pages = {80–96},
numpages = {17},
location = {Huntsville, Ontario, Canada},
series = {SOSP '19}
}

@article{Low2025Fast,
 author = {Low, Guang Hao and King, Robbie and Berry, Dominic W and Han, Qiushi and DePrince III, A Eugene and White, Alec F and Babbush, Ryan and Somma, Rolando D and Rubin, Nicholas C},
 journal = {Physical Review X},
 number = {4},
 pages = {041016},
 publisher = {APS},
 title = {Fast Quantum Simulation of Electronic Structure by Spectral Amplification},
 url = {https://journals.aps.org/prx/abstract/10.1103/pb2g-j9cw},
 volume = {15},
 year = {2025}
}

@misc{Marlinspike2016x3dh,
 author = {Marlinspike, Moxie and Perrin, Trevor},
 month = {August},
 title = {The {X3DH} Key Agreement Protocol},
 url = {https://signal.org/docs/specifications/x3dh/x3dh.pdf},
 year = {2016}
}

@techreport{Martini2009bad,
 address = {New York, NY},
 author = {{Martini, Luca and Heuser, Matthias and Fest, Martin and Stegemann, Uwe and Schneider, Sebastian and Brenna, Gabriel and Windhagen, Eckart and Poppensieker, Thomas}},
 institution = {McKinsey Working Papers on Risk},
 month = {aug},
 number = {12},
 title = {{Bad Banks: Finding the Right Exit from the Financial Crisis}},
 type = {Working Paper},
 url = {https://www.mckinsey.com/~/media/McKinsey/Business{\%}20Functions/Risk/Our{\%}20Insights/Bad{\%}20banks{\%}20finding{\%}20the{\%}20right{\%}20exit{\%}20from{\%}20the{\%}20financial{\%}20crisis/Bad{\%}20banks.pdf},
 year = {2009}
}

@misc{Marzougui2023feasibility,
 author = {Soundes Marzougui and Ievgan Kabin and Juliane Kr{\"a}mer and Thomas Aulbach and Jean-Pierre Seifert},
 howpublished = {Cryptology {ePrint} Archive, Paper 2023/142},
 title = {On the Feasibility of Single-Trace Attacks on the Gaussian Sampler using a {CDT}},
 url = {https://eprint.iacr.org/2023/142},
 year = {2023}
}

@book{Maurushat2013disclosure,
 author = {Maurushat, Alana},
 doi = {10.1007/978-1-4471-5004-6},
 isbn = {9781447150046},
 issn = {2193-9748},
 journal = {SpringerBriefs in Cybersecurity},
 publisher = {Springer London},
 title = {Disclosure of Security Vulnerabilities: Legal and Ethical Issues},
 url = {http://dx.doi.org/10.1007/978-1-4471-5004-6},
 year = {2013}
}

@techreport{Mazieres2015stellar,
 author = {Mazi{\`e}res, David},
 title = {The {Stellar} Consensus Protocol: {A} Federated Model for Internet-level Consensus},
 institution = {Stellar Development Foundation},
 year = {2015},
 url = {https://www.stellar.org/papers/stellar-consensus-protocol.pdf},
 urldate = {2026-03-24},
}

@misc{McCormack2018bitcoins,
 author = {McCormack, Peter and Held, Dan},
 day = {1},
 howpublished = {Medium: HackerNoon},
 month = {nov},
 note = {Interview Date: Oct 25, 2018. Discusses the philosophical significance of {Satoshi Nakamoto}'s disappearance and the "Immaculate Conception" theory of {Bitcoin}'s launch.},
 title = {{Bitcoin's} Immaculate Conception: {Audio} interview transcription --- {WBD043}},
 url = {https://medium.com/hackernoon/bitcoins-immaculate-conception-with-dan-held-905540eb15fa},
 urldate = {2026-03-23},
 year = {2018}
}

@article{McEliece1978Public,
 author = {McEliece, Robert J.},
 title = {A Public-Key Cryptosystem Based on Algebraic Coding Theory},
 journal = {The Deep Space Network Progress Report},
 volume = {44},
 pages = {114--116},
 year = {1978},
 url = {https://ipnpr.jpl.nasa.gov/progress_report2/42-44/44N.PDF},
}

@article{Merkle1978secure,
 address = {New York, NY, USA},
 author = {Merkle, Ralph C.},
 doi = {10.1145/359460.359473},
 issn = {0001-0782},
 issue_date = {April 1978},
 journal = {Commun. ACM},
 month = {April},
 number = {4},
 numpages = {6},
 pages = {294--299},
 publisher = {Association for Computing Machinery},
 title = {Secure communications over insecure channels},
 url = {https://doi.org/10.1145/359460.359473},
 volume = {21},
 year = {1978}
}

@phdthesis{Merkle1979Secrecy,
 author = {Merkle, Ralph C.},
 title = {Secrecy, Authentication, and Public Key Systems},
 school = {Stanford University},
 address = {Stanford, CA, USA},
 year = {1979},
 month = {June},
 url = {https://www.ralphmerkle.com/papers/Thesis1979.pdf}
}

@inproceedings{Merkle1990certified,
 address = {New York, NY},
 author = {Merkle, Ralph C.},
 booktitle = {Advances in Cryptology --- CRYPTO' 89 Proceedings},
 editor = {Brassard, Gilles},
 isbn = {978-0-387-34805-6},
 pages = {218--238},
 publisher = {Springer New York},
 title = {A Certified Digital Signature},
 year = {1990}
}

@misc{MerriamWebsterLaches,
 author = {{Merriam-Webster}},
 howpublished = {Merriam-Webster.com Dictionary},
 note = {defines "laches" as "undue delay in asserting a legal right or privilege"},
 title = {Definition of {Laches}},
 url = {https://www.merriam-webster.com/dictionary/laches},
 urldate = {2026-03-22}
}

@inproceedings{Miller1986Ecc,
 address = {Berlin, Heidelberg},
 author = {Miller, Victor S.},
 booktitle = {Advances in Cryptology --- CRYPTO '85 Proceedings},
 editor = {Williams, Hugh C.},
 isbn = {978-3-540-39799-1},
 pages = {417--426},
 publisher = {Springer Berlin Heidelberg},
 title = {Use of Elliptic Curves in Cryptography},
 year = {1986}
}

@misc{Miller2017announcing,
 author = {Miller, Andrew and Bowe, Sean},
 day = {11},
 howpublished = {Zcash Foundation Blog},
 month = {nov},
 note = {Official announcement of the {Powers of Tau} {MPC} ceremony for the {Sapling} upgrade.},
 title = {Announcing the world's largest multi-party computation ceremony},
 url = {https://zfnd.org/announcing-the-worlds-largest-multi-party-computation-ceremony/},
 urldate = {2026-03-23},
 year = {2017}
}

@techreport{Milton2025bitcoin,
 address = {New York, NY},
 author = {Milton, Anthony and Shikhelman, Clara},
 institution = {Chaincode Labs},
 month = {may},
 note = {Accessed: 2026-03-22},
 title = {{Bitcoin and Quantum Computing: Current Status and Future Directions}},
 type = {Research Report},
 url = {https://chaincode.com/bitcoin-post-quantum.pdf},
 year = {2025}
}

@article{Montanaro2016,
 author = {Montanaro, Ashley},
 day = {12},
 doi = {10.1038/npjqi.2015.23},
 issn = {2056-6387},
 journal = {npj Quantum Information},
 month = {Jan},
 number = {1},
 pages = {15023},
 title = {Quantum algorithms: an overview},
 url = {https://doi.org/10.1038/npjqi.2015.23},
 volume = {2},
 year = {2016}
}

@article{Montgomery1987speeding,
 author = {Montgomery, Peter L.},
 doi = {10.1090/S0025-5718-1987-0866113-7},
 journal = {Mathematics of Computation},
 number = {359},
 pages = {243-264},
 publisher = {American Mathematical Society},
 title = {{Speeding the Pollard and elliptic curve methods of factorization}},
 url = {https://pubs.ams.org/journals/mcom/1987-48-177/S0025-5718-1987-0866113-7/S0025-5718-1987-0866113-7},
 volume = {95},
 year = {1987}
}

@article{Moore1965cramming,
 author = {Moore, Gordon E.},
 day = {19},
 journal = {Electronics},
 month = {apr},
 note = {Reprinted in: {IEEE} Solid-State Circuits Newsletter, Vol. 11, No. 3, Sept. 2006.},
 number = {8},
 pages = {114--117},
 title = {{Cramming More Components onto Integrated Circuits}},
 url = {https://ieeexplore.ieee.org/document/4785860},
 volume = {38},
 year = {1965}
}

@misc{Musk2025grok,
 author = {Musk, Elon},
 day = {2},
 howpublished = {X (formerly Twitter)},
 month = {aug},
 title = {{@grok estimate the probability of quantum computing cracking SHA-256}},
 url = {https://x.com/elonmusk/status/1951596018438373740},
 urldate = {2026-03-22},
 year = {2025}
}

@article{Nadler2023tornado,
 author = {Nadler, Matthias and Sch{\"a}r, Fabian},
 title = {Tornado Cash and Blockchain Privacy: A Primer for Economists and Policymakers},
 journal = {Federal Reserve Bank of St. Louis Review},
 year = {2023},
 month = apr,
 day = {10},
 url = {https://www.stlouisfed.org/publications/review/2023/02/03/tornado-cash-and-blockchain-privacy-a-primer-for-economists-and-policymakers},
 note = {First Quarter 2023}
}

@misc{Nakamoto2008bitcoin,
 author = {Nakamoto, Satoshi},
 howpublished = {\url{https://bitcoin.org/bitcoin.pdf}},
 title = {{Bitcoin: A Peer-to-Peer Electronic Cash System}},
 year = {2008}
}

@article{Neagle2025BitcoinQC,
 author = {Neagle, Shane},
 title = {How Should {B}itcoiners View Quantum Computing?},
 journal = {Bitcoin Magazine},
 year = {2025},
 month = jan,
 day = {15},
 url = {https://bitcoinmagazine.com/technical/how-should-bitcoiners-view-quantum-computing},
 note  = {Technical Analysis. Accessed March 29, 2026}
}

@article{Nerem2023conditions,
 author = {Nerem, Robert R. and Gaur, Daya R.},
 doi = {10.1016/j.bcra.2023.100141},
 issn = {2096-7209},
 journal = {Blockchain: Research and Applications},
 month = {September},
 number = {3},
 pages = {100141},
 publisher = {Elsevier BV},
 title = {Conditions for advantageous quantum Bitcoin mining},
 url = {http://dx.doi.org/10.1016/j.bcra.2023.100141},
 volume = {4},
 year = {2023}
}

@misc{Nichols2024marathon,
 author = {Nichols, Spencer},
 day = {22},
 howpublished = {Bitcoin Magazine (via {Nasdaq})},
 month = {feb},
 title = {Marathon Launches {Slipstream} Tech Stack to Process Non-Standard {Bitcoin} Transactions},
 url = {https://www.nasdaq.com/articles/marathon-launches-slipstream-tech-stack-to-process-non-standard-bitcoin-transactions},
 urldate = {2026-03-23},
 year = {2024}
}

@techreport{Nick2020liquid,
 author = {Nick, Jonas and Poelstra, Andrew and Sanders, Gregory},
 day = {22},
 institution = {Blockstream},
 month = {may},
 title = {{Liquid: A Bitcoin Sidechain}},
 type = {Protocol White Paper},
 url = {https://blockstream.com/assets/downloads/pdf/liquid-whitepaper.pdf},
 year = {2020}
}

@inproceedings{Nils2004,
 address = {Berlin, Heidelberg},
 author = {Gura, Nils
and Patel, Arun
and Wander, Arvinderpal
and Eberle, Hans
and Shantz, Sheueling Chang},
 booktitle = {Cryptographic Hardware and Embedded Systems - CHES 2004},
 editor = {Joye, Marc
and Quisquater, Jean-Jacques},
 isbn = {978-3-540-28632-5},
 pages = {119--132},
 publisher = {Springer Berlin Heidelberg},
 title = {Comparing Elliptic Curve Cryptography and {RSA} on 8-bit {CPUs}},
 url = {https://link.springer.com/chapter/10.1007/978-3-540-28632-5_9},
 year = {2004}
}

@article{Nino2018survey,
 author = {Lara-Nino, Carlos Andres and Diaz-Perez, Arturo and Morales-Sandoval, Miguel},
 doi = {10.1109/ACCESS.2018.2881444},
 journal = {IEEE Access},
 number = {},
 pages = {72514-72550},
 title = {Elliptic Curve Lightweight Cryptography: A Survey},
 volume = {6},
 year = {2018}
}

@misc{NISTGlossary_defense_in_depth,
  author = {{National Institute of Standards and Technology}},
  title = {Glossary: {D}efense-in-Depth},
  year = {2026},
  howpublished = {\url{https://csrc.nist.gov/glossary/term/defense_in_depth}},
  note = {{CSRC Computer Security Resource Center.}},
  urldate = {2026-03-24}
}

@book{NISTUS2015sha3,
 doi = {10.6028/nist.fips.202},
 institution = {National Institute of Standards and Technology (U.S.)},
 title = {SHA-3 standard: permutation-based hash and extendable-output functions},
 url = {http://dx.doi.org/10.6028/NIST.FIPS.202},
 year = {2015}
}

@book{NISTUS2024modulelatticebased_dsa204,
 doi = {10.6028/nist.fips.204},
 institution = {National Institute of Standards and Technology (U.S.)},
 month = {August},
 title = {Module-lattice-based digital signature standard},
 url = {http://dx.doi.org/10.6028/NIST.FIPS.204},
 year = {2024}
}

@book{NISTUS2024modulelatticebased_kem203,
 doi = {10.6028/nist.fips.203},
 institution = {National Institute of Standards and Technology (U.S.)},
 month = {August},
 title = {Module-lattice-based key-encapsulation mechanism standard},
 url = {http://dx.doi.org/10.6028/NIST.FIPS.203},
 year = {2024}
}

@book{NISTUS2024stateless,
 doi = {10.6028/nist.fips.205},
 institution = {National Institute of Standards and Technology (U.S.)},
 month = {August},
 title = {Stateless hash-based digital signature standard},
 url = {http://dx.doi.org/10.6028/NIST.FIPS.205},
 year = {2024}
}

@book{North1990political,
 author={North, Douglass C.},
 title={Institutions, Institutional Change and Economic Performance},
 place={Cambridge},
 series={Political Economy of Institutions and Decisions},
 publisher={Cambridge University Press},
 year={1990},
 collection={Political Economy of Institutions and Decisions}
}

@article{North1991institutions,
 author = {North, Douglass C.},
 title = {Institutions},
 journal = {Journal of Economic Perspectives},
 volume = {5},
 number = {1},
 year = {1991},
 month = {March},
 pages = {97–112},
 doi = {10.1257/jep.5.1.97},
 url = {https://www.aeaweb.org/articles?id=10.1257/jep.5.1.97}
}

@misc{Notes2024new,
 author = {{Officer's Notes}},
 day = {1},
 howpublished = {Medium: Coinmonks},
 month = {dec},
 title = {New Theory by {BTCparser}: Who Might Be the Real {Satoshi}? Answering the Question of Why He Never Touched His 2009 {Bitcoins}!},
 url = {https://medium.com/coinmonks/new-theory-by-btcparser-who-might-be-the-real-satoshi-98d617c6e7e7},
 urldate = {2026-03-23},
 year = {2024}
}

@article{o2017quantum,
 author = {O'Gorman, Joe and Campbell, Earl T},
 journal = {Physical Review A},
 number = {3},
 pages = {032338},
 publisher = {APS},
 title = {Quantum computation with realistic magic-state factories},
 url = {https://journals.aps.org/pra/abstract/10.1103/PhysRevA.95.032338},
 volume = {95},
 year = {2017}
}

@misc{OBrien2023Chromium,
 author = {O'Brien, Devon},
 title = {Protecting {C}hrome Traffic with {H}ybrid {K}yber {KEM}},
 howpublished = {Chromium Blog},
 year = {2023},
 month = aug,
 day = {10},
 url = {https://blog.chromium.org/2023/08/protecting-chrome-traffic-with-hybrid.html},
 note = {Accessed March 26, 2026}
}

@techreport{OCP2023GPUFW,
 author = {{AMD} and {Google} and {Meta} and {Microsoft} and {NVIDIA}},
 title = {{OCP} {GPU} {FW} Update Specification},
 institution = {Open Compute Project},
 year = {2023},
 version = {1.0},
 url = {https://www.opencompute.org/documents/external-ocp-gpu-fw-update-specification-v1-0-1-pdf}
}

@Article{Olle2024Simultaneous,
author={Olle, Jan
and Zen, Remmy
and Puviani, Matteo
and Marquardt, Florian},
title={Simultaneous discovery of quantum error correction codes and encoders with a noise-aware reinforcement learning agent},
journal={npj Quantum Information},
year={2024},
month={Dec},
day={03},
volume={10},
number={1},
pages={126},
issn={2056-6387},
doi={10.1038/s41534-024-00920-y},
url={https://doi.org/10.1038/s41534-024-00920-y}
}

@techreport{Palatinus2013bip,
 author = {Palatinus, Marek and Rusnak, Pavol and Voisine, Aaron and Bowe, Sean},
 day = {10},
 institution = {Bitcoin Core},
 month = {sep},
 note = {Source: \url{https://github.com/bitcoin/bips/blob/master/bip-0039.mediawiki}. Accessed: 2026-03-22},
 number = {39},
 title = {{BIP 39: Mnemonic code for generating deterministic keys}},
 type = {Bitcoin Improvement Proposal},
 url = {https://bips.dev/39/},
 year = {2013}
}

@techreport{Palladino2019erc1967,
 author = {Palladino, Santiago and Giordano, Francisco and Croubois, Hadrien},
 day = {24},
 institution = {Ethereum Foundation},
 month = {apr},
 note = {Accessed: 2026-03-23},
 number = {1967},
 title = {{ERC-1967: Proxy Storage Slots}},
 type = {Ethereum Improvement Proposal},
 url = {https://eips.ethereum.org/EIPS/eip-1967},
 year = {2019}
}

@misc{Parker2024stepping,
 author = {Parker, Luke},
 day = {30},
 howpublished = {GitHub Gist},
 month = {nov},
 note = {Technical manifesto calling for a moratorium on quantum-vulnerable {R\&D} and the immediate development of a fallback post-quantum {Monero} protocol.},
 title = {Stepping Back},
 url = {https://gist.github.com/kayabaNerve/e5b262c5efefcfcfa32748a0d99bc0e1},
 urldate = {2026-03-23},
 year = {2024}
}

@article{Pattison2024fast,
 author = {Pattison, Christopher A. and Baranes, Gefen and Ataides, J. Pablo Bonilla and Lukin, Mikhail D. and Zhou, Hengyun},
 journal = {arXiv:2408.15936},
 title = {Fast quantum interconnects via constant-rate entanglement distillation},
 url = {https://arxiv.org/abs/2408.15936},
 year = {2024}
}

@inproceedings{Pedersen1992noninteractive,
 address = {Berlin, Heidelberg},
 author = {Pedersen, Torben Pryds},
 booktitle = {Advances in Cryptology --- CRYPTO '91},
 editor = {Feigenbaum, Joan},
 isbn = {978-3-540-46766-3},
 pages = {129--140},
 publisher = {Springer Berlin Heidelberg},
 title = {Non-Interactive and Information-Theoretic Secure Verifiable Secret Sharing},
 year = {1992}
}

@misc{Peikert2015decade,
 author = {Chris Peikert},
 howpublished = {Cryptology {ePrint} Archive, Paper 2015/939},
 title = {A Decade of Lattice Cryptography},
 url = {https://eprint.iacr.org/2015/939},
 year = {2015}
}

@misc{Perlner2025fips,
 author = {Perlner, Ray},
 howpublished = {Technical Presentation, \url{https://csrc.nist.gov/csrc/media/presentations/2025/fips-206-fn-dsa-(falcon)/images-media/fips_206-perlner_2.1.pdf}},
 month = {sep},
 note = {Presented at the 6th {PQC} Standardization Conference in Gaithersburg, Maryland},
 title = {{{FIPS} 206 Status Update}},
 year = {{2025}}
}

@techreport{Pertsev2019Tornado,
 author = {Pertsev, Alexey and Semenov, Roman and Storm, Roman},
 title = {Tornado {C}ash Privacy Solution},
 institution = {Tornado Cash},
 year = {2019},
 month = dec,
 day = {17},
 type = {Technical Whitepaper},
 number = {Version 1.4},
 url = {https://berkeley-defi.github.io/assets/material/Tornado%20Cash%20Whitepaper.pdf},
 note = {Accessed via Berkeley DeFi educational resources}
}

@misc{Pointcheval2017commitment,
 author = {Pointcheval, David},
 howpublished = {Course Material, {Master Parisien de Recherche en Informatique} ({MPRI})},
 institution = {{D{\'e}partement d'Informatique, {\'E}cole Normale Sup{\'e}rieure (ENS) Paris, Universit{\'e} PSL}},
 note = {Tutorial covering formal properties of {Pedersen} and {ElGamal} commitments.},
 title = {Commitment Schemes ({TD 16})},
 url = {https://www.di.ens.fr/david.pointcheval/enseignement/mpri2/td16.pdf},
 urldate = {2026-03-22},
 year = {2017}
}

@article{Pont2024downtime,
 author = {Pont, Jamie J. and Kearney, Joseph J. and Moyler, Jack and Perez-Delgado, Carlos A.},
 journal = {arXiv:2410.16965},
 title = {Downtime Required for Bitcoin Quantum-Safety},
 url = {https://arxiv.org/abs/2410.16965},
 year = {2024}
}

@techreport{Poon2016bitcoin,
 author = {Poon, Joseph and Dryja, Thaddeus},
 day = {14},
 institution = {Lightning Network},
 month = {jan},
 number = {DRAFT Version: 0.5.9.2},
 title = {{The {Bitcoin} Lightning Network: Scalable Off-Chain Instant Payments}},
 type = {Protocol White Paper},
 url = {https://lightning.network/lightning-network-paper.pdf},
 year = {2016}
}

@book{Poundstone1992prisoners,
 author = {Poundstone, W.},
 isbn = {9780385415675},
 lccn = {91013906},
 publisher = {Doubleday},
 title = {Prisoner's Dilemma},
 url = {https://books.google.com/books?id=9uruAAAAMAAJ},
 year = {1992}
}

@misc{PQShield2026postquantum,
 author = {{PQShield}},
 howpublished = {GitHub Pages},
 title = {Post-Quantum Signatures Zoo},
 type = {Technical Reference Registry},
 url = {https://pqshield.github.io/nist-sigs-zoo/},
 urldate = {2026-03-22},
 year = {2026}
}

@misc{Proos2003shors,
 author = {Proos, John and Zalka, Christof},
 copyright = {Assumed arXiv.org perpetual, non-exclusive license to distribute this article for submissions made before January 2004},
 doi = {10.48550/ARXIV.QUANT-PH/0301141},
 publisher = {arXiv},
 title = {{S}hor's discrete logarithm quantum algorithm for elliptic curves},
 url = {https://arxiv.org/abs/quant-ph/0301141},
 year = {2003}
}

@article{Quan2022improving,
 author = {Quan, Yunjia},
 doi = {10.1109/access.2022.3227394},
 issn = {2169-3536},
 journal = {IEEE Access},
 pages = {132472--132482},
 publisher = {Institute of Electrical and Electronics Engineers (IEEE)},
 title = {{Improving Bitcoin's Post-Quantum Transaction Efficiency With a Novel Lattice-Based Aggregate Signature Scheme Based on CRYSTALS-Dilithium and a STARK Protocol}},
 url = {http://dx.doi.org/10.1109/ACCESS.2022.3227394},
 volume = {10},
 year = {2022}
}

@inproceedings{Quisquater1989to,
 author = {Jean-Jacques Quisquater and Myriam Quisquater and Muriel Quisquater and Micha{"e}l Quisquater and Louis C. Guillou and Marie Annick Guillou and Ga{"i}d Guillou and Anne-Claire Guillou and Gwenol{'e} Guillou and Soazig Guillou and Thomas A. Berson},
 booktitle = {Annual International Cryptology Conference},
 title = {How to Explain Zero-Knowledge Protocols to Your Children},
 url = {https://api.semanticscholar.org/CorpusID:30932582},
 year = {1989}
}

@inproceedings{Regev2005lattices,
 address = {New York, NY, USA},
 author = {Regev, Oded},
 booktitle = {Proceedings of the Thirty-Seventh Annual ACM Symposium on Theory of Computing},
 doi = {10.1145/1060590.1060603},
 isbn = {1581139608},
 location = {Baltimore, MD, USA},
 numpages = {10},
 pages = {84--93},
 publisher = {Association for Computing Machinery},
 series = {STOC '05},
 title = {On lattices, learning with errors, random linear codes, and cryptography},
 url = {https://doi.org/10.1145/1060590.1060603},
 year = {2005}
}

@article{Reiher2017Elucidating,
 author = {Reiher, Markus and Wiebe, Nathan and Svore, Krysta M and Wecker, Dave and Troyer, Matthias},
 journal = {Proceedings of the national academy of sciences},
 number = {29},
 pages = {7555--7560},
 publisher = {National Academy of Sciences},
 title = {Elucidating reaction mechanisms on quantum computers},
 url = {https://www.pnas.org/doi/10.1073/pnas.1619152114},
 volume = {114},
 year = {2017}
}

@techreport{Reitwiessner2017eip196,
 author = {Reitwiessner, Christian},
 day = {2},
 institution = {Ethereum Foundation},
 month = {feb},
 note = {Accessed: 2026-03-23},
 number = {196},
 title = {{EIP-196: Precompiled contracts for addition and scalar multiplication on the elliptic curve altbn128}},
 type = {Ethereum Improvement Proposal},
 url = {https://eips.ethereum.org/EIPS/eip-196},
 year = {2017}
}

@misc{Reuters2021us,
 author = {{Reuters}},
 day = {13},
 howpublished = {Reuters},
 month = {oct},
 title = {{US} becomes largest {Bitcoin} mining centre following {China} ban},
 url = {https://www.reuters.com/technology/us-becomes-largest-bitcoin-mining-centre-following-china-ban-2021-10-13/},
 urldate = {2026-03-23},
 year = {2021}
}

@misc{rfc5656,
 author = {Douglas Stebila and Jon Green},
 doi = {10.17487/RFC5656},
 howpublished = {RFC 5656},
 month = {December},
 number = {5656},
 publisher = {RFC Editor},
 series = {Request for Comments},
 title = {{Elliptic Curve Algorithm Integration in the Secure Shell (SSH) Transport Layer}},
 url = {https://www.rfc-editor.org/info/rfc5656},
 year = {2009}
}

@misc{rfc6605,
 author = {Paul Hoffman and Wouter Wijngaards},
 doi = {10.17487/RFC6605},
 howpublished = {RFC 6605},
 month = {April},
 number = {6605},
 publisher = {RFC Editor},
 series = {Request for Comments},
 title = {{Elliptic Curve Digital Signature Algorithm (DSA) for DNSSEC}},
 url = {https://www.rfc-editor.org/info/rfc6605},
 year = {2012}
}

@misc{rfc8446,
 author = {Eric Rescorla},
 doi = {10.17487/RFC8446},
 howpublished = {RFC 8446},
 month = {August},
 number = {8446},
 pagetotal = {160},
 publisher = {RFC Editor},
 series = {Request for Comments},
 title = {{The Transport Layer Security (TLS) Protocol Version 1.3}},
 url = {https://www.rfc-editor.org/info/rfc8446},
 year = {2018}
}

@misc{Ripple2026documentation,
 author = {{Ripple}},
 howpublished = {XRP Ledger Dev Portal},
 title = {Documentation -- Best Practices -- Key Management -- Remove a Regular Key Pair},
 url = {https://xrpl.org/docs/tutorials/how-tos/manage-account-settings/change-or-remove-a-regular-key-pair},
 urldate = {2026-03-23},
 year = {2026}
}

@techreport{RippleX2025future,
 author = {{Team RippleX}},
 institution = {Ripple Labs, Inc.},
 title = {{The Future of Asset Tokenization: A New Token Standard for Institutional-Grade Finance on XRP Ledger}},
 type = {Industry Research Report},
 url = {https://xrpl.org/static/pdf/Whitepaper_the_future_of_asset_tokenization.pdf},
 year = {2025}
}

@article{rivest1978method,
 author = {Rivest, Ronald L and Shamir, Adi and Adleman, Leonard},
 journal = {Communications of the ACM},
 number = {2},
 pages = {120--126},
 publisher = {ACM New York, NY, USA},
 title = {A method for obtaining digital signatures and public-key cryptosystems},
 url = {https://dl.acm.org/doi/10.1145/359340.359342},
 volume = {21},
 year = {1978}
}

@article{rocca2024reducing,
 author = {Rocca, Dario and Cortes, Cristian L and Gonthier, J{\'e}r{\^o}me F and Ollitrault, Pauline J and Parrish, Robert M and Anselmetti, Gian-Luca and Degroote, Matthias and Moll, Nikolaj and Santagati, Raffaele and Streif, Michael},
 journal = {Journal of Chemical Theory and Computation},
 number = {11},
 pages = {4639--4653},
 publisher = {ACS Publications},
 title = {Reducing the runtime of fault-tolerant quantum simulations in chemistry through symmetry-compressed double factorization},
 url = {https://pubs.acs.org/doi/10.1021/acs.jctc.4c00352},
 volume = {20},
 year = {2024}
}

@article{Roetteler2017quantum,
 author = {Roetteler, Martin and Naehrig, Michael and Svore, Krysta M. and Lauter, Kristin},
 journal = {arXiv:1706.06752},
 title = {Quantum resource estimates for computing elliptic curve discrete logarithms},
 url = {https://arxiv.org/abs/1706.06752},
 year = {2017}
}

@misc{Ruffing2017switch,
 author = {Tim Ruffing and Giulio Malavolta},
 howpublished = {Cryptology {ePrint} Archive, Paper 2017/237},
 title = {Switch Commitments: A Safety Switch for Confidential Transactions},
 url = {https://eprint.iacr.org/2017/237},
 year = {2017}
}

@misc{Ruffing2018switch,
 author = {Ruffing, Tim},
 day = {22},
 howpublished = {MimbleWimble Mailing List Archive},
 month = {apr},
 note = {Proposes to commit with $C=vG+bH$ where $b=b'+\text{hash}(vG+b'H, b'J)$, so $C$ can be treated as either Pedersen or ElGamal commitment.},
 title = {Switch Commitments (Again)},
 url = {https://archive.lists.launchpad.net/mimblewimble/msg00479.html},
 urldate = {2026-03-23},
 year = {2018}
}

@misc{Ruffing2025postquantum,
 author = {Tim Ruffing},
 howpublished = {Cryptology {ePrint} Archive, Paper 2025/1307},
 title = {The Post-Quantum Security of Bitcoin's Taproot as a Commitment Scheme},
 url = {https://eprint.iacr.org/2025/1307},
 year = {2025}
}

@article{Russell2006rough,
 address = {USA},
 author = {Russell, Andrew L.},
 doi = {10.1109/MAHC.2006.42},
 issn = {1058-6180},
 issue_date = {July 2006},
 journal = {IEEE Ann. Hist. Comput.},
 month = {July},
 number = {3},
 numpages = {14},
 pages = {48--61},
 publisher = {IEEE Educational Activities Department},
 title = {{'Rough Consensus and Running Code' and the Internet-OSI Standards War}},
 url = {https://doi.org/10.1109/MAHC.2006.42},
 volume = {28},
 year = {2006}
}

@techreport{Ryan2024eip7594,
 author = {Ryan, Danny and Feist, Dankrad and D'Amato, Francesco and Wang, Hsiao-Wei and Stokes, Alex},
 day = {12},
 institution = {Ethereum Foundation},
 month = {jan},
 note = {Accessed: 2026-03-23},
 number = {7594},
 title = {EIP-7594: PeerDAS - Peer Data Availability Sampling},
 type = {Ethereum Improvement Proposal},
 url = {https://eips.ethereum.org/EIPS/eip-7594},
 year = {2024}
}

@techreport{Saggese2025towards,
 author = {Pietro Saggese and Michael Fr{\"o}wis and Stefan Kitzler and Bernhard Haslhofer and Raphael Auer},
 doi = {None},
 institution = {Bank for International Settlements},
 month = {May},
 note = {Available at: \url{https://www.bis.org/publ/work1268.pdf}},
 number = {1268},
 title = {Towards verifiability of total value locked (TVL) in decentralized finance},
 type = {BIS Working Papers},
 url = {https://ideas.repec.org/p/bis/biswps/1268.html},
 year = {2025}
}

@article{Sattath2020insecurity,
 author = {Sattath, Or},
 doi = {10.1007/s10207-020-00493-9},
 issn = {1615-5270},
 journal = {International Journal of Information Security},
 month = {March},
 number = {3},
 pages = {291--302},
 publisher = {Springer Science and Business Media LLC},
 title = {On the insecurity of quantum Bitcoin mining},
 url = {http://dx.doi.org/10.1007/s10207-020-00493-9},
 volume = {19},
 year = {2020}
}

@article{Schafer2018fast,
 author = {Sch{\"a}fer, V. M. and Ballance, C. J. and Thirumalai, K. and Stephenson, L. J. and Ballance, T. G. and Steane, A. M. and Lucas, D. M.},
 doi = {10.1038/nature25737},
 issn = {1476-4687},
 journal = {Nature},
 month = {March},
 number = {7694},
 pages = {75--78},
 publisher = {Springer Science and Business Media LLC},
 title = {Fast quantum logic gates with trapped-ion qubits},
 url = {http://dx.doi.org/10.1038/nature25737},
 volume = {555},
 year = {2018}
}

@book{Schleich2025quantum,
 address = {Washington, DC, USA},
 author = {Schleich, Philipp and Calder{\'o}n, Luis Mantilla  and Sun, Chong and Bagherimehrab, Mohsen and Aldossary, Abdulrahman and Kottmann, Jakob S. and Aspuru-Guzik, Al{\'a}n},
 doi = {10.1021/acsinfocus.7e9012},
 publisher = {American Chemical Society},
 title = {Quantum Computing for Quantum Chemistry},
 url = {https://pubs.acs.org/doi/abs/10.1021/acsinfocus.7e9012},
 year = {2025}
}

@article{Schmidhuber2025quartic,
 author = {Schmidhuber, Alexander and O'Donnell, Ryan and Kothari, Robin and Babbush, Ryan},
 doi = {10.1103/physrevx.15.021077},
 issn = {2160-3308},
 journal = {Physical Review X},
 month = {June},
 number = {2},
 publisher = {American Physical Society (APS)},
 title = {Quartic Quantum Speedups for Planted Inference},
 url = {http://dx.doi.org/10.1103/PhysRevX.15.021077},
 volume = {15},
 year = {2025}
}

@misc{Schneier2011new,
 author = {Schneier, Bruce},
 day = {18},
 howpublished = {\url{https://www.schneier.com/blog/archives/2011/08/new_attack_on_a_1.html}},
 month = {aug},
 note = {Accessed: 2026-03-22},
 title = {{New Attack on AES}},
 year = {2011}
}

@article{Schneider2025Fault,
 author = {Schneider, Kevin and Auer, Lukas and Wagner, Alexander},
 title = {Fault Attacks on {ECC} Signature Verification},
 journal = {{IACR} Transactions on Cryptographic Hardware and Embedded Systems},
 year = {2025},
 volume = {2025},
 number = {4},
 pages = {1010--1052},
 doi = {10.46586/tches.v2025.i4.1010-1052},
 issn = {2569-2925},
 publisher = {Ruhr-Universit{\"a}t Bochum}
}

@techreport{Schoedon2018eip999,
 author = {Schoedon, Afri},
 day = {4},
 institution = {Ethereum Foundation},
 month = {apr},
 note = {Accessed: 2026-03-23},
 number = {999},
 title = {{EIP-999: Restore Contract Code at 0x863DF6BFa4469f3ead0bE8f9F2AAE51c91A907b4 [DRAFT]}},
 type = {Ethereum Improvement Proposal},
 url = {https://eips.ethereum.org/EIPS/eip-999},
 year = {2018}
}

@article{Scott2023timing,
 author = {Scott, John R. and Balram, Krishna C.},
 doi = {10.1103/physrevapplied.20.024019},
 issn = {2331-7019},
 journal = {Physical Review Applied},
 month = {August},
 number = {2},
 publisher = {American Physical Society (APS)},
 title = {Timing Constraints Due to Real-Time Graph-Traversal Algorithms on Incomplete Cluster States in Photonic Measurement-Based Quantum Computing},
 url = {http://dx.doi.org/10.1103/PhysRevApplied.20.024019},
 volume = {20},
 year = {2023}
}

@techreport{Securities2017report,
 author = {{U.S. Securities and Exchange Commission}},
 day = {25},
 institution = {U.S. Securities and Exchange Commission},
 month = {jul},
 number = {Release No. 81207},
 title = {{Report of Investigation Pursuant to Section 21(a) of the Securities Exchange Act of 1934: The DAO}},
 type = {Report of Investigation},
 url = {https://www.sec.gov/files/litigation/investreport/34-81207.pdf},
 year = {2017}
}

@misc{Shen2021chinas,
 author = {Shen, Samuel and Galbraith, Andrew},
 day = {2r54},
 howpublished = {Reuters},
 month = {jun},
 title = {{China}'s ban forces some bitcoin miners to flee overseas, others sell out},
 url = {https://www.reuters.com/technology/chinas-ban-forces-some-bitcoin-miners-flee-overseas-others-sell-out-2021-06-25/},
 urldate = {2026-03-23},
 year = {2021}
}

@inproceedings{Shor1994algorithms,
 author = {Shor, P.W.},
 booktitle = {Proceedings 35th Annual Symposium on Foundations of Computer Science},
 doi = {10.1109/SFCS.1994.365700},
 number = {},
 pages = {124-134},
 title = {Algorithms for quantum computation: discrete logarithms and factoring},
 volume = {},
 year = {1994}
}

@article{Shor1997Polynomial,
 author = {Shor, Peter W.},
 doi = {10.1137/S0097539795293172},
 journal = {SIAM Journal on Computing},
 number = {5},
 pages = {1484-1509},
 title = {Polynomial-Time Algorithms for Prime Factorization and Discrete Logarithms on a Quantum Computer},
 url = {https://doi.org/10.1137/S0097539795293172},
 volume = {26},
 year = {1997}
}

@article{Shor1997polynomialtime,
 author = {Shor, Peter W.},
 doi = {10.1137/S0097539795293172},
 journal = {SIAM Journal on Computing},
 number = {5},
 pages = {1484-1509},
 title = {Polynomial-Time Algorithms for Prime Factorization and Discrete Logarithms on a Quantum Computer},
 url = {https://doi.org/10.1137/S0097539795293172},
 volume = {26},
 year = {1997}
}

@article{Singh2020sidechain,
 address = {GBR},
 author = {Singh, Amritraj and Click, Kelly and Parizi, Reza M. and Zhang, Qi and Dehghantanha, Ali and Choo, Kim-Kwang Raymond},
 doi = {10.1016/j.jnca.2019.102471},
 issn = {1084-8045},
 issue_date = {Jan 2020},
 journal = {J. Netw. Comput. Appl.},
 month = {January},
 number = {C},
 numpages = {16},
 publisher = {Academic Press Ltd.},
 title = {Sidechain technologies in blockchain networks: An examination and state-of-the-art review},
 url = {https://doi.org/10.1016/j.jnca.2019.102471},
 volume = {149},
 year = {2020}
}

@article{Sklavos2005implementation,
 author = {Sklavos, N. and Koufopavlou, O.},
 doi = {10.1007/s11227-005-0086-5},
 issn = {1573-0484},
 journal = {The Journal of Supercomputing},
 month = {March},
 number = {3},
 pages = {227--248},
 publisher = {Springer Science and Business Media LLC},
 title = {Implementation of the SHA-2 Hash Family Standard Using FPGAs},
 url = {http://dx.doi.org/10.1007/s11227-005-0086-5},
 volume = {31},
 year = {2005}
}

@article{Snyder1971``prisoners,
 author = {Snyder, Glenn H.},
 doi = {10.2307/3013593},
 eprint = {https://academic.oup.com/isq/article-pdf/15/1/66/5096260/15-1-66.pdf},
 issn = {0020-8833},
 journal = {International Studies Quarterly},
 month = {03},
 number = {1},
 pages = {66-103},
 title = {``Prisoner's Dilemma'' and ``Chicken'' Models in International Politics},
 url = {https://doi.org/10.2307/3013593},
 volume = {15},
 year = {1971}
}

@misc{sro2023understanding,
 author = {Vellum Labs s.r.o.},
 day = {18},
 howpublished = {Cexplorer.io},
 month = {aug},
 note = {Technical analysis of {Cardano}'s stake pool and delegation certificates, detailing the separation of payment and staking credentials.},
 title = {Understanding {Cardano} Certificates},
 url = {https://cexplorer.io/article/understanding-cardano-certificates},
 urldate = {2026-03-23},
 year = {2023}
}

@article{Stano2022review,
 author = {Stano, Peter and Loss, Daniel},
 doi = {10.1038/s42254-022-00484-w},
 issn = {2522-5820},
 journal = {Nature Reviews Physics},
 month = {August},
 number = {10},
 pages = {672--688},
 publisher = {Springer Science and Business Media LLC},
 title = {Review of performance metrics of spin qubits in gated semiconducting nanostructures},
 url = {http://dx.doi.org/10.1038/s42254-022-00484-w},
 volume = {4},
 year = {2022}
}

@article{Stewart2018committing,
 author = {Stewart, I. and Ilie, D. and Zamyatin, A. and Werner, S. and Torshizi, M. F. and Knottenbelt, W. J.},
 doi = {10.1098/rsos.180410},
 issn = {2054-5703},
 journal = {Royal Society Open Science},
 month = {June},
 number = {6},
 pages = {180410},
 publisher = {The Royal Society},
 title = {Committing to quantum resistance: a slow defence for Bitcoin against a fast quantum computing attack},
 url = {http://dx.doi.org/10.1098/rsos.180410},
 volume = {5},
 year = {2018}
}

@article{Stutz2026reuse,
 author = {St\"utz, Rainer and Stifter, Nicholas and Dragaschnig, Melitta and Haslhofer, Bernhard and Judmayer, Aljosha},
 journal = {arXiv:2601.19500},
 title = {Reuse of Public Keys Across UTXO and Account-Based Cryptocurrencies},
 url = {https://arxiv.org/abs/2601.19500},
 year = {2026}
}

@misc{Swayne2025solana,
 author = {Swayne, Matt},
 day = {4},
 howpublished = {The Quantum Insider},
 month = {jan},
 title = {{Solana} Takes A Step Toward {PQC} Era With Quantum-Resistant Vault},
 url = {https://thequantuminsider.com/2025/01/04/solana-takes-a-step-toward-pqc-era-with-quantum-resistant-vault/},
 urldate = {2026-03-23},
 year = {2025}
}

@misc{SyntheticBird2024discussion,
 author = {{SyntheticBird}},
 day = {10},
 howpublished = {Monero Research Lab ({MRL}), {GitHub} Issue \#131},
 month = {dec},
 note = {Formalizes the proposal for a 5-year post-quantum migration timeline, citing ethical responsibilities and the {Google Willow} quantum processor announcement.},
 title = {Discussion: Post-quantum security and ethical considerations over elliptic curve cryptography},
 url = {https://github.com/monero-project/research-lab/issues/131},
 urldate = {2026-03-23},
 year = {2024}
}

@inproceedings{Szabo2018smart,
 author = {Nick Szabo},
 title = {Smart Contracts: Building Blocks for Digital Markets},
 url = {https://api.semanticscholar.org/CorpusID:198956172},
 year = {2018}
}

@misc{Team2026pasta,
 author = {{Zcash Team}},
 howpublished = {GitHub Repository},
 note = {Technical specification and implementation of a cycle of prime-order elliptic curves designed for highly efficient recursive proof composition in {Halo 2}.},
 title = {The {Pasta} Curves: {Pallas} and {Vesta}},
 url = {https://github.com/zcash/pasta_curves},
 urldate = {2026-03-23},
 year = {2026}
}

\appendix

\section{Appendix: Zero-Knowledge Proof of Resource Costs}
\label{app:zkp}

In this Appendix, we provide a Zero-Knowledge (ZK) proof to substantiate the resource estimates for breaking the 256-bit Elliptic Curve Discrete Log Problem (ECDLP) described in the main text, without disclosing our improved logical circuits. We use the SP1 Zero-Knowledge Virtual Machine (zkVM)~\cite{Labs2026sp1} to generate a Groth16 Zero Knowledge Succinct Non-interactive ARgument of Knowledge (SNARK)~\cite{Groth2016size} proof of the fact that we possess a quantum circuit that correctly implements elliptic curve point addition on the secp256k1 curve~\cite{Jancar2026secp256k1, Brown2010sec} and satisfies certain resource constraints. In the subsequent sections, we also describe how these resources relate to the resources for running the full protocol for solving ECDLP on a quantum computer. Below, we state our precise ZK statements for the two circuit variants optimized for low qubits and low gate counts. For each ZK statement, we provide the verification key corresponding to our compiled Reduced Instruction Set Computer V (RISC-V) Executable and Linkable Format (ELF) binary, SHA-256 hash of our committed circuits and Groth16 proof bytes generated using SP1. The binary file containing the proof binary with public outputs can be found in our Zenodo upload~\cite{ZKP_Zenodo}. 

\subsection{Low-Qubit Circuit Variant}
\label{aps:zkp_low_qubit}
\begin{zkstatement}[Low-Qubit Variant]
We possess a quantum kickmix circuit $C_{\text{low-qubit}}$ (uniquely committed to via its cryptographic hash) with resource counts of at most:
\begin{itemize}
    \item $2,700,000$ non-Clifford gates (CCX + CCZ)
    \item $1,175$ logical qubits
    \item $17,000,000$ total operations
\end{itemize}
that correctly computes point addition on the elliptic curve \texttt{secp256k1} across all 9,024 pseudo-random inputs deterministically derived from the circuit's own hash.
\end{zkstatement}

\noindent \textbf{Circuit SHA-256 Hash:} \\
\texttt{0x5373e67ca5e900819747f8c37a4a7fa9a3ea28986835436eaa9825b12a082ff2}

\vspace{1em}
\noindent \textbf{Groth16 Proof Bytes:} \\
\begin{small}
\texttt{\seqsplit{0x0e78f4db0000000000000000000000000000000000000000000000000000000000000000008cd56e10c2fe24795cff1e1d1f40d3a324528d315674da45d26afb376e867000000000000000000000000000000000000000000000000000000000000000000e7e4e086c9f9f4e47318d5b4925cefa0efa4853719b7c5786b5bcc4272c8c132ef1b3c2193d8ad2912a81915c8789863ba3e24bf50c88963543cba35085b1ef17eba3e7eaf1e3d628171f307bc9b2b390a297625d14df336ade99fc482a232f1b3ebf5e82075429c0d55834d1e05f555f3db6174603700e0c1275a50ee029861a098d42a49655aac19ba69cdec70d87f41d56c30711683d48cb838dbbe352cc0ffe49497d676aee03fc9e11636f456014aebb15add03831c9b624ace73dd2e9025776ccc3d1de9d8eb934f0d21eea8beb450f9544046e343d5ae83e6601763d0453613c2b1c511323c75fda5192382cbcd18902551cedd849d3125af8469fad}}
\end{small}

\vspace{1em}
\noindent \textbf{Verification Key:} \\
\texttt{0x005e287a2654d72d3c9b25ecb40772be4d3b60c2c2e009535599273746390686}

\subsection{Low-Gate Circuit Variant}

\begin{zkstatement}[Low-Gate Variant]
We possess a quantum kickmix circuit $C_{\text{low-gate}}$ (uniquely committed to via its cryptographic hash) with resource counts of at most:
\begin{itemize}
    \item 2,100,000 non-Clifford gates (CCX + CCZ)
    \item 1,425 logical qubits
    \item 17,000,000 total operations
\end{itemize}
that correctly computes point addition on the elliptic curve \texttt{secp256k1} across all 9,024 pseudo-random inputs deterministically derived from the circuit's own hash.
\end{zkstatement}

\noindent \textbf{Circuit SHA-256 Hash:} \\
\texttt{0x04f17175a034cade07b0350481aab02ec4ad08254aa5d4dfd53ba217afca4f0c}

\vspace{1em}
\noindent \textbf{Groth16 Proof (Hex):} \\
\begin{small}
\texttt{\seqsplit{0x0e78f4db0000000000000000000000000000000000000000000000000000000000000000008cd56e10c2fe24795cff1e1d1f40d3a324528d315674da45d26afb376e867000000000000000000000000000000000000000000000000000000000000000001387201c4d8f17a4582424224cf57e7df14680fc14474411475f312e65b06206112b2f47088cc51c2c924fc0008eb4ade18cb371c32211143f39b0c36b216b7b11cabe1fd8faec5702b3eabba3a306fd008cfa1c61111a47541aa233271366f51f3b47d04c9f2be8cc8427ea8052ef6ec41c24747bba26c143780d7af5873d5d20be2236503b2b7af6769f48bdd72ecf243dc650c39fe080edda195e7eaadd2b055f4262ee94c5cac9c9ad26e6072500952fbf5a48e08d07dff7790a7e7c3a0e1e7982791e47a0d8b231a9e731d890092c6b9929d987d668bd06210233d10e3411a1f4be17a13e75b0aaf29d0f5b4f05ec7cd1a8c3ef4d9d7aa16a6ad1295f3e}}
\end{small}

\vspace{1em}
\noindent \textbf{Verification Key:} \\
\texttt{0x005e287a2654d72d3c9b25ecb40772be4d3b60c2c2e009535599273746390686}

\subsection{Circuit Architecture and Approximate Correctness}

First note that in this Appendix we sometimes say ``Toffoli count'', instead of ``non-Clifford count'', despite including CCX and CCZ gates in the count (technically Toffoli refers specifically to a CCX). Furthermore, in our quantum circuits, some instructions do not execute because they are conditioned upon classical bits. Because the exact number of executed CCX and CCZ gates depends on the runtime input, we report the average executed Toffoli count as measured across the 9024 evaluated test cases.

Similar to prior works~\cite{Gouzien2023performance, Litinski2023to}, our elliptic curve circuits perform a sequence of in-place windowed elliptic curve point additions, each requiring three table lookups. The core operation evaluates
\begin{equation*}
    				Q \gets Q + P[k]
\end{equation*}
where P is a (classically) pre-computed table of elliptic curve points ($P[0]$, $P[1]$, \dots, $P[2^w - 1]$), $w$ is the window size, $k$ is a $w$-qubit quantum register storing an integer in two's complement representation, and $Q$ is a $2n$ qubit target accumulator register holding a superposition of points on the elliptic curve. 
Because it is well known how to build the entire algorithm out of these $Q \mathrel{{+}{=}} P[k]$ operations, we focus on proving that this operation is approximately correct. 
This is helpful because while the overarching ECDLP algorithm relies on quantum interference, this specific point addition subroutine evaluates a purely classical boolean function. 
While it could theoretically be synthesized entirely from a standard classical reversible gate set (NOT, CNOT, and CCNOT/Toffoli gates), we use measurement based uncomputation~\cite{Jones2013lowoverhead, Gidney2019verifying} to reduce the non-Clifford gate counts.

Given the cost of a point addition, we can derive the cost of the entire ECDLP algorithm. The algorithm performs phase estimation across two variables~\cite{Shor1994algorithms, Proos2003shors}, each requiring $n=256$ controlled elliptic curve point additions. 
Similar to~\cite{Gouzien2023performance}, we can merge $w$ point additions into one windowed point addition at the cost of $3 \times 2^w$ Toffolis. Also similar to~\cite{Litinski2023to}, we replace the first windowed point addition with a direct lookup of its output and skip the last three windowed point additions by using additional classical post-processing~\cite{Ekera2019revisiting}. Thus, the total non-Clifford gate cost of the $n$-bit ECDLP circuit is

\begin{equation}
    \text{ECDLP}^{n}_\text{Toff} = \left(\text{PA}^{n}_\text{Toff} +3 \times 2^w\right)\times \left(\frac{2n}{w} - 4\right)
\end{equation}
where $\text{ECDLP}^{n}_\text{Toff}$ is the total number of Toffoli gates needed by the $n$-bit ECDLP circuit, $\text{PA}^{n}_\text{Toff}$ is the number of Toffoli gates used by the $n$-bit point addition circuit and $w$ is the window size.

Naively, we would require $2n$ additional qubits for the two phase estimation registers to obtain the qubit costs of the final ECDLP circuit. But using qubit-recycled Quantum Fourier Transform~\cite{Griffiths1996semiclassical}, it is sufficient to only maintain a single $w$ length phase register used to control the windowed point additions. Thus, the qubit cost of the final ECDLP circuit scales as 

\begin{equation}
    \text{ECDLP}^{n}_\text{Qubits} = (\text{PA}^n_\text{Qubits} + w) 
\end{equation}
where $\text{ECDLP}^{n}_\text{Qubits}$ is the total number of logical qubits needed by the $n$-bit ECDLP circuit, $\text{PA}^{n}_\text{Qubits}$ is the number of logical qubits used by the $n$-bit point addition circuit and $w$ is the window size.

Given the costs of our point addition circuit, the optimal window size is $w=16$. At this window size, a total of
\begin{equation}
    \left(\frac{2\times 256}{16} - 4\right) =  28
\end{equation} windowed point additions are performed for the $256$-bit ECDLP algorithm.

Compared to the millions of gates used by the windowed lookup point additions, there are negligible additional costs in the final circuit. For example, we use measurement based uncomputation to uncompute the table lookups~\cite{Berry2019qubitization} which has a small cost associated with it. 
Similarly, each phase estimation must be accompanied by a small amount of work associated with performing a qubit-recycled Quantum Fourier Transform~\cite{Griffiths1996semiclassical}. 
We note again that these details are not expected to materially help an adversary as they have been common ingredients of all papers on using a quantum computer to break ECDLP since 2019.

\subsection{Measurement Based Uncomputation and Kickmix Circuits}

We use measurement based uncomputation (MBUC)~\cite{Jones2013lowoverhead, Gidney2018halving, Gidney2019verifying} to clear intermediate ancilla qubits. Traditional reversible computation requires executing the inverse circuit $U^\dagger$ to cleanly disentangle the intermediate ancilla qubits used as temporary scratchpad, effectively doubling the logical gate counts. MBUC offers a cheaper, measurement driven alternative. Again, this is a common thread in all recent work and so the fact we use MBUC should not be surprising to an adversary. Suppose, in every branch of the superposition, an ancilla qubit $q$ holds a deterministic boolean function $f(k)$ of the data register, represented by the state $\sum \alpha_k\ket{k}\ket{f(k)}$. Measuring $q$ in the Pauli X basis yields two equally probable outcomes:

\begin{itemize}
    \item $\ket{+}$ (False): The qubit $q$ is successfully disentangled and erased. The system collapses to $\sum \alpha_k\ket{k}$. No additional uncomputation operations are needed.
    \item $\ket{-}$ (True): The qubit $q$ is erased but a phase kickback occurs, negating the amplitudes of the states where $f(k) = 1$. The system collapses to $\sum \alpha_k(-1)^{f(k)}\ket{k}$.
\end{itemize}
In the latter case, the uncomputation is incomplete until a phase correction is applied. This is possible because the data register $\ket{k}$ is preserved and the phase correction is a function of $k$. The cost of this phase correction is usually slightly smaller~\cite{Gidney2018halving}, and sometimes much smaller~\cite{Berry2019qubitization}, than the cost of computing $f(k)$.

We define kickmix circuits as the specific subclass of quantum circuits composed entirely of classical reversible logic gates, measurement-based uncomputation via Pauli-X basis measurements, and diagonal phasing gates for phase correction. By restricting the architecture exclusively to this gate set, we ensure the circuit never generates intractable quantum entanglement and can be efficiently simulated classically~\cite{Gidney2019verifying}. The etymology of \emph{kickmix} derives from the circuit's ability to support phase \emph{kick}back and \emph{mix}ing classical states.

\subsection{Verifiable Fuzz Testing via the Fiat-Shamir Heuristic}

Shor's algorithm functions well even if a small percentage of values in the superposition are wrong  ---  a superposition with 1\% of points in the wrong place will cause the algorithm to fail at most 1\% of the time~\cite{Proos2003shors, Zalka2006shors, Gidney2025to, Gidney2025to}. Consequently, it is sufficient for us to prove our circuits are approximately correct rather than exactly correct.

For the proof of approximate correctness, we use fuzz testing.
We prove 9024 random inputs, chosen by the Fiat-Shamir heuristic~\cite{Fiat1987to}, are mapped to the correct output.
This is sufficient for 128 bits of cryptographic security that the circuit is acting correctly on at least 99\% of inputs, because if the circuit mapped more than 1\% of inputs to the wrong output then the probability of it passing 9024 independent random tests without a single failure is at most $(1 - 0.01)^{9024} \approx  2^{-130}$. To implement the Fiat-Shamir heuristic, we use the SHAKE256 extendable-output function (XOF)~\cite{NISTUS2015sha3} seeded with the bytes of the secret circuit as a Cryptographically Secure Pseudo Random Number Generator (CSPRNG).
Each test point is generated as $G \cdot k$ where $G$ is the curve's generator and $k$ is a random 256 bit integer sampled using this CSPRNG.

\subsection{Cryptographic Attestation via SP1 and Groth16 SNARK}

To provide a publicly verifiable proof that this Fiat-Shamir fuzz testing was executed honestly on a circuit $C$ satisfying our claimed resource bounds, without publishing $C$ itself, we simulate $C$ inside the SP1 Zero-Knowledge Virtual Machine (zkVM)~\cite{Labs2026sp1}. SP1 acts as a standard RISC-V processor that mathematically arithmetizes its own execution. We implement a guest program to be run inside the SP1 zkVM consisting of our kickmix circuit simulator, the Fiat-Shamir testing framework, and the secp256k1 elliptic curve point addition logic in standard Rust, which SP1 compiles directly into RISC-V bytecode. 

Inside the zkVM, the guest program performs four critical tasks. 
First, it computes a SHA-256 hash of the input circuit and commits it as a public output, along with the demanded resource counts that the input circuit must satisfy. 
Second, it uses the SHAKE256 extendable-output function (XOF)~\cite{NISTUS2015sha3} initialized with the bytes of the secret circuit to generate 9024 test inputs via the Fiat-Shamir transform, as described in the section above. 
Third, it simulates the parsed circuit using the kickmix simulator on the 9024 Fiat-Shamir derived test cases to prove functional correctness of the parsed circuit. 
The kickmix simulator uses the XOF initialized above for input test generation, as a pseudo-random number generator (PRNG) to simulate quantum measurements in MBUC. 
The simulator also records the total number of non-Clifford gates (CCX+CCZ) executed by the circuit across the 9024 test runs. 
Finally, the program asserts that the input private circuit satisfies the demanded resource counts  ---  verifying that $C_{\text{low-qubit}}$ executes at most $2,700,000$ Toffoli gates and requires at most $1175$ logical qubits and $C_{\text{low-gate}}$ executes at most $2,100,000$ Toffoli gates and requires at most $1425$ logical qubits.

As SP1 simulates the guest program, it records the state of the CPU at every clock cycle into an execution trace matrix. To cryptographically attest to this multi-billion-cycle execution, SP1 divides the computation into shards of roughly $2^{22}$ RISC V instructions, generates a STARK proof~\cite{BenSasson2018scalable} of each shard, and leverages recursive proof composition~\cite{Valiant2008incrementally, Goldberg2021cairo} to generate a final compressed STARK proof. However, since individual STARK proofs in SP1 do not currently satisfy the Zero Knowledge Property~\cite{Labs2026groth16}, we invoke the SP1 prover in Groth16 mode which wraps the compressed hash-based STARK proof into a Groth16 zk-SNARK~\cite{Groth2016size} and ensures Zero Knowledge and Succinctness. The resulting zero-knowledge proof artifact is published alongside our prover and verifier code in the accompanying Zenodo dataset~\cite{ZKP_Zenodo}.

Finally, we note a practical irony in our choice of cryptographic attestation. Because the Groth16 SNARK relies on pairing-friendly elliptic curves, its soundness is inherently vulnerable to the very quantum attacks analyzed in this work. In principle, one could use a sufficiently large, fault-tolerant quantum computer to forge this zero-knowledge proof. However, since cryptographically relevant quantum computers do not exist as of today, the soundness of our proof remains intact.

\subsection{Code Availability for Zero-Knowledge Proof}
The source code for generating the Zero-Knowledge (ZK) proofs, along with all associated artifacts, is archived and publicly available in our Zenodo repository~\cite{ZKP_Zenodo}. 
We note that the SHA256 hash values of our circuits changed between v1 and v2 of this paper. This is because we switched from computing the hash of the serialized bytes of the parsed circuit object (v1) to hashing the contents of the circuit files directly (v2). The underlying circuits themselves remain identical, and the old hashes can still be verified by rolling back to v1 of the repository. We believe the new approach is a more intuitive choice, and we refer readers to our Zenodo uploads to view the difference in implementation
\end{document}